\documentclass[10pt,twocolumn,letterpaper]{article}

\usepackage[]{cvpr}      % To produce the REVIEW version
\usepackage{algorithm}
\usepackage{algorithmic}
\usepackage{array}
\usepackage{xcolor}
\usepackage[table]{xcolor}
\usepackage{adjustbox}
\usepackage{amsthm}
\usepackage{booktabs}
\usepackage{multirow}

\def\bth{{\boldsymbol{\boldsymbol \theta}}}

\def\bxi{{\boldsymbol{\xi}}}

\def\b1{{\boldsymbol{1}}}
\def\c1{{\textcircled{a}}}

\def\bb{{\boldsymbol{b}}}

\def\bd{{\boldsymbol{d}}}

\def\bt{{\boldsymbol{t}}}

\def\bx{{\boldsymbol{x}}}
\def\by{{\boldsymbol{y}}}
\def\bz{{\boldsymbol{z}}}
\def\bA{{\mathbf{A}}}
\def\bC{{\boldsymbol{C}}}
\def\bD{{\boldsymbol{D}}}
\def\bE{{\boldsymbol{E}}}
\def\bF{{\mathbf{F}}}

\def\bI{{\mathbf{I}}}

\def\bB{{\mathbf{B}}}

\def\bS{{\boldsymbol{S}}}

\definecolor{cvprblue}{rgb}{0.21,0.49,0.74}
\usepackage[pagebackref,breaklinks,colorlinks,allcolors=cvprblue]{hyperref}

\def\paperID{17982} % *** Enter the Paper ID here
\def\confName{CVPR}
\def\confYear{2026}

\title{Annealed Langevin Posterior Sampling (ALPS): A Rapid Algorithm for Image Restoration with  Multiscale Energy Models}
\author{Jyothi Rikhab Chand\\
University of Virginia\\
{\tt\small jyothi-rikhabchand@virginia.edu}
\and
Mathews Jacob\\
University of Virginia\\
{\tt\small mjacob@virginia.edu}
}

\begin{document}
\maketitle
\begin{abstract}
Solving inverse problems in imaging requires models that support efficient inference, uncertainty quantification, and principled probabilistic reasoning. Energy-Based Models (EBMs), with their interpretable energy landscapes and compositional structure, are well-suited for this task but have historically suffered from high computational costs and training instability. To overcome the historical shortcomings of EBMs, we introduce a fast distillation strategy to transfer the strengths of pre-trained diffusion models into multi-scale EBMs. These distilled EBMs enable efficient sampling and preserve the interpretability and compositionality inherent to potential-based frameworks. Leveraging EBM compositionality, we propose Annealed Langevin Posterior Sampling (ALPS) algorithm for Maximum-A-Posteriori (MAP), Minimum Mean Square Error (MMSE), and uncertainty estimates for inverse problems in imaging. Unlike diffusion models that use complex guidance strategies for latent variables, we perform annealing on static posterior distributions that are well-defined and composable. Experiments on image inpainting and MRI reconstruction demonstrate that our method matches or surpasses diffusion-based baselines in both accuracy and efficiency, while also supporting MAP recovery. Overall, our framework offers a scalable and principled solution for inverse problems in imaging, with potential for practical deployment in scientific and clinical settings. ALPS code is available at the GitHub repository \href{https://github.com/JyoChand/ALPS}{ALPS}.
\end{abstract}    
\section{Introduction}
\label{sec:intro}
Inverse problems in imaging—such as inpainting, super-resolution, and MRI reconstruction—require models that support efficient, accurate, and interpretable inference. These tasks often demand posterior sampling, uncertainty quantification, and Maximum-A-Posteriori (MAP) estimation, which are challenging for current generative models. Diffusion models have achieved state-of-the-art performance in image generation \cite{song2020score,kingma2021variational} and have been adapted for inverse problems \cite{dps,daps,daras2024survey}. However, the lack of a global energy landscape makes compositional reasoning and controlled generation difficult. In particular, the conditional probabilities $p(\bx_t|\bx_{t+\delta})$ and noise schedules in diffusion models are optimized for prior sampling; naively adding likelihood terms destabilizes dynamics in inverse problems. Current methods design step-wise guidance for stability~\cite{dps,daps,daras2024survey}, but require backpropagation or ODE integration, increasing complexity. Likelihood evaluation is also costly, involving high-dimensional reverse-time ODEs~\cite{song2021score,kingma2021variational}.

Energy-Based Models (EBMs), with their explicit energy landscapes and conservative score fields, offer a principled probabilistic framework well-suited for inverse problems ~\cite{lecun2006tutorial, song2021score}. Their compositionality enables modular integration of priors and likelihoods \cite{du2019implicit,lecun2006tutorial}, allowing direct posterior sampling and MAP estimation without modifying the underlying dynamics \cite{hurault2022gradient,song2023potential,muse}. However, EBMs have historically faced challenges in training stability and sampling efficiency, limiting their scalability and adoption in large-scale inverse problems. For instance, maximum likelihood training requires Markov Chain Monte Carlo (MCMC) sampling ~\cite{song2021score, du2019implicit}, resulting in slow convergence and instability associated with  adversarial training. Recent approaches use Denoising Score Matching (DSM) to overcome the stability issues \cite{hurault2022gradient,song2023potential,muse,habring2025energy}. However, the  evaluation of the gradient of the EBMs (a.k.a. score) involves backpropagation, which translates to higher computational costs than diffusion models. 

\begin{figure*}[t!]
\centering
  \includegraphics[width=\linewidth]{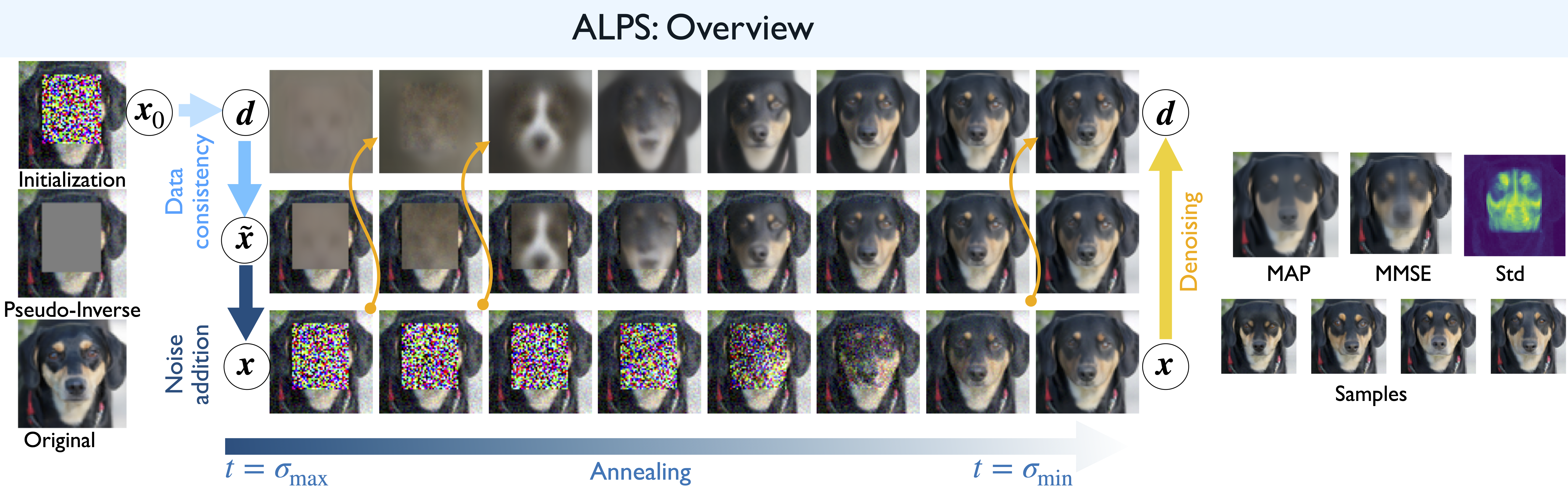}\vspace{-1em}
   \caption {Overview of the proposed Annealed Langevin Posterior Sampling (ALPS) algorithm for inverse problems using multi-scale EBM regularizers. Please see Algorithm 1 and text for details. Current diffusion-based inverse problem solvers traverse through samples from annealed prior distributions $p_t(\bx)$, which correspond to noise corrupted images. These schemes carefully add data consistency terms so that the trajectory does not deviate from the learned paths. The compositional property of EBMs enable us to \textbf{define} a family of time-dependent posterior distributions, each of which are sampled using preconditioned Langevin dynamics. The samples in these distributions are NOT noise perturbed versions from the time-dependent priors, but from well defined posterior distributions (bottom row marked by $\bx_t \sim p_t(\bx|\by) :\propto p(\by|\bx)p_t(\bx)$). Breaking from the conventional diffusion sampling setting enables simpler algorithms that alternate between denoising using the score model at a specific scale to obtain ${\bd_t}$ from $\bx_t$, data consistency enforcement using quadratic optimization to obtain $\tilde{\bx_t}$ from $\bd_t$, and forward-model dependent noise addition (more noise in the null-space than range space of $\bA$) to yield the posterior samples $\bx_t$. The posterior samples $\bx_{t}$ derived after $K$ iterations at one scale are used as initialization for Langevin dynamics at the next scale $p_{t+\delta}$, denoted by the curved yellow arrows. This approach results in a smoother trajectory from a good initialization $\bx_0$ (top left corner) related to the least square solution to the final one, unlike most diffusion inverse solvers that start with pure noise. The EBM in study was distilled from a diffusion model, trained on the AFHQ ($64 \times 64$) dataset. }\vspace{-0.5em}
  \label{overview}
\end{figure*}

We introduce a distillation strategy to transfer the strength of diffusion models to potential models. Specifically, the score distillation strategy learns the conservative component of a pre-trained diffusion score field using an EBM.This approach results in competitive generative performance, while retaining the probabilistic interpretations of EBM and compositionality. On CIFAR-10, we obtain FID = 3.35 with 35 NFEs. 

Leveraging the compositionality of EBMs, we introduce a novel framework for inverse problems that enables principled posterior sampling and MAP estimation. We define a family of static posterior distributions parameterized by $t$, which converges to the true posterior as $t \to 0$.  These posteriors are sampled using Annealed Langevin Posterior Sampling (ALPS), which alternates between denoising via the EBM score, enforcing data consistency through quadratic optimization, and adding noise. This annealing strategy yields efficient inference with fewer steps and avoids backpropagation through the score or forward model. For MAP estimation, we employ a Majorization Minimization (MM) approach that sequentially minimizes the posterior energy using surrogate functions, ensuring monotonic descent and further reducing computational cost. As shown in Fig.\ref{overview} and Algorithm.1, both ALPS and MAP estimation operate on well-defined static posteriors, enabling smooth trajectories from initial estimates to final solutions. Our method achieves competitive performance on inverse tasks such as MRI reconstruction and inpainting, while supporting MMSE recovery, uncertainty estimation, and MAP inference. Compared to diffusion-based approaches that rely on stepwise data-consistency guidance and costly reverse-time dynamics, our method offers a more natural and scalable alternative for scientific and medical imaging.

% \paragraph{Contributions.}
% \begin{enumerate}
%     \item We introduce distillation-based training strategies to significantly accelerate the training of potential models. By transfering the information from diffusion models, the distilled models can achieve state-of-the-art FID, outperforming prior EBM approaches. On CIFAR-10, we reach \textbf{FID = 3.89} with \textbf{11 NFEs}.
%     \item We introduce a novel and principled algorithm for image recovery in inverse problems. Unlike diffusion models that operate on latent variables $\bx \sim p_t(\bx)$ and impose data consistency through guidance terms, we perform annealing on static posterior distributions that are well-defined and composable. This results in fast and conceptually simpler algorithms for posterior sampling and MAP estimation. 
% \end{enumerate}

%-------------------------------------------------------------------------

\section{Related Work}
\subsection{Energy-Based Models } 
Unnormalized EBMs define an explicit energy function $E_\bth(\bx)$ over data $\bx$, inducing a distribution ~\cite{lecun2006tutorial}:
\[
p_\bth(\bx) \propto \exp(-E_\bth(\bx))
\]
Its gradient $\nabla_x E_\bth(\bx)$ forms a conservative score field, enabling probabilistic inference and uncertainty estimation. Such models have been introduced in inverse problems \cite{hurault2022gradient,song2023potential}, while multi-scale EBMs were introduced to improve convergence \cite{muse}. Despite the above advantages, EBMs are rarely used at scale due to challenges in training them. Maximum likelihood requires MCMC sampling, which is slow and unstable~\cite{du2019implicit}. 
DSM-based training for single and multi-scale EBMs~\cite{song2023potential,hurault2022gradient,muse} improves stability but remains costly because score evaluation needs backpropagation and suffers from high variance of DSM loss. Current EBM approaches for inverse problems \cite{hurault2022gradient,song2020score,muse} often relied on smaller models and datasets, limiting their performance compared to larger diffusion models. Recent works explore implicit sampling and cooperative training~\cite{coopnets}, yet EBMs still lag behind diffusion models in sample quality and efficiency.

\subsection{Inverse Problems with Generative Models.} Generative priors have been widely used for inverse problems such as inpainting, super-resolution, and MRI reconstruction. Diffusion-based approaches dominate recent literature, but they require solving high-dimensional ODEs and backpropagating through forward models, leading to high computational cost. EBMs offer a promising alternative due to their exact compositionality: given a prior $E_\bth(\bx)$ and likelihood $E_\ell(\bx)$, the posterior energy is simply $E(\bx) = E_\bth(\bx) + E_\ell(\bx)$, allowing Langevin sampling without modifying the dynamics. Our work leverages this property to design efficient posterior sampling and MAP optimization algorithms, achieving competitive results with fewer steps and reduced complexity.

\subsection{Diffusion Models for inverse problems} Diffusion models learn the score field $\nabla_\bx \log p_t(\bx)$ for noise-perturbed distributions $\{p_t(\bx)\}_{t\in[\sigma_{max},\sigma_{min}]}$ via DSM ~\cite{song2020score}. They approximate reverse Markov transitions $p(\bx_{t-\delta}\!\mid\!\bx_t)$ under a Gaussian assumption~\cite{song2020score,kingma2021variational}, with iterative sampling corresponding to a reverse-time SDE. The noise schedule provides an optimized path from $p_{\sigma_{max}}(\bx)$ to $p_{\sigma_{min}}(\bx)$. Diffusion models achieve state-of-the-art image quality and have been widely applied to inverse problems~\cite{dps,daps,daras2024survey}. However, adding data-consistency terms often destabilizes reverse dynamics, requiring complex guidance strategies, many iterations, and backpropagation through score or forward models~\cite{daras2024survey}. Moreover, exact likelihood evaluation is computationally prohibitive, involving high-dimensional reverse-time ODEs and intractable divergence measures~\cite{kingma2021variational}.
\subsection{Diffusion distillation methods}
Several works accelerate sampling by distilling multi-step diffusion into one- or few-step generators~\cite{salimans2022progressive,song2023consistency,meng2023distillation,feng2024rdd}, enabling fast image generation. However, these methods lose the probabilistic interpretation of diffusion models, hindering uncertainty estimation and Bayesian reasoning. This makes them less suitable for inverse problems, where posterior sampling and likelihood evaluation are critical. Moreover, incorporating data-consistency or likelihood guidance is non-trivial since the learned mapping is deterministic. Our score distillation (Sec.~\ref{score_distillation}) shares conceptual similarities with~\cite{thornton2025energy}, but differs in formulation and focus: their work targets temperature-controlled sampling and composition, while ours addresses inverse problems. A comparison of current models in probabilistic inference and inverse problems is provided in Table 1. \begin{table}[ht]
\label{table1}
\centering
\caption{Utility of generative models in the inverse problem setting}
\scriptsize
\begin{tabular}{|c|c|c|c|c|}
\hline
\textbf{Distilled} & \textbf{Prior} & \textbf{Likelihood} &\textbf{Guidance in} & \textbf{Composit-}\\
\textbf{Models} & \textbf{Sampling} & \textbf{Evaluation} & \textbf{Inverse} & \textbf{ionality}\\
\textbf{} & \textbf{NFE} & \textbf{} & \textbf{problems}& \\
\hline
Diffusion Models & 50--1000 & Expensive & Expensive& No \\
Consistency Models & 1--4 & No & Non-trivial& No \\
Progressive Distill. & 1--4 & No & Non-trivial& No \\
Rectified Flow & 1--4 & No & Non-trivial& No \\
Traditional EBMs & 100--500 & Yes & Yes& Yes \\
\textbf{Distilled EBMs} & 11--35 & Yes & Yes& Yes \\
\hline
\end{tabular}
\end{table}

\section{Methodology}
\label{sec:method}
The main objective of this paper is to use EBMs for inverse problems. As discussed previously, prior methods \cite{hurault2022gradient,song2023potential,muse} rely on smaller models, which were not competitive with respect to large diffusion models. To make EBMs competitive, we first distill EBMs from pre-trained diffusion models. This approach preserves the sample quality and efficiency of diffusion models, while retaining the benefits of potential models, including compositional reasoning and principled probabilistic sampling. 

\subsection{Multi-scale Energy Based Model}
We define a multi-scale EBM prior over $\bx$ at noise level $t$ as:
\begin{equation}
    p_\bth(\bx;t) = \frac{1}{Z_\bth} \exp\!\left(-\frac{E_\bth(\bx;t)}{t^2}\right),
    \label{eq:prior}
\end{equation}
where $Z_\bth$ is the partition function and $E_\bth(\bx;t)$ is the energy function. We parameterize the energy as:
\begin{equation}
    E_\bth(\bx;t) = \frac{1}{2}\|\bx - D_\bth(\bx;t)\|^2,
    \label{eq:energy}
\end{equation}
where $D_\bth:\mathbb{R}^{n+1}\!\rightarrow\!\mathbb{R}^n$ is a CNN. In this work, we ignore the impact of the partition function $Z_{\theta}$. This model has conceptual similarities to \cite{li2023multiscale}. 

The unnormalized $E_\bth$ may be learned from data using DSM~\cite{song2020score}. For a noisy input $\tilde{\bx} = \bx + t \bz$ with $\bz \sim \mathcal{N}(0,I)$, DSM minimizes:
\begin{align}
\theta^* %&= \arg\min_\bth \mathbb{E}_{\sigma,\bx,\bz}\!\Big[w(\sigma)\big\|-\frac{\nabla_{\tilde{\bx}}E_\bth(\tilde{\bx};\sigma)}{\sigma^2} + \frac{\bz}{\sigma}\big\|^2\Big] \nonumber\\
&= \arg\min_\bth \mathbb{E}_{t,\bx,\bz}\!\Big[w(t)\big\|\nabla_{\tilde{\bx}}E_\bth(\tilde{\bx};t) - t\bz\big\|^2\Big]
\label{eq:dsm}
\end{align}
where $\tilde{\bx} = \bx+ t \bz$, $\bz \sim \mathcal{N}(0,\bI)$, and $w(t)$ are the weights. 

One can define a new denoiser using the gradient of the EBM:
\begin{align}\label{eq:jvp}
\tilde{D}_\bth(\bx;t) &= \bx - \nabla_\bx E_\bth(\bx;t) \\\nonumber
&= \bx - \big[(\bx - D_\bth(\bx;t)) - J_D^\top(\bx - D_\bth(\bx;t))\big],
\end{align}
where $J_D$ is the Jacobian of $D_\bth$. The vector-Jacobian product is computed using PyTorch's \texttt{autograd}. %Substituting \eqref{eq:jvp} into \eqref{eq:dsm} yields an equivalent loss:
%\begin{equation}
%\theta^* = \arg\min_\bth \mathbb{E}_{t,\bx,\bz}\!\Big[w(t)\|\tilde{D}_\bth(\tilde{\bx};t) - \bx\|^2\Big]
%\label{eq:loss}
%\end{equation}

\subsection{Preconditioning for Stability}
We adapt the preconditioning strategy from \cite{karras2022edm}), which improves numerical stability and sampling performance. Note that the vector $\bx -\bD_\bth(\bx;t)=t \bz $ has variance $t^{2}$, which when used to compute the vector-Jacobian product in \eqref{eq:jvp} can cause numerical instability. Using Jacobian linearity, we precondition as:
\begin{equation}
\tilde{D}_\bth(\bx;t) = \bx - \Big[(\bx - D_\bth(\bx;t)) - t J_D^\top\!\Big(\frac{\bx - D_\bth(\bx;t)}{t}\Big)\Big]
\label{eq:denoiser_energy}
\end{equation}
We parameterize $D_\bth$ with scale-dependent normalization \cite{karras2022edm}:
\begin{equation}
D_\bth(\bx;t) = c_{\rm skip}(t)\bx + c_{\rm out}(t)\bF_\bth(\cdot)(c_{\rm in}(\sigma)\bx; c_{\rm noise}(t)),
\label{eq:denoiser_score}
\end{equation}
where $\bF_\bth(\cdot)$ is a deep neural network, $ c_{\rm{in}}(t)$ and  $ c_{\rm{out}}(t)$ ensures that the input and output signal magnitude is of unit variance, $ c_{\rm{skip}}(t)$ is chosen such that the error made by $F_\bth(\cdot)$ is not amplified. The values of these parameters are chosen as in \cite{karras2022edm}. Substituting \eqref{eq:denoiser_energy} in \eqref{eq:dsm} results in:
\begin{eqnarray}\label{eq:loss1}
  \theta^{*}  
  &=& \arg \min_{\theta} \mathbb{E}_{t,\bx, \bz}\bigg[ w(t) \| \tilde{\bD}_\bth(\tilde{\bx};t) - \bx\|^{2}\bigg]
\end{eqnarray}
 
\subsection{Distilling multiscale energy models}\label{score_distillation}
 Because the score involves an auto-grad, the direct use of \eqref{eq:loss1} to train energy models is more computationally expensive that diffusion models. In addition, the high variance of the DSM objective necessitates large number of iterations, which makes it challenging to train competitive EBM models for large datasets. We propose two distillation strategies to reduce the variance and thus overcome the above limitations. 
Diffusion score vector fields are not constrained to be conservative. By contrast, we know that the true score as well as the one defined by the EBM $\nabla E_{\bth}(\bx;t)$ are conservative vector fields. We propose to learn the potential by matching its score to that of the diffusion model; this approach amounts to learning the conservative (curl-free) component of the diffusion model:
\begin{eqnarray}\label{eq:loss2}
       \arg \min_{\theta} \mathbb{E}_{t,\bx, \bz}\bigg[ w(t) \| \tilde{\bD}_\bth(\tilde{\bx};t) -{\bD}_{\rm{teacher}}(\tilde{\bx};t) \|^{2}\bigg]
\end{eqnarray}
Note that compared to loss \eqref{eq:loss1}, the regression target in \eqref{eq:loss2} is changed. In particular, \eqref{eq:loss2} learns a denoiser $\tilde{\bD}_\bth(\tilde{\bx};t)$ derived using a conservative score and matches it with the diffusion model which may not be conservative. The loss in \eqref{eq:loss2} ensures that the rotation-only component of a vector field is orthogonal to the gradient of the energy \cite{thornton2025energy}. 

Once trained, prior samples can be generated by solving the following ODE:
\begin{eqnarray}
    \label{eq:ODE}\nonumber
    \dfrac{d\bx}{d t} = -t \nabla_\bx \log  p_\bth \left(\bx;t \right) = \dfrac{ \nabla_\bx  E_\bth (\bx;t)}{t} 
\end{eqnarray}
Following the EDM parameterization, we choose $\sigma_{\rm max}=t_0 > t_1 > \cdots t_{N} = \sigma _{\rm{min}}\approx 0$.  We use Heun's method to solve the above ODE.

The prior samples can be generated using transport distillation by solving the following ODE:
\subsection{EBM regularization of inverse problems}
%\begin{eqnarray}\label{mm_up}
%        \hat{\bx}_{n+1} = \left(\dfrac{\bA^{H}\bA}{\eta^{2}}+\dfrac{L}{\sigma^{2}}\bI\right)^{-1}\left(\dfrac{\bA^{H}\bb}{\eta^{2}} + \dfrac{L \bx_{n} - \nabla_{\bx} E_{\bth}(\bx_{n},  \sigma)}{\sigma^{2}}\right)\\\nonumber
%        \bx_{n+1} = \hat{\bx}_{n+1} +\sigma  \bz_{n+1} 
%\end{eqnarray}   
%we choose $\eta^{2}=\sigma^{2}$, $L=1$.

 We will use the compositionality property of EBMs to introduce a \emph{modular} approach for regularizing inverse problems using multi-scale EBMs. Let $\by$ be measurements obtained from $\bx$ via a known forward model with likelihood $p(\by|\bx)$. Ideally, we could combine $\log p(\bx)$ with the log-likelihood term to obtain the true posterior. However, when the data is localized to a low-dimensional manifold in high dimensional space, the gradient of the posterior will be zero, away from the manifold. To improve convergence of the algorithm, we use an annealing approach to start with a smoothed approximation of the posterior to the true one. 

Using the compositionality of EBMs, we \textbf{define} a sequence of posterior distribution:
\begin{eqnarray}
\label{tdeppost}
 -\log p_{t}(\bx|\by) = C_{t}(\bx)=\underbrace{\frac{\|\bA \bx-\by\|^2}{2\eta^2}~ }_{-\log p(\by|\bx)}+\underbrace{\frac{E_{\theta}(\bx;t)}{2t^2}}_{-\log p_{t}(\bx)} 
\end{eqnarray}
which converges to the desired posterior as $t \rightarrow \sigma_{\rm min}$. 

Th annealing of the posteriors is a key conceptual difference with current diffusion-based inverse problem solvers, which add carefully designed guidance term to the reverse-time dynamics specified  $p(\bx_t|\bx_{t+\delta})$ such that the samples stay in $\bx \sim p_t(\bx)$. These approach often requires backpropagation through the score models or careful approximations \cite{daras2024survey}. By contrast, our approach samples directly from $p_{t}(\bx|\by)$, which are well-defined posterior distributions. We note that the samples from $p_{t}(\bx|\by)$ are not noisy versions of $\bx$ learned by the EBM, but rather posterior-consistent estimates at each scale. See Fig. \ref{overview} for an example. We use annealing/continuations strategies \cite{kirkpatrick1983optimization,seguin2023continuation} that are widely used in optimization and sampling, where we start from $p_{\sigma_{\rm max}}(\bx|\by)$ and converge to the desired posterior $p_{\sigma_{\rm min}}(\bx|\by)$. 

We note that at the highest noise scale i.e., $\sigma_{\rm max}$, the prior can be approximated as $p_{\sigma_{\rm max}}(\bx)=\mathcal N\big(\bx|0,\sigma_{\rm max}^2 \bI\big)$, and hence $C_{\sigma_{\rm max}}(\bx) = \|\bA\bx-\by\|^2/2\eta^2 + \|\bx\|^2/2\sigma_{\rm max}^2$.
The samples from this smooth quadratic posterior distribution 
\begin{equation}
\label{initial}
    p_{\sigma_{\rm max}}(\bx|\by) \approx \mathcal N\left(\left(\bA^T\bA\right)^{\dag}\bA^T \by,~ \eta^2\left(\bA^T\bA\right)^{\dag}\right)
\end{equation}
are close approximations of the true samples; see Fig. \ref{overview} for an example. In particular, the mean of the distribution is the pseudo-inverse solution. The noisy samples can be derived either analytically or can be derived from the quadratic posterior using a short-run Langevin iteration. This is another key difference from diffusion models, where the samples are often initialized with $\mathcal N(0,\sigma_{\rm max}^2\bI)$, translating to longer
inference times.

\subsection{Annealed Langevin Posterior Sampling (ALPS)}
\label{sec:alps}
We propose to use annealed preconditioned Langevin dynamics \cite{baldassari2025preconditioned} to derive samples from $p_{{t}}(\bx|\by)$. The general preconditioned Langevin dynamics step is specified by:
\[
\bx_{k+1} = \bx_k - \bB ~\nabla_{\bx_{k}} C_{{t}}(\bx_k) + \sqrt{2\bB}\,\xi_k; \quad \forall k=0,\cdots K
\] 
where $\xi_k \sim \mathcal{N}(0,\bI)$ and $\bB$ is an appropriate preconditioner. 
The gradient of $C_{{t}}(\bx)$ is specified by: 
\begin{eqnarray}
\label{gradt}
\nabla_{\bx} C_{{t}}(\bx) = \frac{1}{\eta^2} \bA^\top(\bA\bx - \by) + \frac{1}{t^2} \nabla_{\bx} E_{\bth}(\bx;t)
\end{eqnarray}
Many of the forward models are diagonal (e.g. inpainting) or diagonizable in the Fourier domain (e.g. deblurring, single channel MRI). In these cases, we choose the preconditioner as:  
\begin{equation}
\label{preconditioner}
\bB = \left( \frac{\bA^\top \bA}{\eta^2} + \frac{\bI}{t^2} \right)^{-1}    
\end{equation}
to obtain the preconditioned Langevin update: 
\begin{eqnarray}\label{noiseaddition}
\bx_{k+1,t} &=& \tilde \bx_{k+1,t}~~ +~~ \underbrace{\mathbf B^{1/2}\,\xi_k}_{\bz_k};~~~\xi_k \sim \mathcal{N}(0,\bI)
\end{eqnarray}
where \vspace{-1em}
\begin{equation}\label{cgsolution}
\tilde \bx_{k+1,t} = \mathbf B \Bigg( \frac{\bA^\top \by}{\eta^2} + \frac{ \bd_{k,t}}{t^2} \Bigg)
\end{equation}
Here, $\bd_{k,t}$ is the denoised version of $\bx_{k,t}$, derived using the score of the distribution at $t$:
\begin{equation}
\label{denoising1}
\bd_{k,t}=\tilde{D}_{\theta}(\bx_{k,t}) = \bx_{k,t} - \nabla E_{\theta}(\bx_{k,t};t)
\end{equation}
is defined in \eqref{eq:jvp}. Here, $\bx_{k,t}$ denotes the k-th iterate at noise scale $t$. We can use conjugate gradients to realize ${\tilde \bx}_{k+1,t}$ in \eqref{cgupdate}:
\begin{equation}
\label{cgupdate}
{\tilde \bx}_{k+1,t} = \arg \min_{\bx} \frac{\|\bA\bx-\by\|^2}{2\eta^2}  + \frac{\|\bx- \bd_{k,t}\|^2}{2~t^2},   
%{\tilde \bx}_{(k+1,i)} = \arg \min_{\bx} \frac{\|\bA\bx-\by\|^2}{2\eta^2}  + \frac{\|\bx- \tilde{D}_{\theta}(\bx_{(k,i)},t_i)\|^2}{2(t_i)^2},   
\end{equation}

% where $\widehat\bx_{(k+1,t)}$ may be seen as the denoised version of $\bx_{k,t}$ using the score $\nabla E_{\theta}(\bx_{(k,t)};t)$
% \begin{equation}
% \label{scoreupdate}
% \widehat\bx_{(k+1,t)}=\bx_{k,t}-\nabla E_{\theta}(\bx_{(k,t)};t)  
% \end{equation}
We note that \eqref{preconditioner} is only one choice for the preconditioner. When the forward model is not diagonalizable (e.g. multichannel MRI), efficient diagonal preconditioners can be used as described in the supplementary material. Another alternative is to use \eqref{preconditioner} with samples $\bz_k \sim \mathcal N(0, (\frac{\bA^\top \bA}{\eta^2} + \frac{\bI}{t^2} )^{-1})$ derived using a short-run preconditioned Langevin scheme, as described in the supplementary material. In the high measurement  noise setting, $\eta^2 >>t^2$, the second term within the square root of \eqref{preconditioner} dominates, when we may also approximate the update as $
\bx_{k+1,t} \approx {\tilde \bx}_{k+1,t} ~+~ t~\bxi_k$. The annealed preconditioned Langevin dynamics algorithm is shown in Algorithm \ref{alg:ALPS} and illustrated in Fig. \ref{overview}.

\begin{algorithm}[t!] % [H] can also be used for exact placement
    \caption{Annealed Langevin Posterior Sampling (ALPS) or maximum-a-posteriori (MAP) estimation}
    \label{alg:ALPS}
    \begin{algorithmic}[1] % [1] enables line numbering
        \STATE \textbf{Input:} Data $\by$, forward operator $\bA$, noise level $\eta$, schedule $t$, iterations $K$ at each scale, score model $\nabla_{\bx} E_\bth(\bx; t)$\\
        
        \STATE \underline{Initialization:} $\bx_{0,t_0}  \sim p_{t_0}(\bx|\by)$ from \eqref{initial}
        \FOR {$i = 0$ to $N$} 
            \FOR{$k = 0$ to $K$}\vspace{0.3em}
                \STATE 
                \underline{Denoising:} $\bd_{k,t_i}\gets \tilde{D}_{\theta}(\bx_{k,t_i},t_i)$ as in \eqref{denoising1}\vspace{0.3em}
                %\widehat{\bx}_{(k+1,t)} \gets \bx_{(k,t)} - \nabla E_\bth(\bx_{(k,t)}; \sigma(t))$
                \STATE \underline{Data-consistency:}  $\tilde{\bx}_{k+1,t_i}$  as in \eqref{cgupdate}\vspace{0.3em}\vspace{0.3em}
                \IF {MAP is true}
                        \STATE   $\bx_{k+1,t_i} \gets \tilde{\bx}_{k+1,t_i}$\vspace{0.3em}
                \ELSIF {ALPS is true}\vspace{0.3em}
                        \STATE \underline{Noise addition:} $\bx_{k+1,t_i} \gets \tilde{\bx}_{k+1,t_i} + \mathbf B^{1/2}\zeta_k$      \vspace{0.3em} 
                \ENDIF\vspace{0.3em}
            \ENDFOR
            \STATE \underline{Initialize next scale:} $\bx_{0,t_{i+1}} \gets \tilde{\bx}_{K,t_i}$
        \ENDFOR
        \STATE Return $\bx_{K,N}$
    \end{algorithmic}
\end{algorithm}

\subsection{Maximum-A-Posterior estimate}
Similar to the sampling approach in the above subsection, we propose to sequentially minimize the $C_t(\bx)$ using a Majorization Minimization (MM) approach. By the descent lemma for an $L$-Lipschitz gradient, we have:
\begin{eqnarray}\nonumber
E_{\theta}(\bx,t) \le E_{\theta}(\bx_k,t)+ \nabla E_{\theta}(\bx_{k},t)^\top (\bx - \bx_{k},t) 
%\\&&\qquad 
+ \frac{L}{2}\|\bx - \bx_{k}\|^2
\end{eqnarray}
which provides a surrogate function $Q(\bx|\bx_{(k,i)})$ 
%Thus, a surrogate function at iteration $k$ for a given noise scale $t_i$:
%\begin{eqnarray}
%Q(\bx|\bx_{(k,i)}) &=& \frac{\|\bA \bx-\by\|^2}{\eta^2} \\\nonumber&&+ \frac{1}{(t_i)^2}\Big[ \nabla E_{\theta}(\bx_{(k,i)};t_i)^\top \bx + \frac{L}{2}\|\bx - \bx_{(k,i)}\|^2 \Big].
%\end{eqnarray}
whose minimization yields:
\begin{align}\label{mm_up}
\bx_{(k+1,i)} 
&= \left(\dfrac{\bA^{H}\bA}{\eta^{2}}+\dfrac{L}{{t}^{2}}\bI\right)^{-1} \nonumber\\\nonumber
&\quad  \left(\dfrac{\bA^{H}\by}{\eta^{2}} + \dfrac{L \bx_{k,t} - \nabla_{\bx_{k,t}} E_{\theta}(\bx_{k,t},{t})}{{t}^{2}}\right)
\end{align}
When $L=1$, the above approach simplies to \eqref{cgupdate}; the MAP approach is similar to the ALPS algorithm as shown in Algorithm \ref{alg:ALPS}, with the exception of the noise addition. This MM algorithm guarantees monotonic decrease of $C_t(\bx)$.

\section{Results}

\subsection{Unconditional image generation}\label{resuts:prior}
%We evaluate the sampling quality of the proposed multi-scale EBMs on two different datasets: CIFAR-10 (with resolution $32 \times 32$) and FFHQ (with resolution $64 \times 64$). For the conditional potential learned using score distillation, we choose ${\bD}_{\rm{teacher}}(\tilde{\bx};t)$ in \eqref{eq:loss2} as the pre-trained diffusion model from \cite{karras2022edm}. Noise and image pairs for the  transport distillation were obtained by solving the ODE  using the pre-trained diffusion model with the optimal inference noise schedule given by the authors in \cite{karras2022edm}. For both the models we used the architecture for $\bD_\bth(\bx;t)$ in \eqref{eq:energy} as that of the pre-trained diffusion model. To realize the single potential discussed in Sec. \ref{single potential}, we combined the multi-scale EBM trained using score distillation with a small noise predictor model that is trained separately using loss \eqref{eq:np}. We realize a noise predictor with two convolutional layers followed by global average pooling and two linear layers to produce a single scalar per image. Fig. \ref{cifar_image} shows the images generated using the three trained potentials for CIFAR-10 dataset. The samples for the FFHQ dataset are available in the supplementary section.

We evaluate the sampling quality of our multi-scale EBMs on CIFAR-10 ($32 \times 32$) and FFHQ ($64 \times 64$). For score distillation, we use the pre-trained diffusion model from \cite{karras2022edm} as the teacher ${\bD}_{\rm{teacher}}(\tilde{\bx};t)$ in \eqref{eq:loss2}. %Transport distillation uses noise-image pairs generated via ODE integration with the same model and its optimal inference noise schedule. 
The EBM adopts the architecture of the teacher model for $\bD_\bth(\bx;t)$ in \eqref{eq:energy}.  Fig.\ref{cifar_image} shows generated samples for CIFAR-10; FFHQ results are provided in the supplementary material.

\begin{figure}[t!]
  \centering
  \begin{subfigure}[b]{0.5\textwidth}
    \includegraphics[]{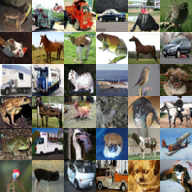}
    \caption{FID 3.35 NFE:35}
  \end{subfigure}
  % \begin{subfigure}[b]{0.26\linewidth}
  %   \includegraphics[width=1\linewidth]{cvpr/results/cifar10_rectified.png}
  %   \caption{FID 3.89 NFE:11}
  % \end{subfigure}
  % \begin{subfigure}[b]{0.26\linewidth}
  %   \includegraphics[width=1\linewidth]{cvpr/results/cifar10_singlepotential.png}
  %   \caption{FID 3.41 NFE:35}
  % \end{subfigure}
  \caption {Unconditional generation of multi-scale EBM  using noise predictor on CIFAR-10 dataset at $32 \times 32$ resolution. }\vspace{-1em}
  \label{cifar_image}
\end{figure}

\begin{table}[t]
\centering
\caption{FID scores on CIFAR-10 for various EBM-based methods.}
\label{tab:fid_cifar_single}
\small
\renewcommand{\arraystretch}{1.05}
\begin{adjustbox}{width=0.75\linewidth} % scales to 90% of text width
\begin{tabular}{@{}l r r@{}}
\toprule
\textbf{Models} & \textbf{NFE} & \textbf{FID} $\downarrow$ \\
\midrule
NT-EBM~\citep{nijkamp2020learning} & - & 78.12 \\
LP-EBM~\citep{pang2020learning} & 40 & 70.15 \\
Adaptive CE~\citep{xiao2022adaptive} & 40 & 65.01 \\
JEM~\citep{grathwohl2019your} & - & 38.40 \\
EBM-IG~\citep{du2019implicit} & - & 38.20 \\
EBM-FCE~\citep{gao2020flow} & - & 37.30 \\
CoopVAEBM~\citep{xie2021learning} & 15 & 36.20 \\
Divergence Triangle~\citep{han2020joint} & - & 30.10 \\
VARA~\citep{grathwohl2020no} & - & 27.50 \\
EBM-CD~\citep{du2020improved} & 40 & 25.10 \\
GEBM~\citep{arbel2020generalized} & - & 19.31 \\
HAT-EBM~\citep{hill2022learning} & 50 & 19.30 \\
CF-EBM~\citep{zhao2020learning} & 60 & 16.71 \\
CoopFlow~\citep{xie2022tale} & 30 & 15.80 \\
CLEL-base~\citep{lee2023guiding} & 600 & 15.27 \\
VAEBM~\citep{xiao2020vaebm} & 16 & 12.16 \\
CDRL~\citep{guo2023egc} & 18 & 9.67 \\
DRL~\citep{gao2020learning} & 180 & 9.58 \\
CLEL-large~\citep{lee2023guiding} & 1200 & 8.61 \\
CDRL~\citep{guo2023egc} & 90 & 4.31 \\
Distilled-EnergyDiffusion~\cite{thornton2025energy} & 35 & 3.01 \\
E-DSM~\cite{thornton2025energy} & 35 & 6.17 \\
\midrule
\textbf{EBM-Score Distillation (Ours)} & \textbf{35} & \textbf{3.35} \\
%\rowcolor{gray!20}
%\textbf{EBM-Transport Distillation (Ours)} & \textbf{11} & \textbf{3.89} \\
%\textbf{EBM-Single Potential (Ours)} & \textbf{35} & \textbf{3.41} \\
\bottomrule
\end{tabular}
\end{adjustbox}

\vspace{-5pt}
\end{table}

\subsection{Application to inverse problems}
We evaluate the recovery performance of multi-scale EBMs on two different inverse problems: (1) inpainting on FFHQ images at a resolution of $64 \times 64$, and (2) reconstruction of MRI images of size $324 \times 324$ from undersampled measurements.
\subsubsection{Image inpainting}
\begin{figure*}[t!]
  \centering
  % Left combined subfigure: (a)
  \begin{subfigure}[t]{0.48\textwidth}
    \centering
    % Top image
    \includegraphics[width=0.75\linewidth]{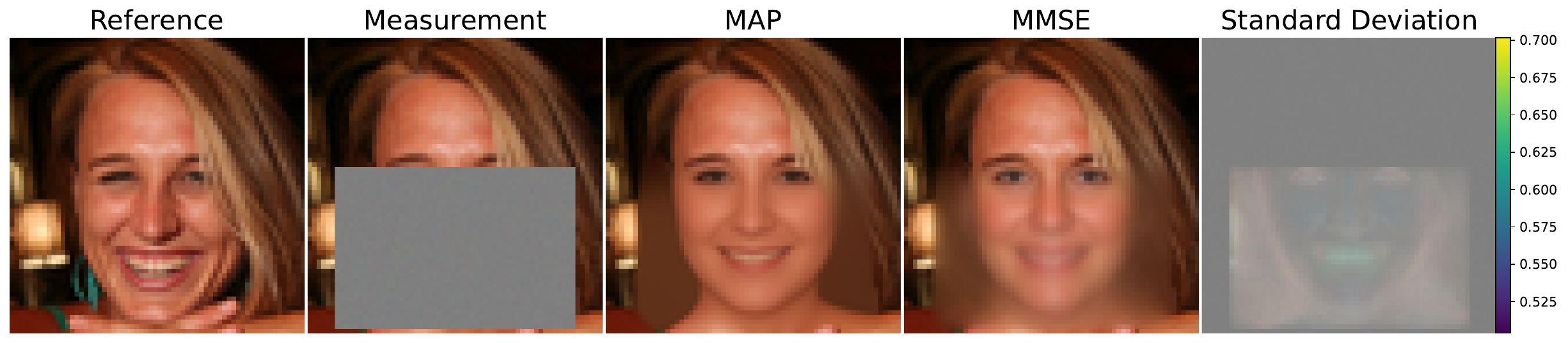}
    \vspace{0.5em}
    % Bottom image
    \includegraphics[width=0.75\linewidth]{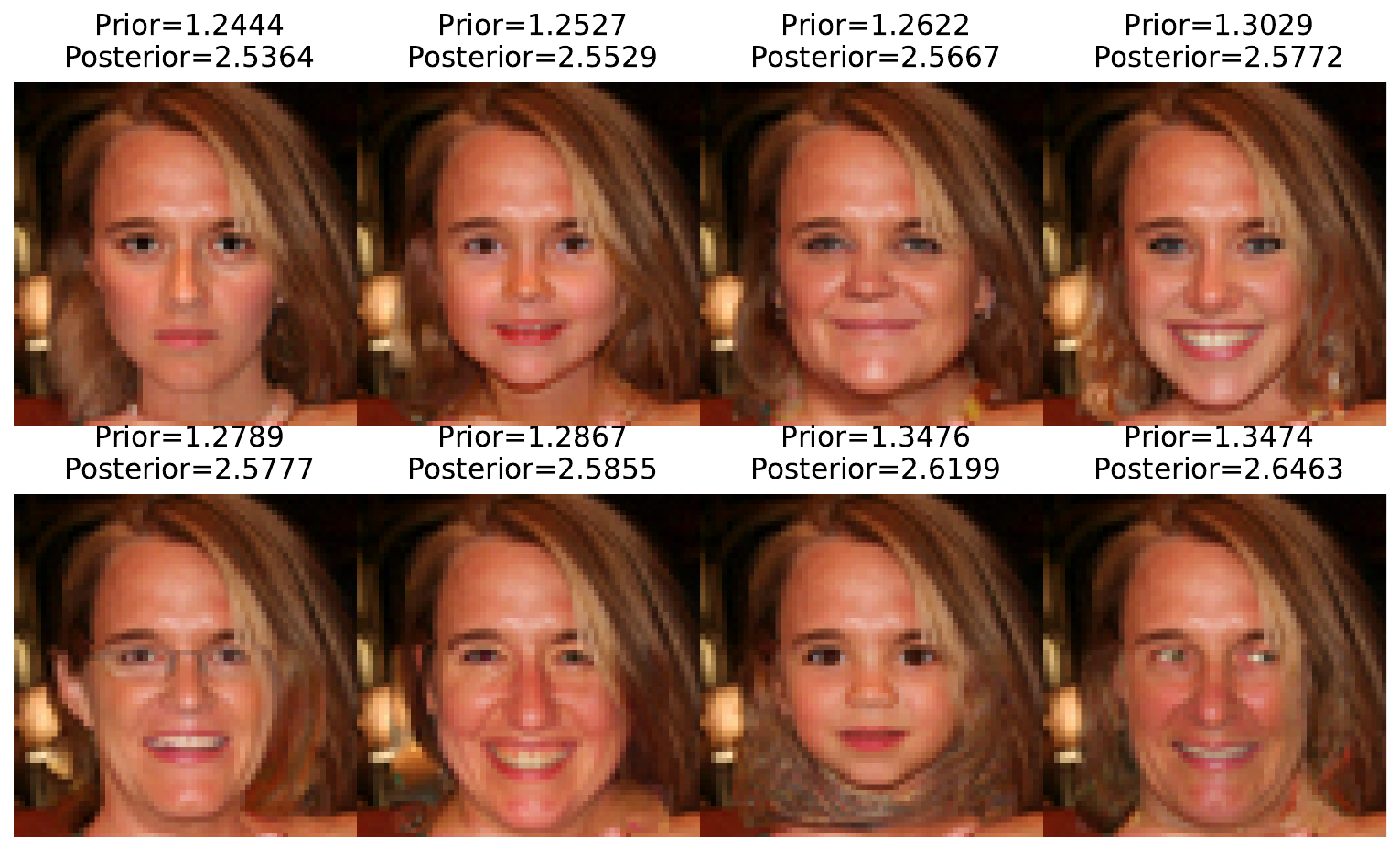}
    \vspace{-1em}
    \label{fig:inpainting_face_full}
  \end{subfigure}
  \hfill
  % Right combined subfigure: (b)
  \begin{subfigure}[t]{0.48\textwidth}
    \centering
    % Top image
    \includegraphics[width=0.72\textwidth]{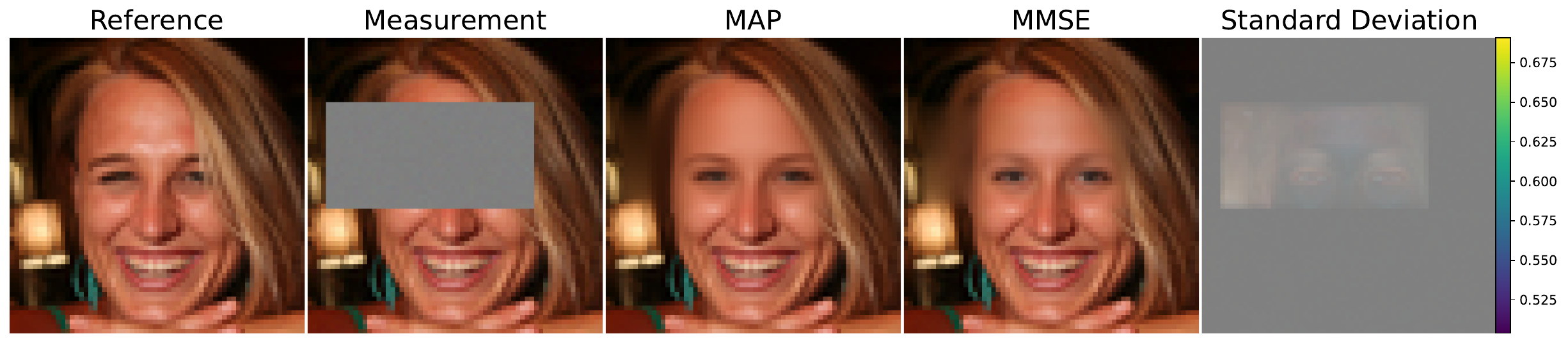}
    \vspace{0.5em}
    % Bottom image
    \includegraphics[width=0.72\textwidth]{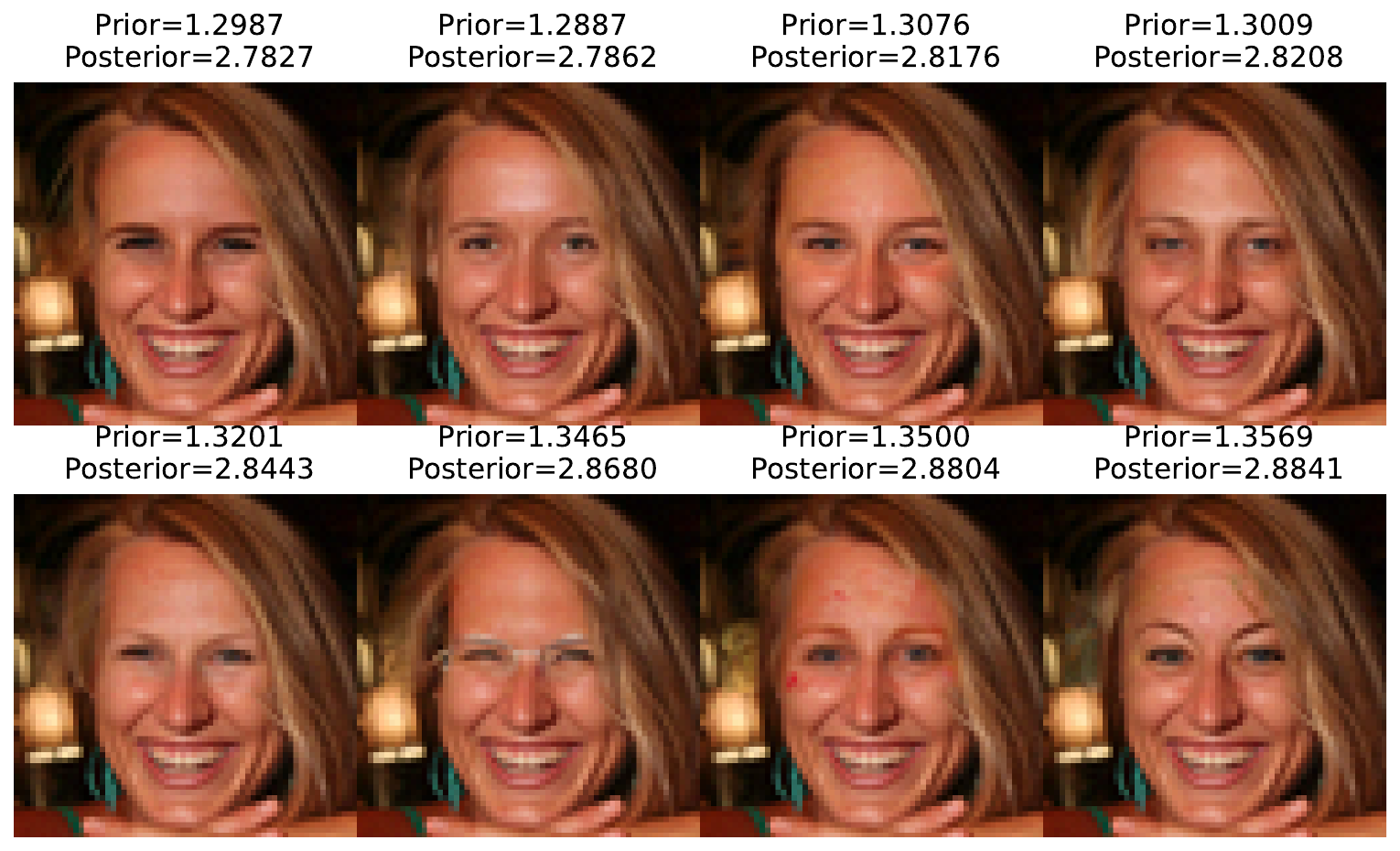}
    \label{fig:inpainting_face_eyes}
  \end{subfigure}\vspace{-1em}
\caption{{Image inpainting using multi-scale EBMs}: Illustration of recovery of images in the context of image inpainting problem for two different masks in (a) and (b). Top row in each figure shows the original image, the corresponding measurements, MAP, MMSE, and uncertainty estimates - whose values are higher in the masked regions. Second and the third rows presents several samples from the corresponding posterior distribution. We also show the negative log-posterior and negative log-prior values on top of each of the samples. Samples in Fig. (a) shows diversity in eyes and lips areas while the samples in Fig. (b) shows diversity in the eyes region.  }
    
  \label{fig:inpainting_multi_scale}
\end{figure*}

\begin{table}[b!]
\centering
\caption{Evaluation of multi-scale EBMs and diffusion-based algorithms to recover MRI images for two different accelerations. We report the Avg. PSNR $+/-$ std in (dB) along with the corresponding NFEs.}
\label{mrirecon}
\small
\setlength{\tabcolsep}{4.5pt}
\renewcommand{\arraystretch}{1.05}
\begin{tabular}{@{}l r r r@{}}
\toprule
\textbf{Method} & \textbf{NFE} & \textbf{4x 1D acceleration} & \textbf{8x 2D acceleration} \\
\midrule
EBM (MMSE) & 50 & \textbf{35.33$\pm$1.31} & \textbf{36.59$\pm$1.61} \\
DPS & 300 & 33.55$\pm$1.58 & 35.47$\pm$1.31 \\
DAPS & 50 & 35.22$\pm$1.30 & 36.44$\pm$1.72 \\
EBM (MAP) & 50 & 35.16$\pm$1.09 & 36.28$\pm$1.15 \\
\bottomrule
\end{tabular}
\vspace{-7pt}
\end{table}
We consider two different box-type mask of varying size with added Gaussian noise (standard deviation $\eta = 0.01$). We used Algorithm 1 to generate the samples with $N=50$, $K=1$, and $10$ CG iterations. This corresponds to one Langevin update per each noise scale and therefore $50$ NFEs. Similar to \cite{karras2022edm}, we choose the noise-scheduler as:
$t_i = \sigma_{\rm max}^{\frac{1}{\rho}} + \dfrac{i}{N-1}\left(\sigma_{\rm min}^{\frac{1}{\rho}}-\sigma_{\rm max}^{\frac{1}{\rho}}\right)^{\rho}$ where $\rho = 5$, $\sigma_{\rm max}=10$, and $\sigma_{\rm min}=0.01$. %The algorithm was initialized randomly from a Gaussian distribution. 
We generated 100 different samples and computed their mean and standard deviation to estimate the MMSE and uncertainty estimates. Fig. \ref{fig:inpainting_multi_scale}.(a) and Fig. \ref{fig:inpainting_multi_scale}.(b) shows the sampling performance using multi-scale EBMs trained via the score distillation loss. In the top row of each figure, we show the MAP, MMSE, and the uncertainty estimates. From the uncertainty estimate it can be observed that the algorithm is highly uncertain in the masked region, while it is confident in its recovery in the unmasked regions. In the second and third row we display the samples from the posterior distribution. Since the image inpainting posterior is highly multi-modal, we observe the diversity in the samples generated. In particular, in Fig. \ref{fig:inpainting_multi_scale}.(a), where the mask cover the face from the noise region, we see diversity in the eyes, lips, and the chin regions of the samples generated. In Fig. \ref{fig:inpainting_multi_scale}.(b), where the masks covers only the eyes region we see different eye expressions among the samples generated. We also report the negative log-prior and negative log-posterior values of the samples generated. The recovery using the single potential is shown in supplementary material. 
%Next, we also evaluate the reconstruction  performance of the single potential wherein we combine a pre-trained noise predictor with score distillation trained EBM. Note that for this model, we do not have to choose a noise schedule $t$ and hence we do not have to tune the hyperparameters $\rho, \sigma_{(max,min)}$. We use Algorithm 1 to generate the samples, where we replace the score $\nabla E_{\bth}(\bx,t)$ in line (5) with the score of the potential $Q (\bx)$ defined in \eqref{eq:sp}. We show the shows the sampling performance of the single potential in the supplementary section. Similar to the multi-scale EBM, the single potential also shows diversity among the samples generated. 

\begin{figure*}[h]
  \centering
  \begin{subfigure}[b]{0.38\linewidth}
    \centering
    \includegraphics[width=1.1\linewidth]{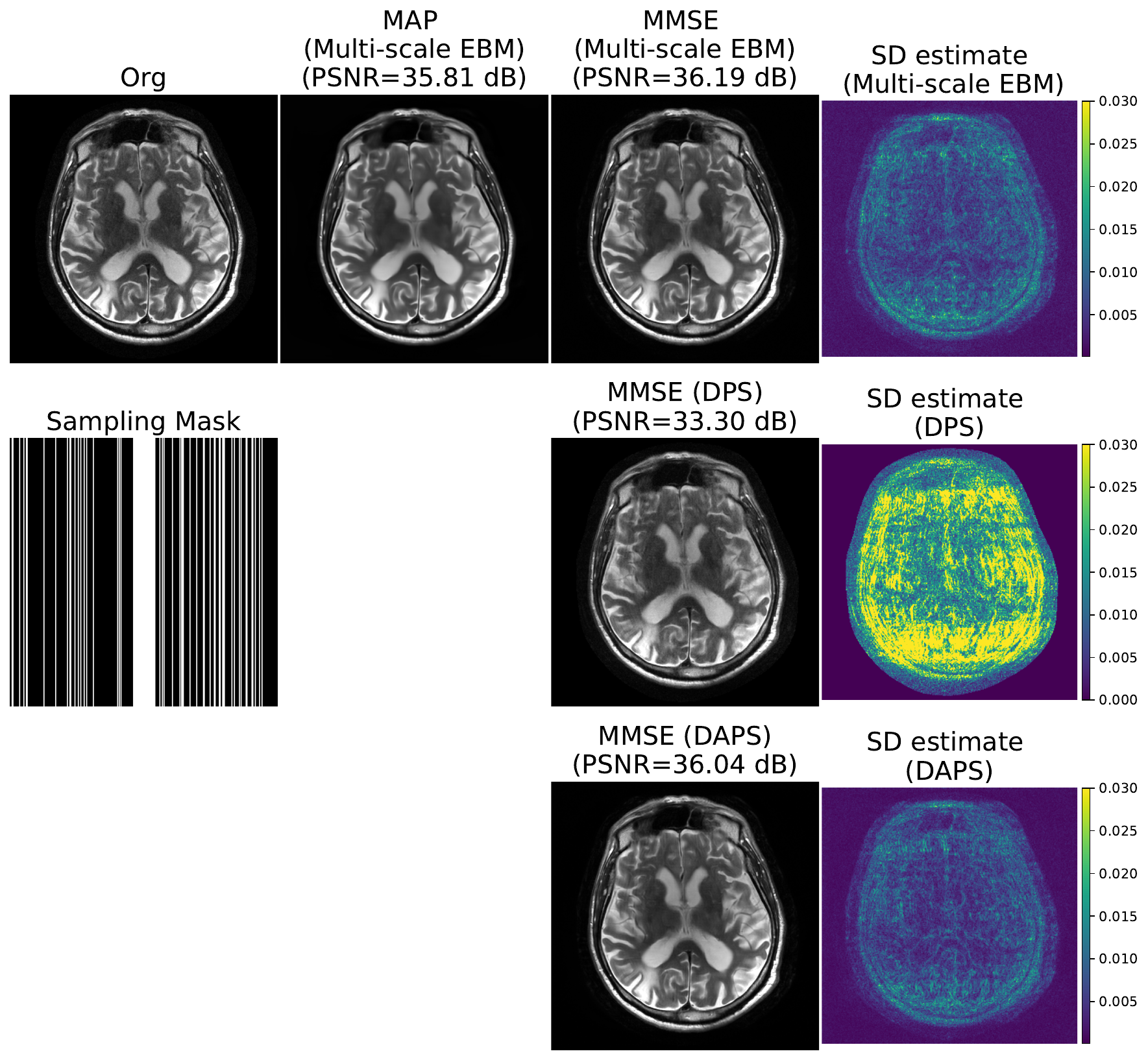}
    \caption{4x one dimensional undersampling}
  \end{subfigure}
  \hfill
  \begin{subfigure}[b]{0.38\linewidth}
    \centering
    \includegraphics[width=1.1\linewidth]{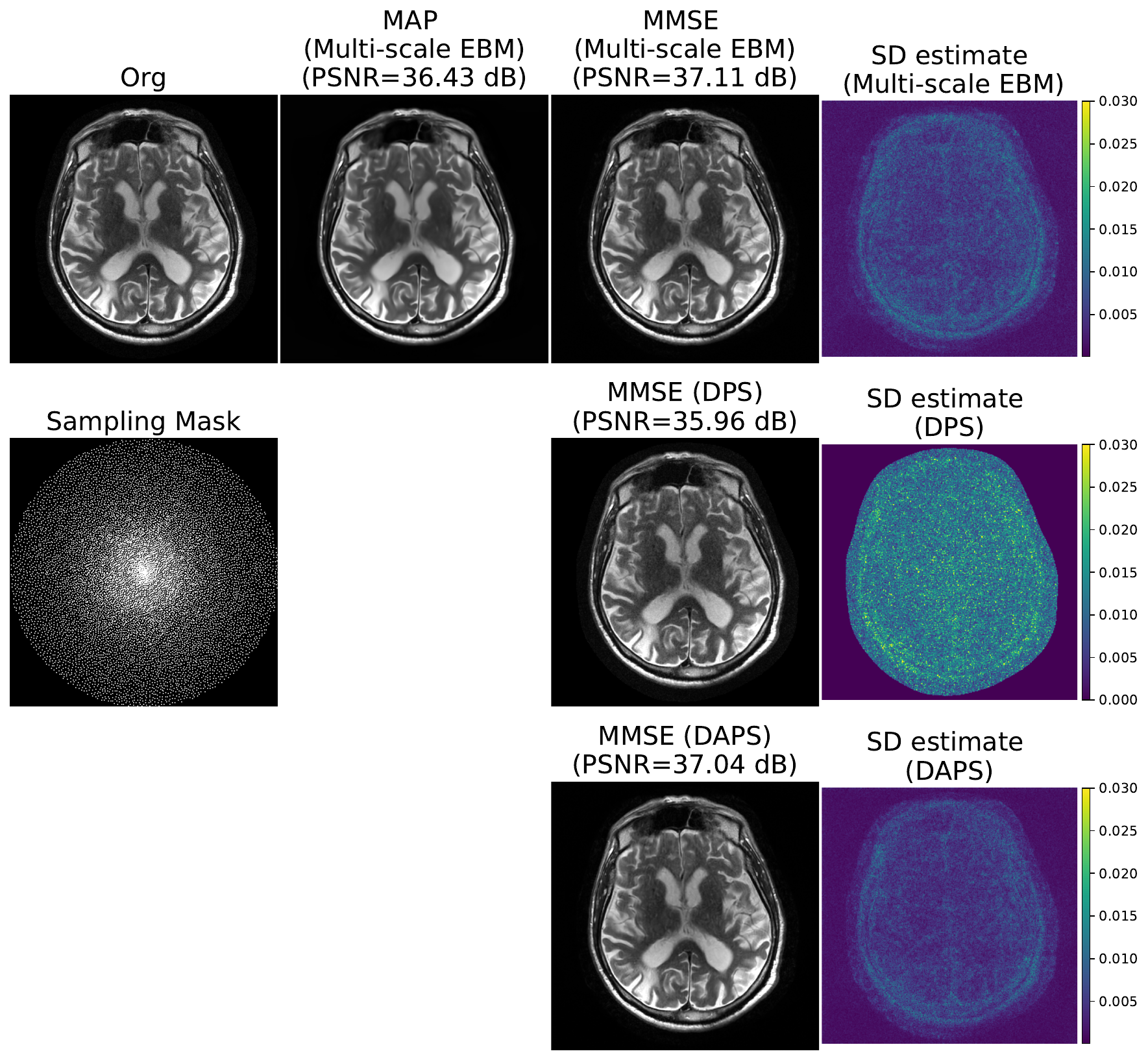}
    \caption{8x two dimensional undersampling}
  \end{subfigure}\vspace{-1em}
  \caption{Comparison of performance of multi-scale energy models with diffusion models on the T2-weighted fastmri brain data set using (a) 4x Cartesian and (b) 8x Poisson undersampling masks. In each figure, first row shows the MAP, MMSE, and uncertainty estimates of the multi-scale EBM. Second and third rows shows the MMSE and uncertainty estimates obtained using DPS and DAPS algorithm, respectively. First image in the second row shows the mask employed for undersampling.}\label{fig:mri_recon}\vspace{-1em}
\end{figure*}

% \begin{figure*}[t!]
%   \centering
%   % Left combined subfigure: (a)
%   \begin{subfigure}[t]{0.48\textwidth}
%     \centering
%     % Top image
%     \includegraphics[width=\linewidth]{cvpr/results/inpainting_face_full_single_potential.pdf}
%     \vspace{0.5em}
%     % Bottom image
%     \includegraphics[width=\linewidth]{cvpr/results/inpainting_samples_face_full_single_potential.pdf}
%     \caption{}
%     \label{fig:inpainting_face_full_single_potential}
%   \end{subfigure}
%   \hfill
%   % Right combined subfigure: (b)
%   \begin{subfigure}[t]{0.48\textwidth}
%     \centering
%     % Top image
%     \includegraphics[width=\linewidth]{cvpr/results/inpainting_face_eyes_single_potential.pdf}
%     \vspace{0.5em}
%     % Bottom image
%     \includegraphics[width=\linewidth]{cvpr/results/inpainting_samples_face_eyes_single_potential.pdf}
%     \caption{}
%     \label{fig:inpainting_face_eyes_single_potential}
%   \end{subfigure}
% \caption{{Image inpainting using single potential}: We repeat the experiment in Fig. \ref{fig:inpainting_multi_scale} using single potential which alleviates the need for noise scheduler during inference.  Top row in each figure shows the original image, the corresponding measurements, MMSE, and uncertainty estimates. Second and the third rows presents several samples from the corresponding posterior distribution along with their negative log-prior and negative log-posterior values. Similar to the multi-scale EBMs, samples generated by single potential also show diversity. }
%   \label{fig:inpainting_single_potential}
% \end{figure*}

\subsubsection{MRI image reconstruction}

We evaluate multi-scale EBMs for MRI reconstruction from undersampled measurements at two acceleration factors. EBMs were distilled from the diffusion model in \cite{aali2025ambient} using a T2-weighted brain dataset of 40,442 slices and tested on 40 slices. We compare against two diffusion-based algorithms: (a) DPS and (b) DAPS. We use the same noise scheduler as in the previous example ($\sigma_{\rm max}=5$, $\sigma_{\rm min}=0.002$, $\rho=7$), $K=1$ Langevin steps, and $N=50$ iterations. MMSE and uncertainty are estimated from five samples; MAP is computed via Algorithm 1 over $N=50$ iterations. Table \ref{mrirecon} shows EBMs achieve performance comparable to DAPS for both accelerations. Fig. \ref{fig:mri_recon}(a,b) illustrates reconstructions at 4× and 8×. Unlike diffusion models, EBMs provide MAP estimates without multiple samples, reducing inference cost and enabling practical clinical deployment.

\section{Conclusion}
In this work, we revisited Energy-Based Models (EBMs) as a principled framework for solving inverse problems in imaging. By using score distillation, we transferred the strengths of pre-trained diffusion models into multi-scale EBMs, enabling efficient sampling while preserving the interpretability and compositionality of potential-based formulations. 
Focusing on inverse problems, we proposed modular algorithms for posterior sampling and MAP estimation. Specifically, we introduced Annealed Langevin Posterior Sampling (ALPS), which enables efficient MMSE inference through static posterior distributions and avoids backpropagation through forward models. For MAP estimation, we developed a majorization minimization strategy. Our results demonstrate that distilled EBMs can match or surpass diffusion models in accuracy and efficiency, making them well-suited for scientific and clinical imaging applications.
%We revisit EBMs as a principled framework for fast image generation and inverse problems. By introducing score and transport distillation, we transfer the strengths of pre-trained diffusion models into EBMs, enabling efficient sampling while preserving interpretability and compositionality. Our strategies cut the usual computational cost of EBMs, achieving state-of-the-art FID on CIFAR-10 with only 11 NFEs.
%We exploit EBM compositionality to design modular algorithms for posterior sampling and MAP estimation. 
%Unlike diffusion methods requiring complex guidance and backpropagation, our approach uses annealed static posteriors for simpler, faster inference. Experiments on inpainting and MRI reconstruction show EBMs match or surpass diffusion models in accuracy and efficiency, while supporting uncertainty quantification and MMSE recovery

\newpage

\maketitlesupplementary

\section{Introduction}
We include additional experiments that were not included in the main paper. The sections are:
\vspace{0.3em}
\begin{enumerate}
    \item Illustration using a toy dataset
    \vspace{0.3em}
    \item Posterior sampling in inverse problems, where we show: 
    \vspace{-0.05em}
    \begin{itemize}
    \item Differences in evolution between  the proposed Annealed Langevin posterior sampling (ALPS) and decoupled annealed posterior sampling (DAPS) algorithms.
    \vspace{0.1em}
    \item Impact of hyperparameters in ALPS and DAPS.
    \vspace{0.1em}
    \item Performance comparisons on inverse problems.
    \vspace{0.1em}
    \item Example reconstructions.
    \vspace{0.2em}
    \end{itemize}
    \vspace{0.3em}
    \item Ability of multi-scale energy-based models (EBMs) to enable out-of-distribution (OOD) detection and model mismatch.
    \vspace{0.3em}
    \item Illustration of  posterior sampling via single potential.
    \vspace{0.3em}
    \item Implementation details.
\end{enumerate}
\section{Illustration using the 2D Moons dataset}

The main focus of this section is to illustrate the proposed ALPS approach using a toy dataset. We trained the multi-scale EBM on the 2D \texttt{moons} dataset. The top row of Fig.\ref{fig:energies_across_t} visualizes the learned negative log-prior (NLPr) or energy $E_{\theta}(\bx;t)$ over several noise levels $t$.  The blue points represent the data samples used to learn the EBMs, and the iso-contours indicate the learned energy landscape. The top row shows the energy at different noise scales $t \in \{0.750, 0.350, 0.200, 0.100\}$. The illustration indicates that the multi-scale EBM captures the manifold geometry at different scales.  At the largest noise level $t = \sigma_{\max}$, the energy is approximately quadratic, corresponding to a Gaussian prior. As $t \to 0$, the energy landscape becomes increasingly representative of the true data distribution, capturing its multi-modal structure. This progressive refinement is a direct consequence of the annealed score-matching approach, which learns the score function across multiple noise scales.

\subsection{Modeling the posterior}
In inverse problems, the true posterior distribution often exhibits complex, multi-modal behavior due to measurement uncertainty and prior constraints. Here, we consider a linear measurement model $\by=\mathbf a^\top \mathbf x + \bb + \eta \boldsymbol\epsilon; \eta=0.2; \epsilon \sim \mathcal N(0,\bI)$, denoted by the red line in Fig. \ref{fig:energies_across_t}. The measurement line intersects the manifold at multiple locations, inducing a multi-modal posterior. Directly sampling from the true posterior is challenging because of the complex multi-modal nature. The composition property of EBMs allows us to combine the learned energy at time $t$ with the negative-log likelihood term to form a family of posterior energies:
\[
E_{\text{posterior}}(\bx; \bt) = E_{\text{prior}}(\bx; t) + E_{\text{likelihood}}(\bx),
\]
where $E_{\text{likelihood}}$ encodes the measurement model. The time-dependent posterior energies at different noise scales are shown in the bottom row of Fig. \ref{fig:energies_across_t}. 

\subsection{ALPS recovery of samples}

\begin{figure}[h!]
    \centering
    \includegraphics[width=0.45\textwidth]{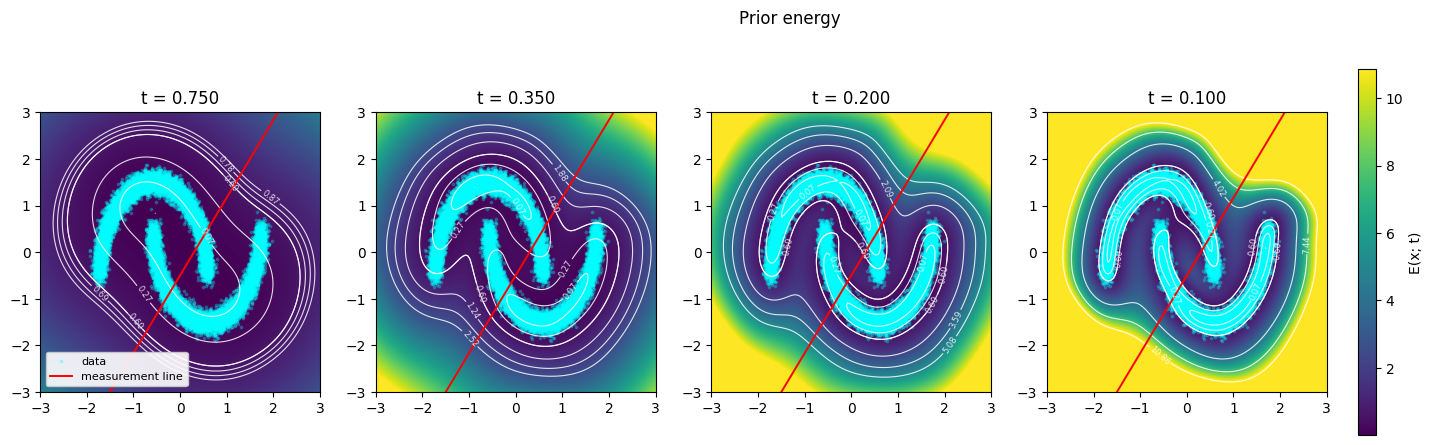}
    \includegraphics[width=0.45\textwidth]{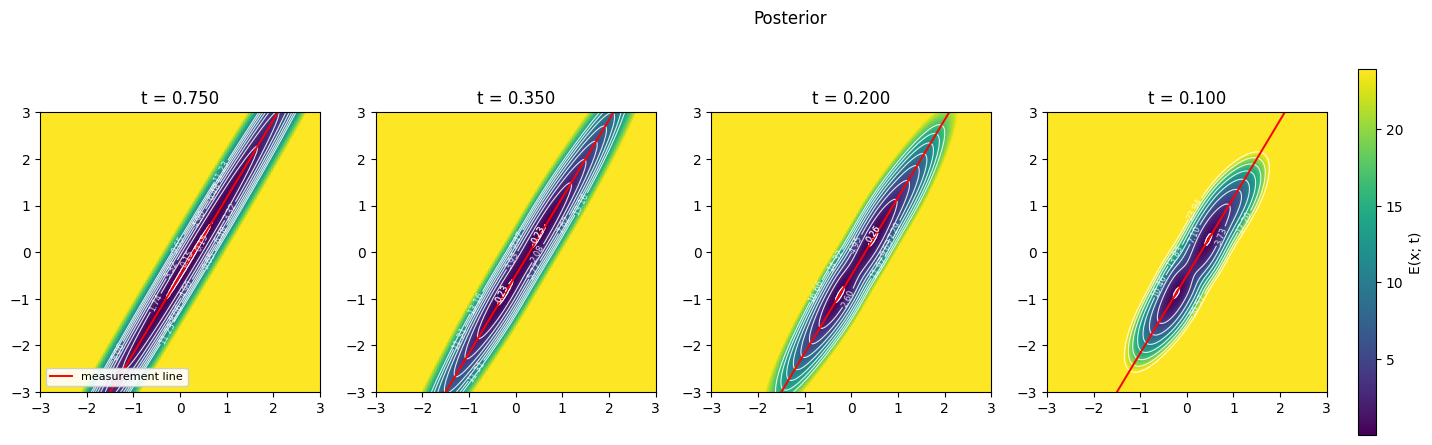}
    \caption{Learned EBM using score-based learning from the moons dataset. In the top row we show the data samples (blue points) and the score-learned energy at different time-scales. At $t=\sigma_{max}$, the energy is roughly quadratic corresponding to a Gaussian prior. As time approaches to $0$, the energy becomes more representative of the true distribution. We also assume the measurements to be defined by the red straight line. The posterior energies at different time scales are shown in the bottom row; the maximum of the negative log-posterior is clipped for improved visualization. As time $t\rightarrow 0$, the posterior evolves from the likelihood to the true multi-modal posterior. The evolution of the samples offered by the ALPS algorithm is shown in Fig. \ref{fig:alps_vs_daps}}
    \label{fig:energies_across_t}
\end{figure}

ALPS uses pre-conditioned Langevin dynamics with $K$ steps to derive samples from the posterior at each scale. It also uses an annealing strategy, which gradually decreases $t$, starting from $t_0=\sigma_{\rm max}$. This approach mitigates the impact of local minima and facilitates exploration of the energy landscape. At high noise levels, the energy surface is smooth and unimodal, allowing easy mixing. As $t$ decreases, the model transitions towards the true multi-modal posterior, while the initialization from previous scales helps avoid getting trapped in spurious modes. This approach is particularly powerful for inverse problems where the posterior is highly non-convex and traditional optimization methods fail. Overall, Fig. \ref{fig:energies_across_t} demonstrates how score-based EBMs, combined with annealed Langevin sampling, provide a principled framework for tackling challenging inverse problems with complex posterior distributions.

\begin{figure}[htbp]
    \centering
    \begin{tabular}{cccc}
    \includegraphics[trim=15mm 15mm 15mm 15mm,clip,width=0.1\textwidth]
    {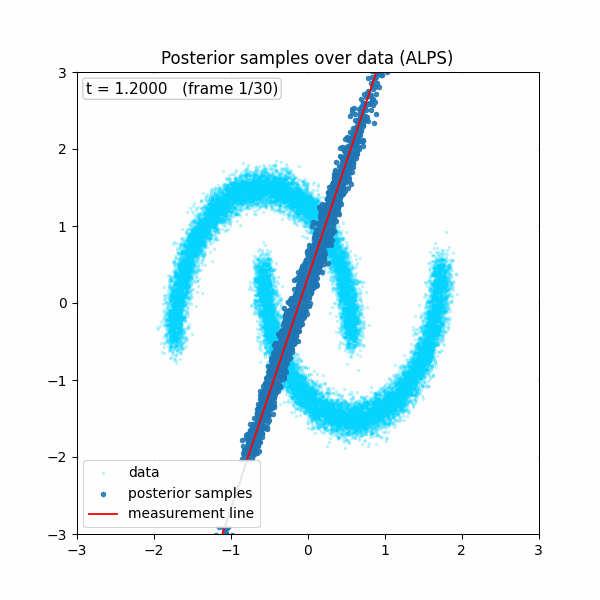} &
    \includegraphics[trim=15mm 15mm 15mm 15mm,clip,width=0.1\textwidth]{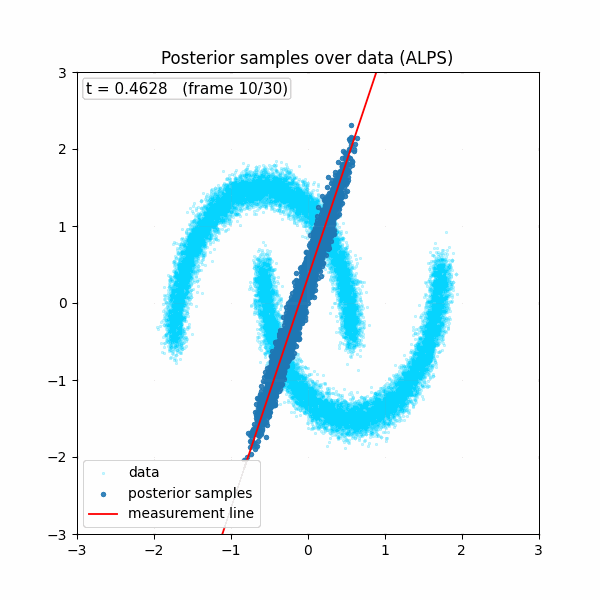} &
    \includegraphics[trim=15mm 15mm 15mm 15mm,clip,width=0.1\textwidth]{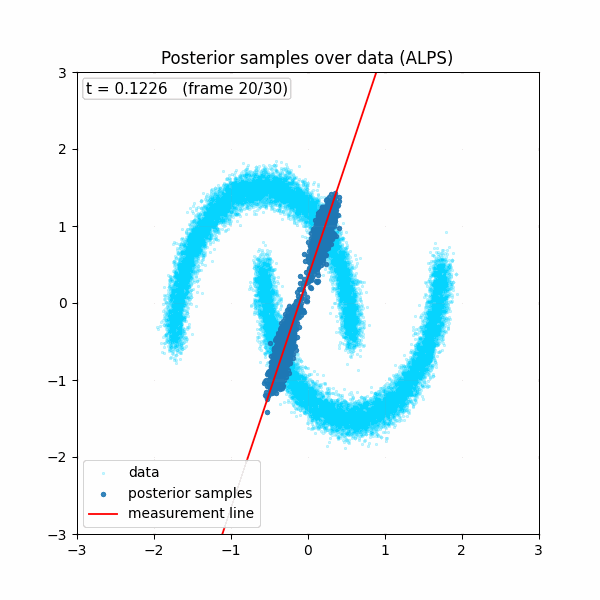} &
    \includegraphics[trim=15mm 15mm 15mm 15mm,clip,width=0.1\textwidth]{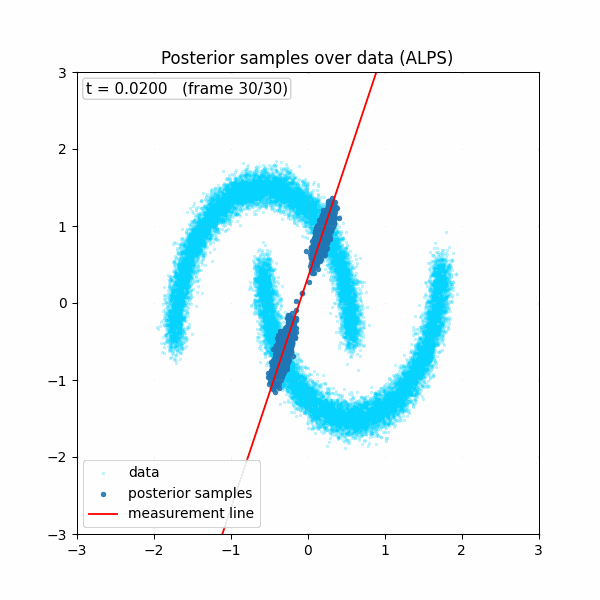}
\\
    \includegraphics[trim=15mm 15mm 15mm 15mm,clip,width=0.1\textwidth]{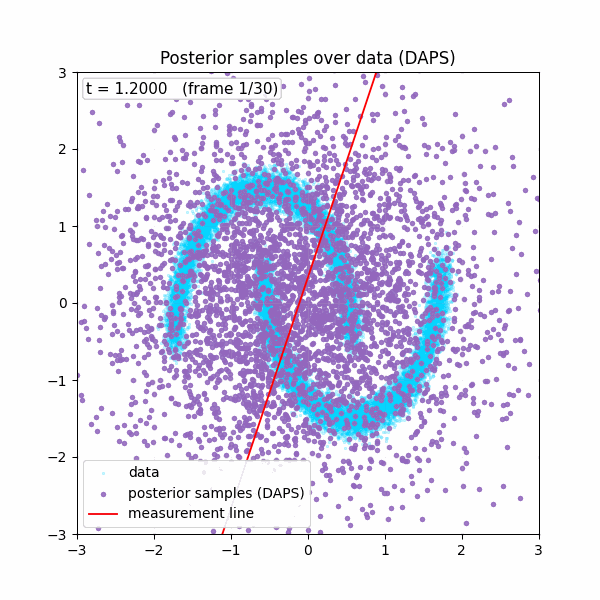} &
    \includegraphics[trim=15mm 15mm 15mm 15mm,clip,width=0.1\textwidth]{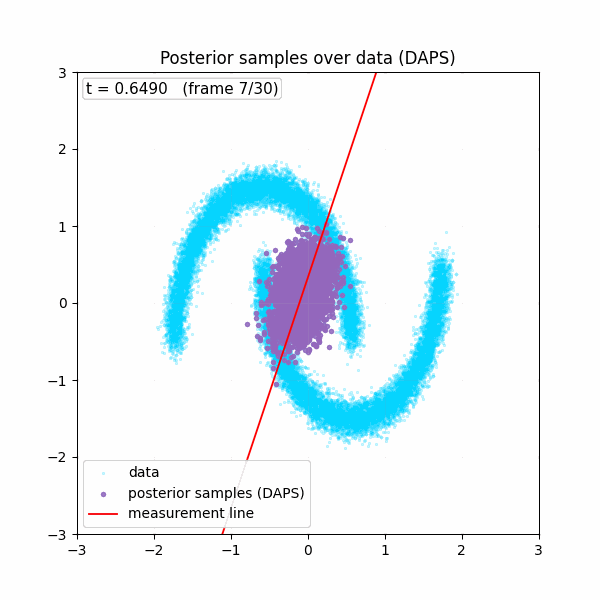} &
    \includegraphics[trim=15mm 15mm 15mm 15mm,clip,width=0.1\textwidth]{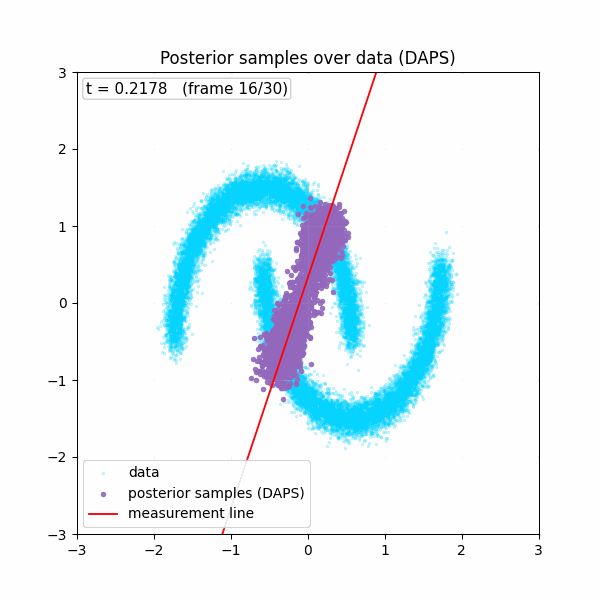} &
    \includegraphics[trim=15mm 15mm 15mm 15mm,clip,width=0.1\textwidth]{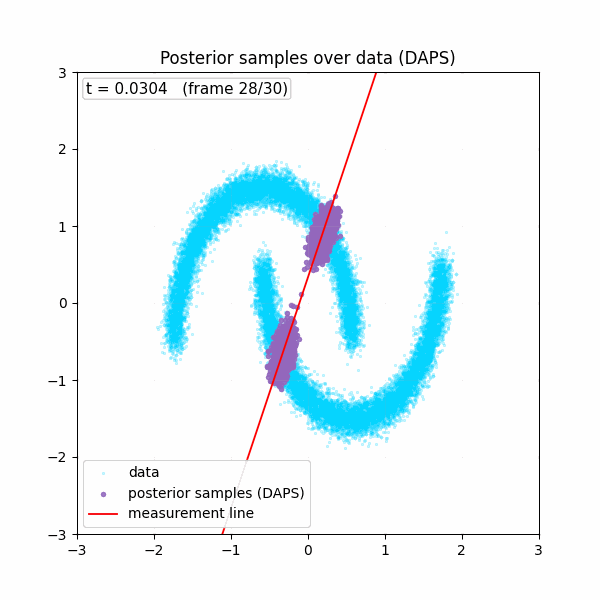}
    \end{tabular}
    \caption{Comparison of posterior samples from ALPS (top row) and DAPS (bottom row), as a function of time. The columns correspond to posterior samples as a function of time, with the left most columns corresponding to the initialization and the right most ones as the final solution. The blue points indicate the data samples on the moons dataset, which is used to learn the prior distributions. We note that ALPS initializes the samples closer to the measurement manifold; the annealing of the posterior distributions smoothly transforms them to the desired posterior samples. By contrast, DAPS are initialized with the prior distribution $p_{\sigma_{\rm max}}$. The guidance term is carefully engineered such that the samples stay in the time dependent prior distributions $p_t$, while becoming data consistent at $t\rightarrow 0$. The smoother evolution of the samples in ALPS translates to faster sampling, and hence improved performance. }
    \label{fig:alps_vs_daps}
\end{figure}

\begin{figure*}[t!]
    \centering
    \begin{subfigure}[b]{\textwidth}
        \includegraphics[width=\textwidth]{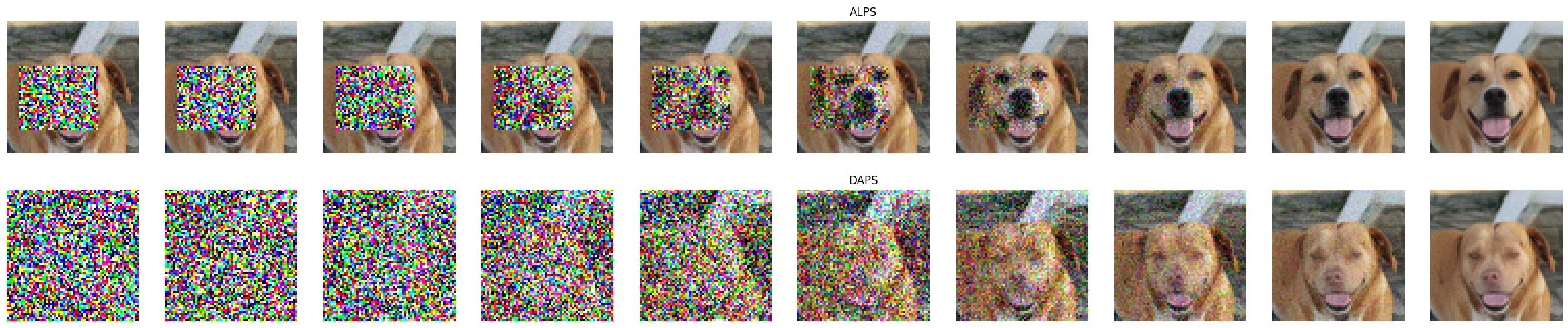}
        \caption{Inpainting with a box}
        \label{fig:subfig1}
    \end{subfigure}
    \begin{subfigure}[b]{\textwidth}
        \includegraphics[width=\textwidth]{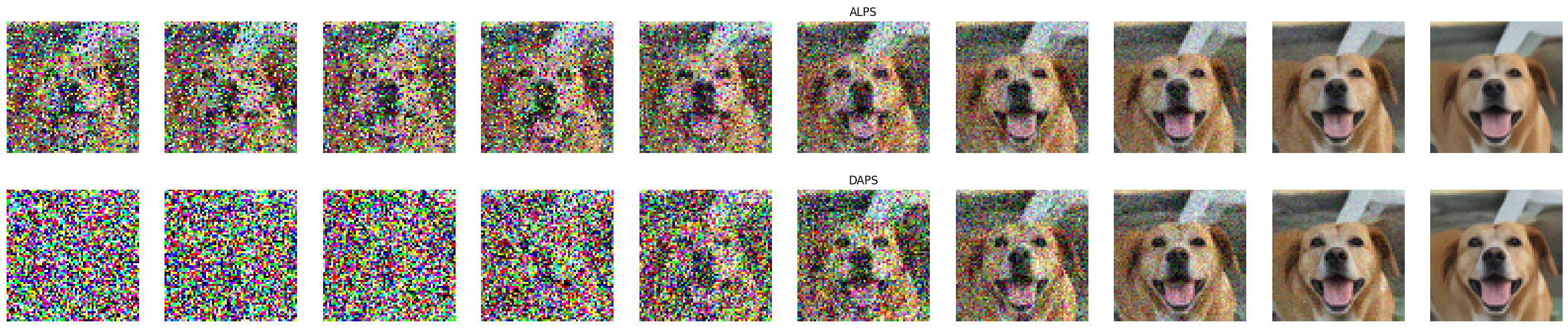}
        \caption{Random inpainting}
        \label{fig:subfig2}
    \end{subfigure}
    \begin{subfigure}[b]{\textwidth}
        \includegraphics[width=\textwidth]{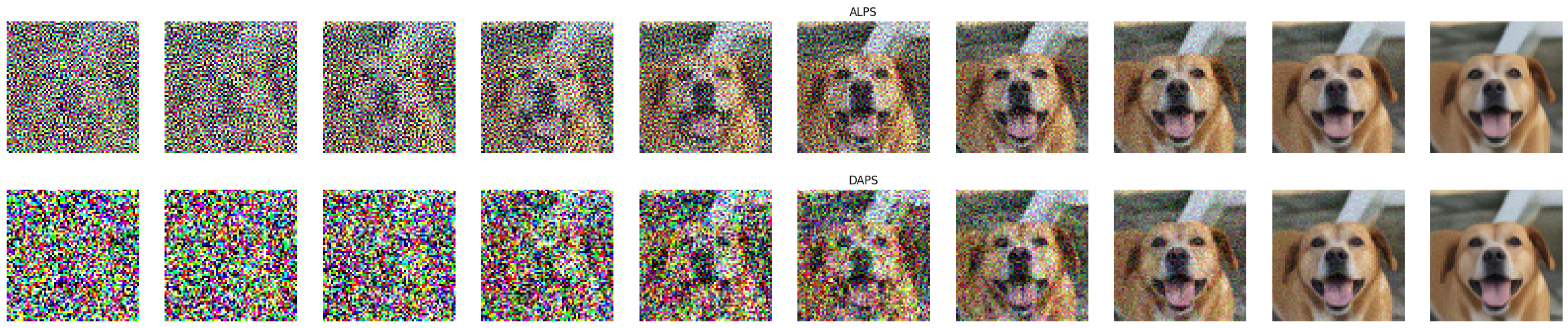}
        \caption{Gaussian blurring}
        \label{fig:subfig3}
    \end{subfigure}
    \begin{subfigure}[b]{\textwidth}
        \includegraphics[width=\textwidth]{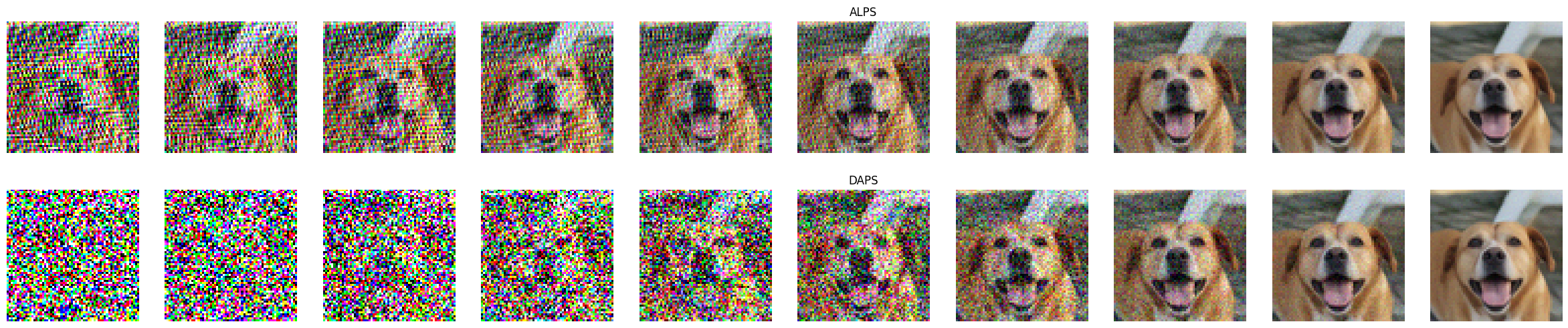}
        \caption{Motion blur}
    \end{subfigure}
    \caption{Evolution of the samples in different inverse problems, where the left-most image being the initialization corresponding to $\sigma_{\rm max}$ and the right-most one the final solution corresponding to $\sigma_{\rm min}$. In each subfigure, the top row is the ALPS evolution from the initialization, while the second row is the corresponding DAPS evolution. We note that the evolution pattern of DAPS, which evolves from noise initialization and stays in the prior distribution is roughly similar in each case. By contrast, the evolution pattern in ALPS is drastically different in each case. It starts with an initialization that is dependent on the pseudo inverse solution, and adds noise predominantly to the null-space. The customization of the noise addition to the specific forward model translates to a smoother path from the initialization to the final results, translating to improved performance and reduced inference time, as seen from the Table \ref{performance}.}
    \label{evolution_inverse}
\end{figure*}

ALPS loop essentially alternates denoising (via the conservative score of the EBM), a quadratic data-consistency update, and noise addition with a scale-aware preconditioner; samples are carried across scales to progressively sharpen the posterior. Figure~\ref{fig:alps_vs_daps} compares ALPS to the state-of-the-art  DAPS. Unlike ALPS that anneals the posterior, most diffusion based inverse solvers (including DAPS) move through samples in time-dependent prior  distributions, specified by the diffusion model. At each time point, DAPS uses an ODE solver to determine the true data samples corresponding to the specific latent, which is used to enforce the data consistency. By contrast, diffusion posterior sampling (DPS) relies on the score to map the latents to the true data samples. As seen from the bottom row of Fig. \ref{fig:alps_vs_daps}, the samples at each time point closely approximate the prior distributions. By contrast, ALPS does not require these mappings between latents and data samples as it works directly with time-dependent posteriors. Note from the top row of Fig. \ref{fig:alps_vs_daps} that the samples are initialized approximately in the null-space of the measurement operator (the line). The noise update in ALPS ensures that they approximately stay in the null space, while evolving to the multimodal posterior distribution. This forward-model-aware update enables ALPS to obtain smoother trajectories across $t$, translating to faster convergence. The smoother nature of the convergence can also be seen from Fig.   \ref{evolution_inverse}.

\begin{figure*}[h!]
   \centering
   \begin{subfigure}[b]{0.32\linewidth}
     \includegraphics[width=1\linewidth]{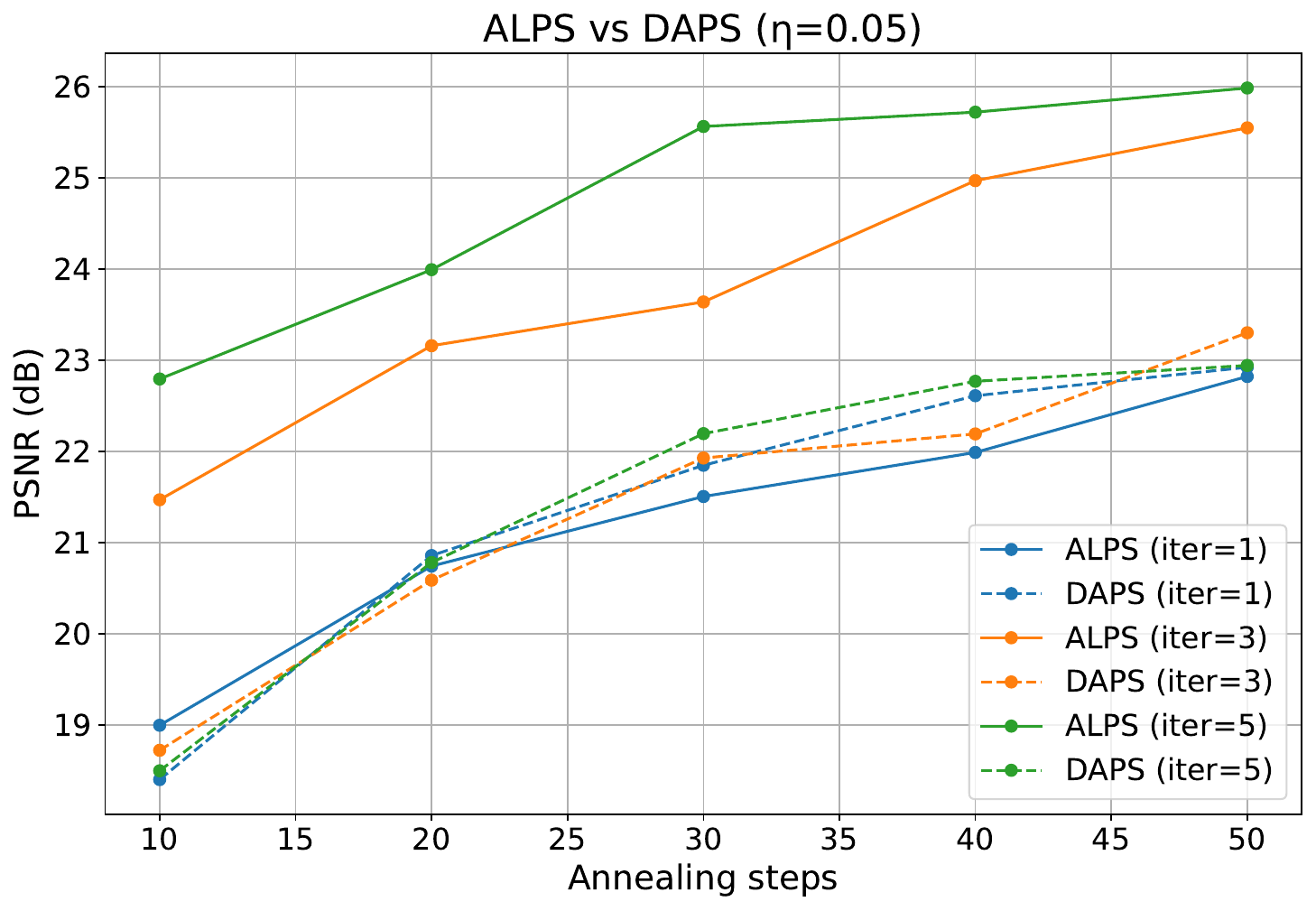}
     \caption{Inpainting $\eta=0.05$}
   \end{subfigure}
   \begin{subfigure}[b]{0.32\linewidth}
     \includegraphics[width=1\linewidth]{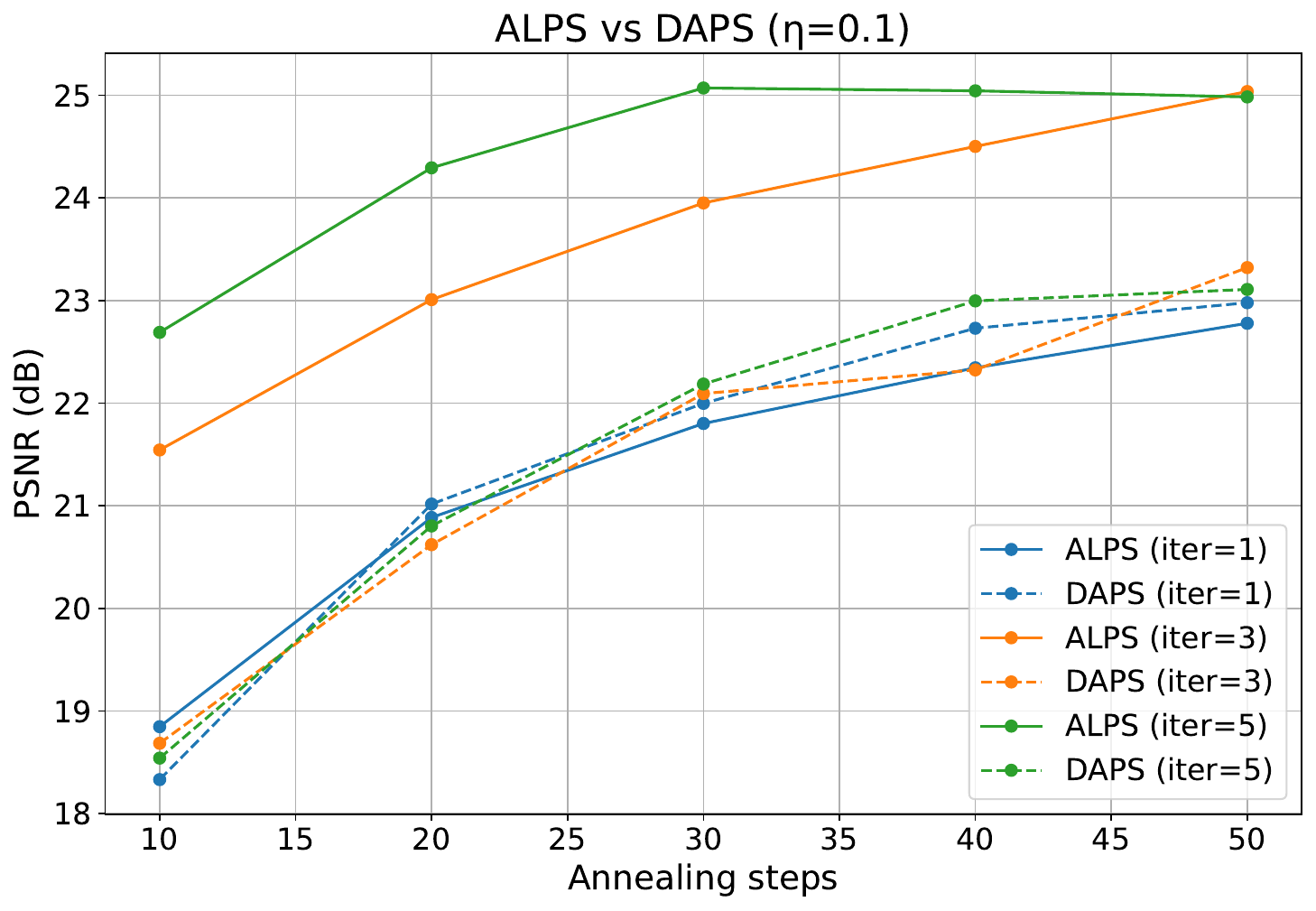}
     \caption{Inpainting $\eta=0.1$}
   \end{subfigure}
   \begin{subfigure}[b]{0.32\linewidth}
     \includegraphics[width=1\linewidth]{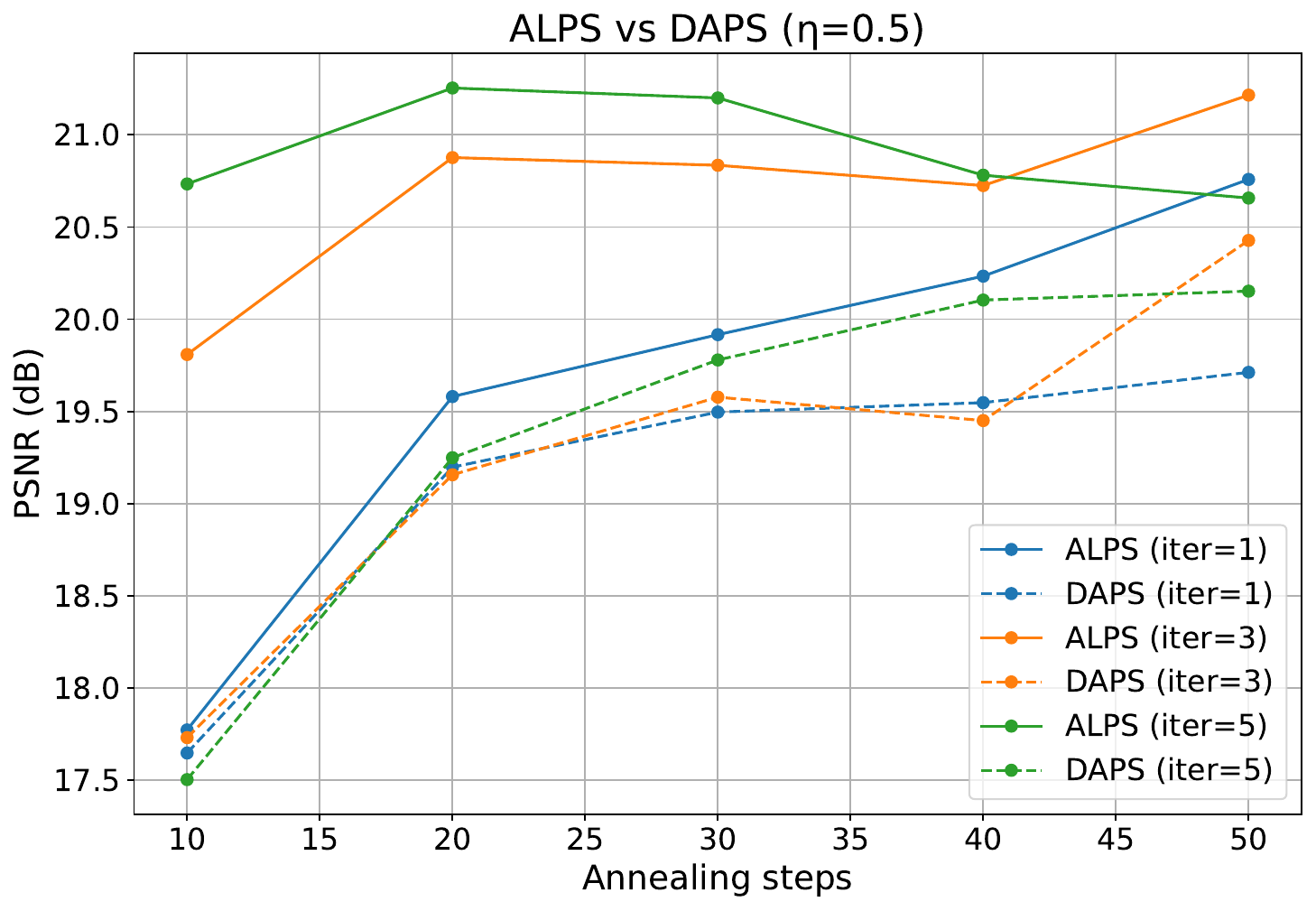}
     \caption{Inpainting $\eta=0.5$}
   \end{subfigure}
 % \begin{figure*}[h!]
 %   \centering
 %   \begin{subfigure}[b]{0.32\linewidth}
 %     \includegraphics[width=1\linewidth]{cvpr/results/blur_eta_005.pdf}
 %     \caption{Blur $\eta=0.05$}
 %   \end{subfigure}
 %   \begin{subfigure}[b]{0.32\linewidth}
 %     \includegraphics[width=1\linewidth]{cvpr/results/blur_eta_01.pdf}
 %     \caption{Blur $\eta=0.1$}
 %   \end{subfigure}
 %   \begin{subfigure}[b]{0.32\linewidth}
 %     \includegraphics[width=1\linewidth]{cvpr/results/blur_eta_05.pdf}
 %     \caption{Blur $\eta=0.5$}
 %   \end{subfigure}
 %   \caption { }
 %   \label{blur_ablation}
 % \end{figure*}

   \centering
   \begin{subfigure}[b]{0.32\linewidth}
     \includegraphics[width=1\linewidth]{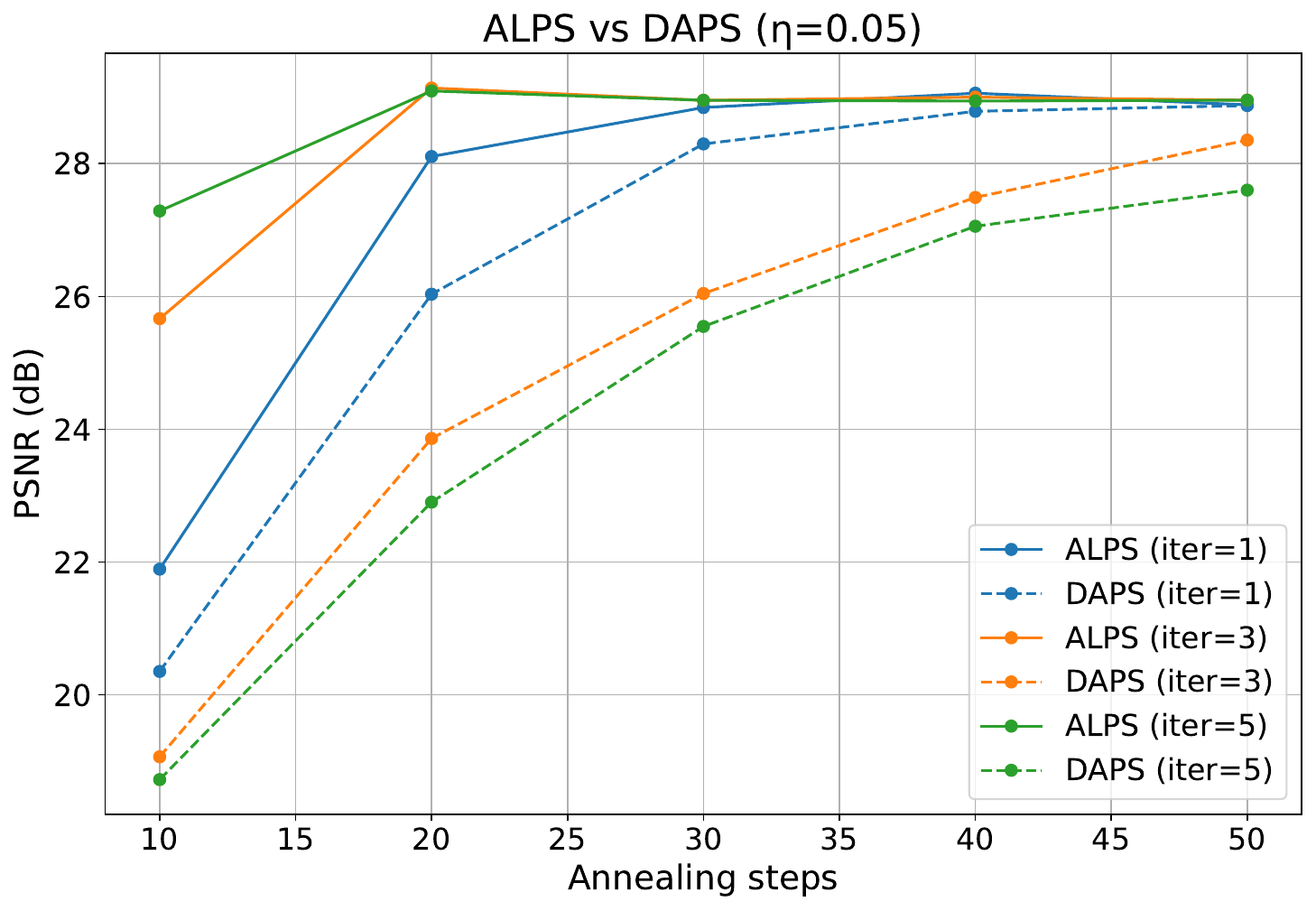}
     \caption{Motion Deblur $\eta=0.05$}
   \end{subfigure}
   \begin{subfigure}[b]{0.32\linewidth}
     \includegraphics[width=1\linewidth]{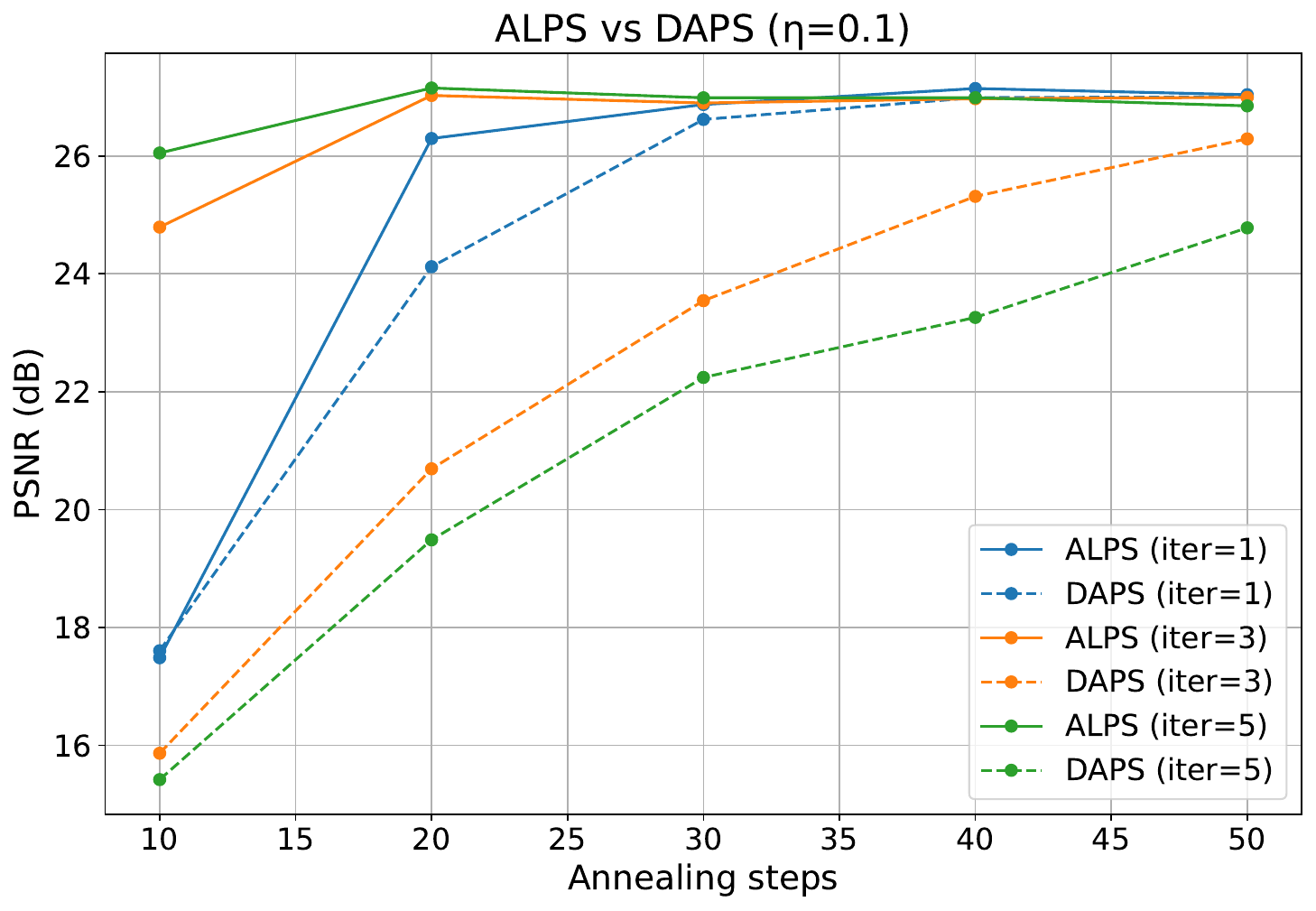}
     \caption{Motion Deblur $\eta=0.1$}
   \end{subfigure}
   \begin{subfigure}[b]{0.32\linewidth}
     \includegraphics[width=1\linewidth]{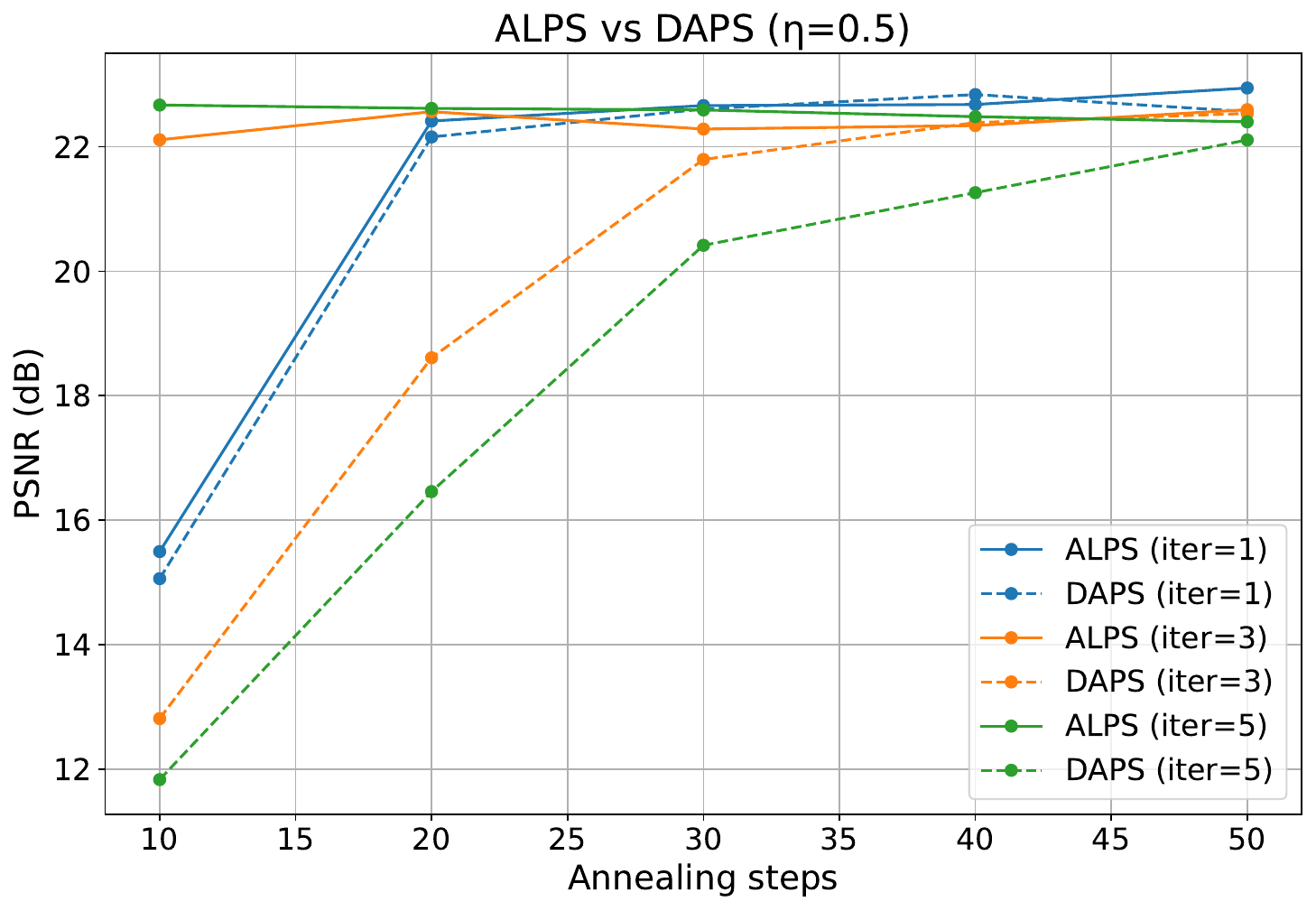}
     \caption{Motion Deblur $\eta=0.5$}
   \end{subfigure}

   \begin{subfigure}[b]{0.32\linewidth}
     \includegraphics[width=1\linewidth]{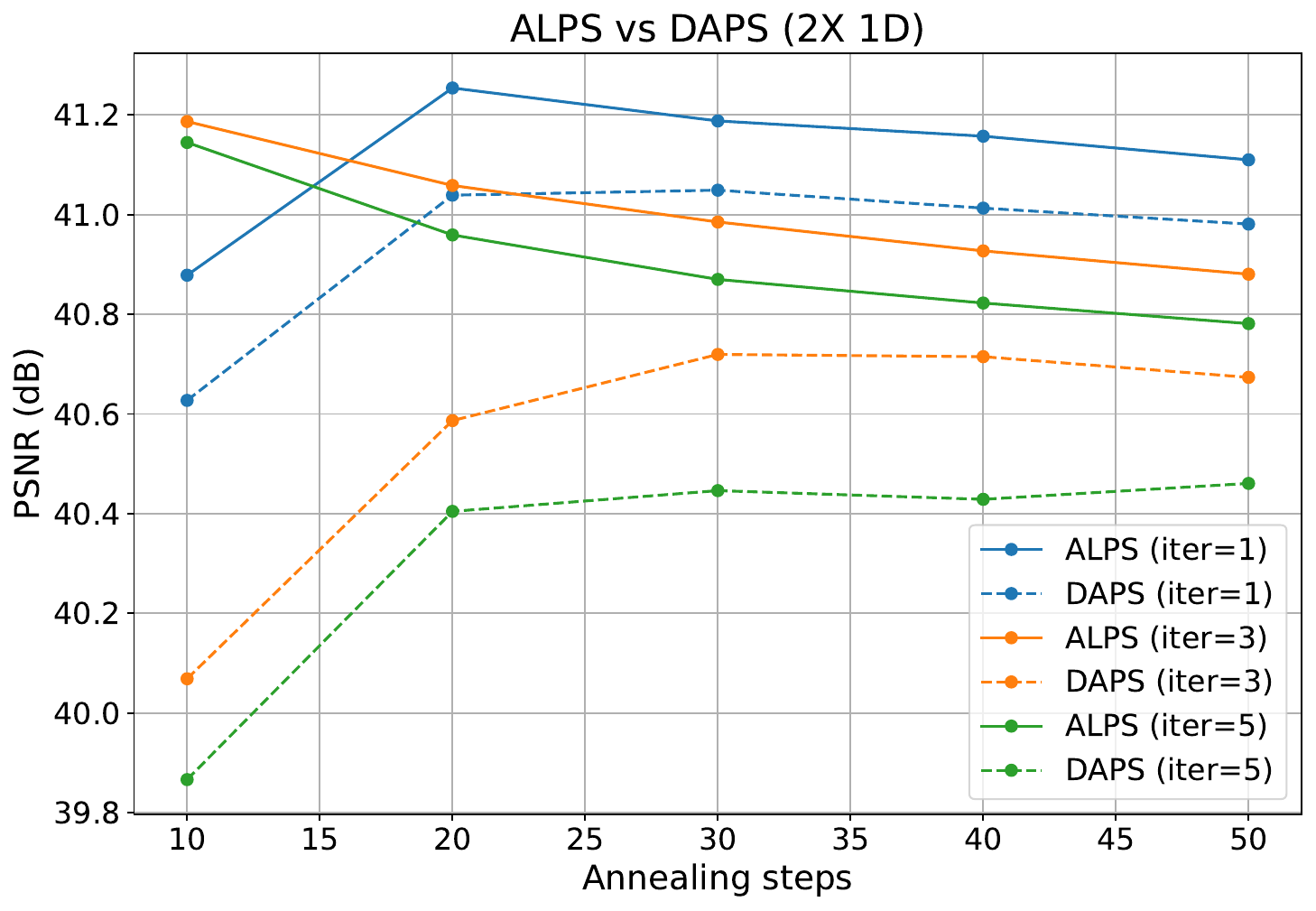}
     \caption{MRI acceleration: 2x 1D}
   \end{subfigure}
   \begin{subfigure}[b]{0.32\linewidth}
     \includegraphics[width=1\linewidth]{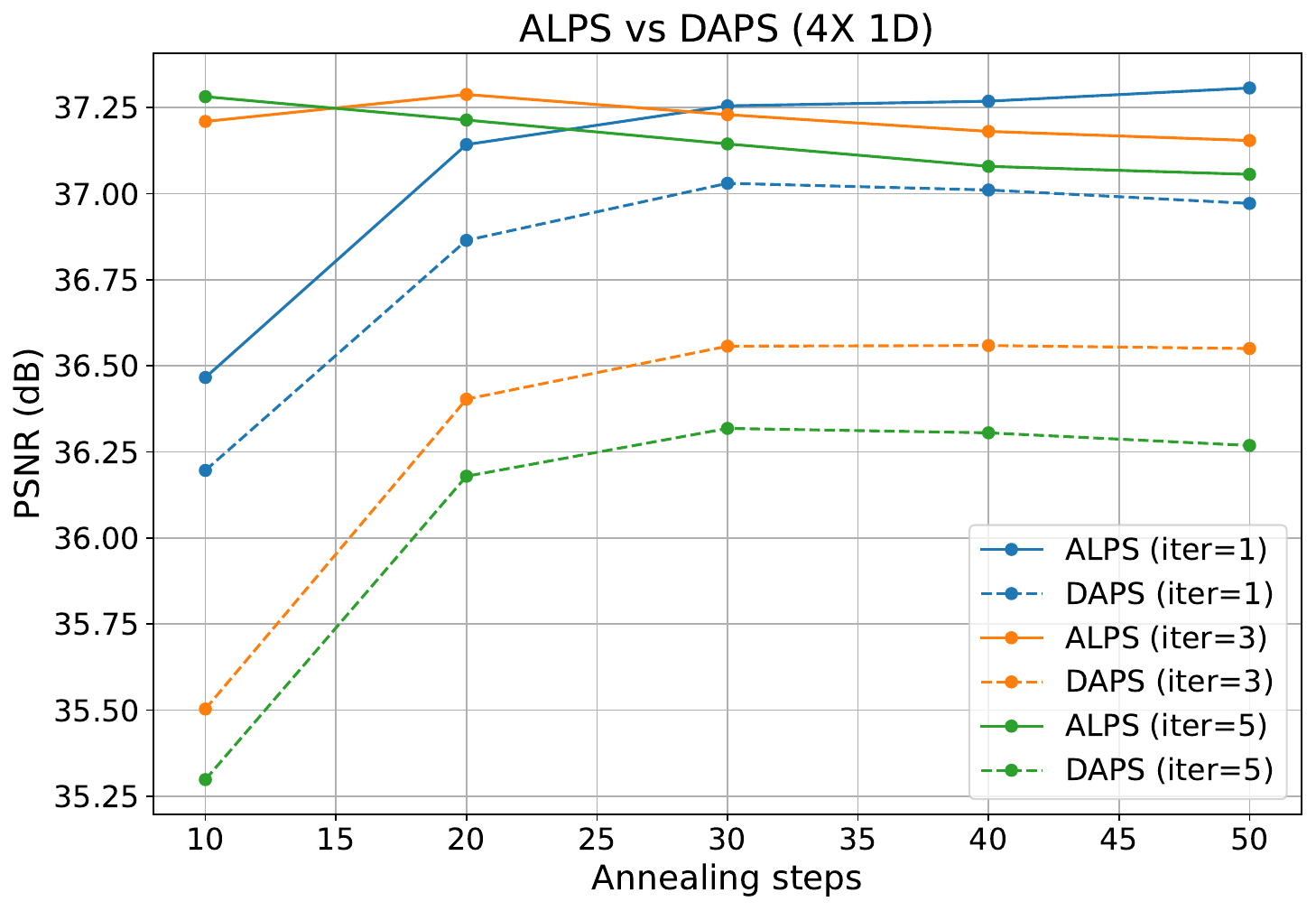}
     \caption{MRI acceleration: 4x 1D}
   \end{subfigure}
   \begin{subfigure}[b]{0.32\linewidth}
     \includegraphics[width=1\linewidth]{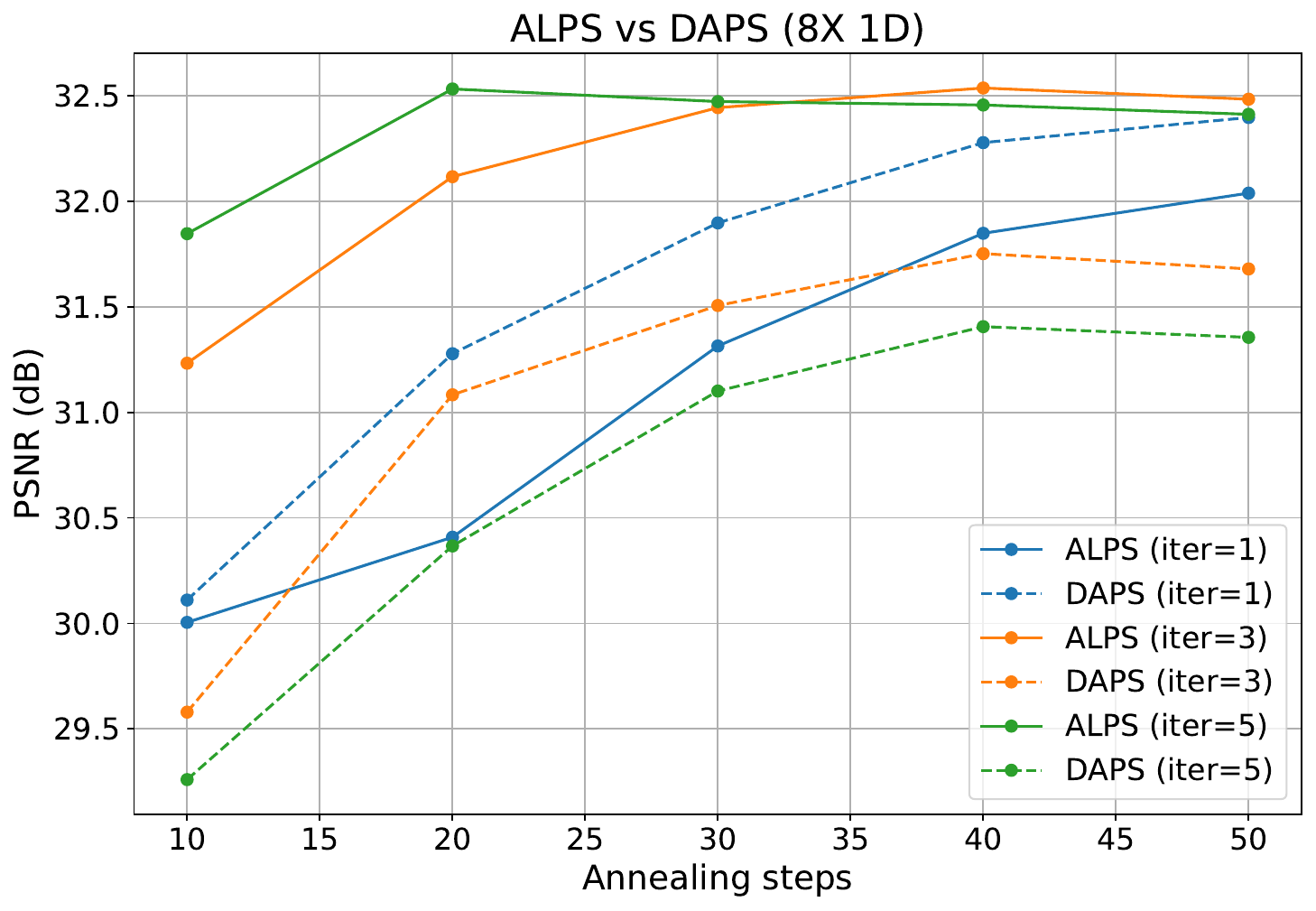}
     \caption{MRI acceleration: 8x 1D}
   \end{subfigure}
   \caption {Comparison of ALPS and DAPS across different inverse problems for varying annealing steps (x-axis). Each curve corresponds to a different value of Langevin iterations per-scale $K$ in ALPS (denoted by iter) and ODE steps per scale in DAPS(denoted by iter). The plots show that that increasing the number of Langevin steps per iteration consistently improves performance, highlighting the benefit of more thorough posterior exploration at each scale. In contrast, the impact of the number of ODE steps used in DAPS to transport the latents at each scale to the data distribution is less consistent across tasks. Overall, ALPS demonstrates faster convergence than DAPS and requires fewer function evaluations to achieve comparable or improved reconstruction quality. }
   \label{ablation_study}
 \end{figure*}
\section{Posterior sampling in imaging inverse problems}
We study the utility of the proposed ALPS algorithm, and compare it with DPS and DAPS in a variety of image-based linear inverse problems. ALPS relies on the multi-scale EBMs distilled from a diffusion model, while DPS and DAPS relies on the original diffusion model. 
%The architecture of the models are discussed in Section \ref{sec:arch}, while the distillation strategy to obtain these models are discussed in Section \ref{sec:distillation}.

\subsection{Comparison of evolution}
 Figure~\ref{evolution_inverse} illustrates the evolution of samples for representative tasks such as box inpainting, random inpainting, Gaussian blurring, and motion blur. A key distinction between DAPS and ALPS lies in their sampling trajectories. The former begins from a pure noise initialization; the guidance term is carefully engineered such that the samples stay in $p_t$, while becoming data consistent as $t\rightarrow 0$. This trajectory remains largely consistent across different inverse problems because the process is governed by the time-dependent prior distributions and the diffusion schedule. In contrast, ALPS adopts a fundamentally different strategy. It uses an initialization based on the \emph{pseudo-inverse solution}, which provides a data-consistent starting point. Noise is then injected \emph{predominantly in the null-space of the forward operator}, ensuring that perturbations do not degrade data fidelity. This customization aligns the sampling trajectory with the geometry of the inverse problem, resulting in a smoother and more direct path from initialization to the final posterior sample. The ability of ALPS to leverage problem-specific structure enable it to reduce unnecessary exploration, reducing inference time compared to the generic noise-to-signal trajectory of diffusion inverse solvers, including DAPS.

\begin{table*}[t!]
    \centering
    \caption{Reconstruction performance (mean $\pm$ std) for various inverse problems using ALPS, DAPS, and DPS. The best hyperparameters for each algorithm were selected for fair comparison. Metrics include PSNR and SSIM, averaged over multiple runs. We also report the MAP estimate. ALPS consistently achieves competitive or superior performance across tasks, particularly in challenging settings such as motion deblurring and MRI reconstruction, while requiring fewer function evaluations compared to diffusion-based approaches.}
    \begin{tabular}{l l c c}
        \toprule
        Task & Algorithm & PSNR (dB) & SSIM \\
        \midrule

        \multirow{2}{*}{\shortstack{Inpaint (random)\\$\eta = 0.05$}}
            & ALPS (ours) & 25.71$\pm$2.01 & 0.84$\pm$0.03\\
            &MAP (ours) &\textbf{25.96$\pm$2.01} & 
            \textbf{0.86$\pm$0.03} \\
            & DAPS & 23.11$\pm$1.41 & 0.75$\pm$0.04 \\
            & DPS & 22.92 $\pm$ 1.66&0.74 $\pm$0.06\\
        \midrule

        \multirow{2}{*}{\shortstack{Inpaint (random)\\$\eta = 0.1$}}
            & ALPS (ours) &25.07$\pm$ 1.58  & 
            0.81$\pm$0.03\\
            &MAP (ours) & \textbf{25.32 $\pm$ 1.79}& \textbf{0.82$\pm$0.03}\\
            & DAPS & 22.78$\pm$ 1.43 & 
            0.73$\pm$ 0.04\\
            & DPS & 22.21 $\pm$ 1.31&0.68 $\pm$0.04\\
        \midrule

        \multirow{2}{*}{\shortstack{Inpaint (random)\\$\eta = 0.5$}}
            & ALPS (ours) & 20.51$\pm$ 1.16 &
            0.57$\pm$0.04\\
            & MAP (ours) &\textbf{20.53$\pm$1.14}& \textbf{0.60 $\pm$ 0.04}\\
            & DAPS & 20.12$\pm$ 1.08  &
            0.56$\pm$0.05\\
            & DPS & 15.81 $\pm$ 0.51&0.27 $\pm$0.02\\
        \midrule

        \multirow{2}{*}{\shortstack{Gaussian Deblurring\\$\eta=0.05$}}
            & ALPS (ours) & \textbf{27.06 $\pm$1.60} &
            \textbf{0.88$\pm$0.02}\\
            & MAP (ours) &26.99 $\pm$ 1.67 &\textbf{0.88 $\pm$ 0.02}\\
            & DAPS & 27.05 $\pm$1.63 & \textbf{0.88$\pm$0.02} \\
            & DPS & 24.61 $\pm$1.38 &0.81 $\pm$ 0.31\\
        \midrule

        \multirow{2}{*}{\shortstack{Motion Deblurring\\$\eta=0.05$}}
            & ALPS (ours) & \textbf{29.49$\pm$1.1}  & 
            \textbf{0.90$\pm$0.02}\\
            &MAP (ours) & 29.35 $\pm$ 1.25 &
            \textbf{0.90$\pm$ 0.02}\\
            & DAPS &29.14$\pm$1.1  & 
            \textbf{0.90$\pm$0.02}\\
            &DPS & 24.92$\pm$ 0.82 & 0.77 $\pm$0.03\\
        \midrule

        \multirow{2}{*}{\shortstack{MRI (2$\times$1D)}}
            & ALPS (ours) & {40.85$\pm$0.51} & 0.94$\pm$0.005 \\
            & MAP (ours) & {40.64$\pm$0.499}&
            0.94 $\pm$ 0.004\\
            & DAPS & 40.66$\pm$0.51 & 0.94$\pm$0.006 \\
            &DPS &\textbf{41.35 $\pm$ 0.51}&\textbf{0.98 $\pm$0.003}\\
        \midrule

        \multirow{2}{*}{\shortstack{MRI (8$\times$1D)}}
            & ALPS (ours) & \textbf{32.09$\pm$0.81} & \textbf{0.88$\pm$0.01} \\
            &MAP (ours)& 
           31.93 $\pm$ 0.97& 0.85$\pm$ 0.01\\
            & DAPS & 31.97$\pm$0.67 & \textbf{0.88$\pm$0.01} \\
            & DPS & 27.39$\pm$1.09 & 0.81$\pm$0.005 \\
        \midrule

        \multirow{2}{*}{\shortstack{MRI (6$\times$2D)}}
            & ALPS (ours) & \textbf{39.51$\pm$0.55} & 0.94$\pm$0.005 \\
            &MAP (ours) &
            39.28 $\pm$0.54 &
            0.94 $\pm$ 0.005\\
            & DAPS & 39.34$\pm$0.55 & 0.94$\pm$0.009 \\
            & DPS & 38.94$\pm$0.6 & \textbf{0.95$\pm$0.007} \\
        \bottomrule
    \end{tabular}
    \label{performance}
\end{table*}

\subsection{Impact of hyperparameters}
Fig. \ref{ablation_study} illustrates the effect of varying the number of annealing steps (shown on the x-axis) and the number of Langevin iterations $K$ per step (denoted by iter in the legend) on the reconstruction performance of ALPS for different inverse problems. Each curve corresponds to a different value of $K$. We also compare the scheme with DAPS, which has a similar nested loop structure. The K ODE steps required by the DAPS algorithm to estimate the true prior samples from the latents is denoted by iter in the legend of Fig. \ref{ablation_study}.

The results in Fig. \ref{ablation_study} shows that increasing the number of Langevin steps at each scale consistently improves performance across all tasks. More Langevin iterations per annealing scale allow the sampler to better approximate the local posterior distribution, before transitioning to the next noise level. By reducing the impact of initialization and discretization errors, this approach improves stability and convergence. 

\begin{figure*}[!h]
\centering
    \begin{subfigure}[b]{0.48\linewidth}
     \includegraphics[width=\linewidth]{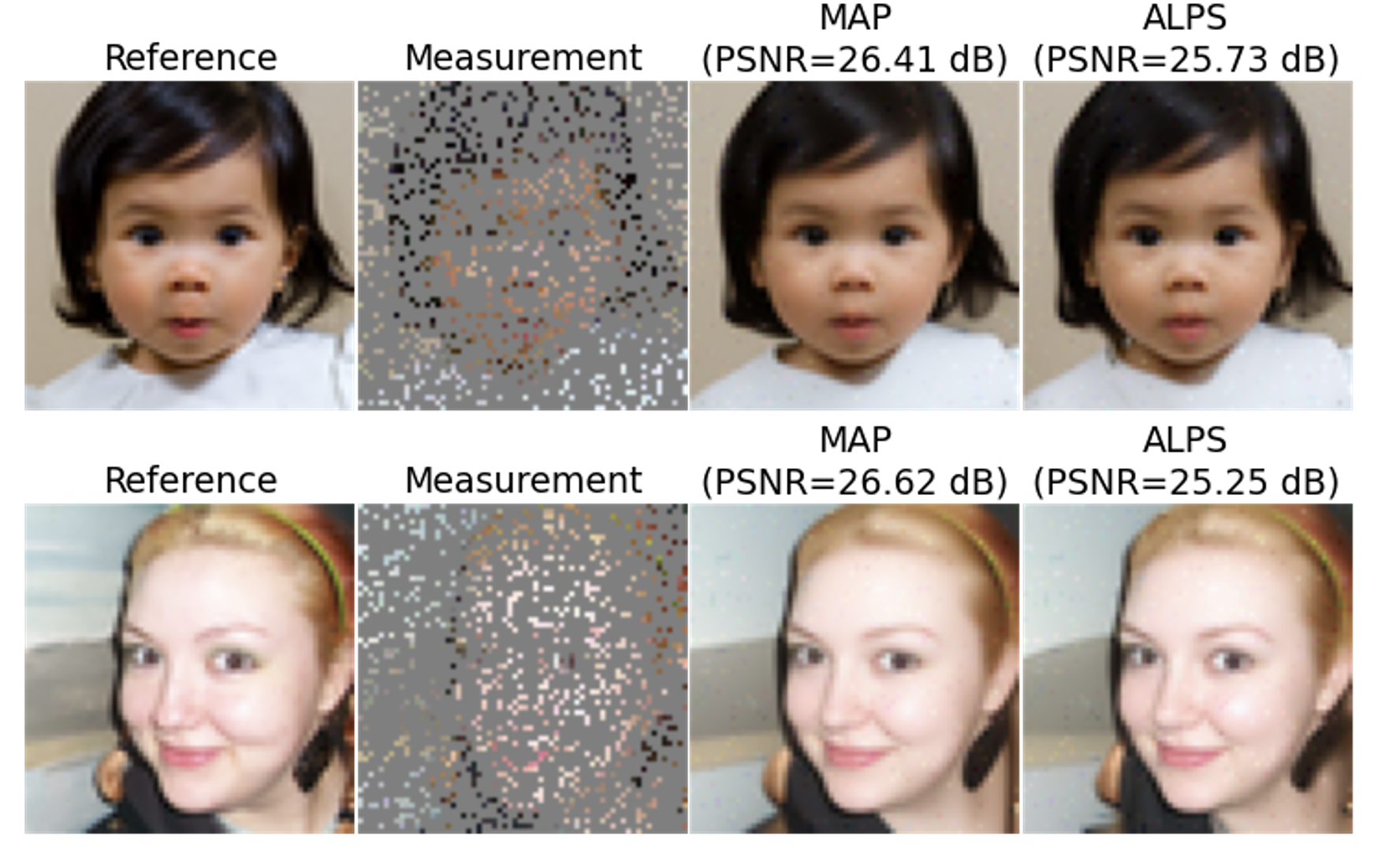}
     \caption{Inpainting}
   \end{subfigure}
       \begin{subfigure}[b]{0.48\linewidth}
     \includegraphics[width=\linewidth]{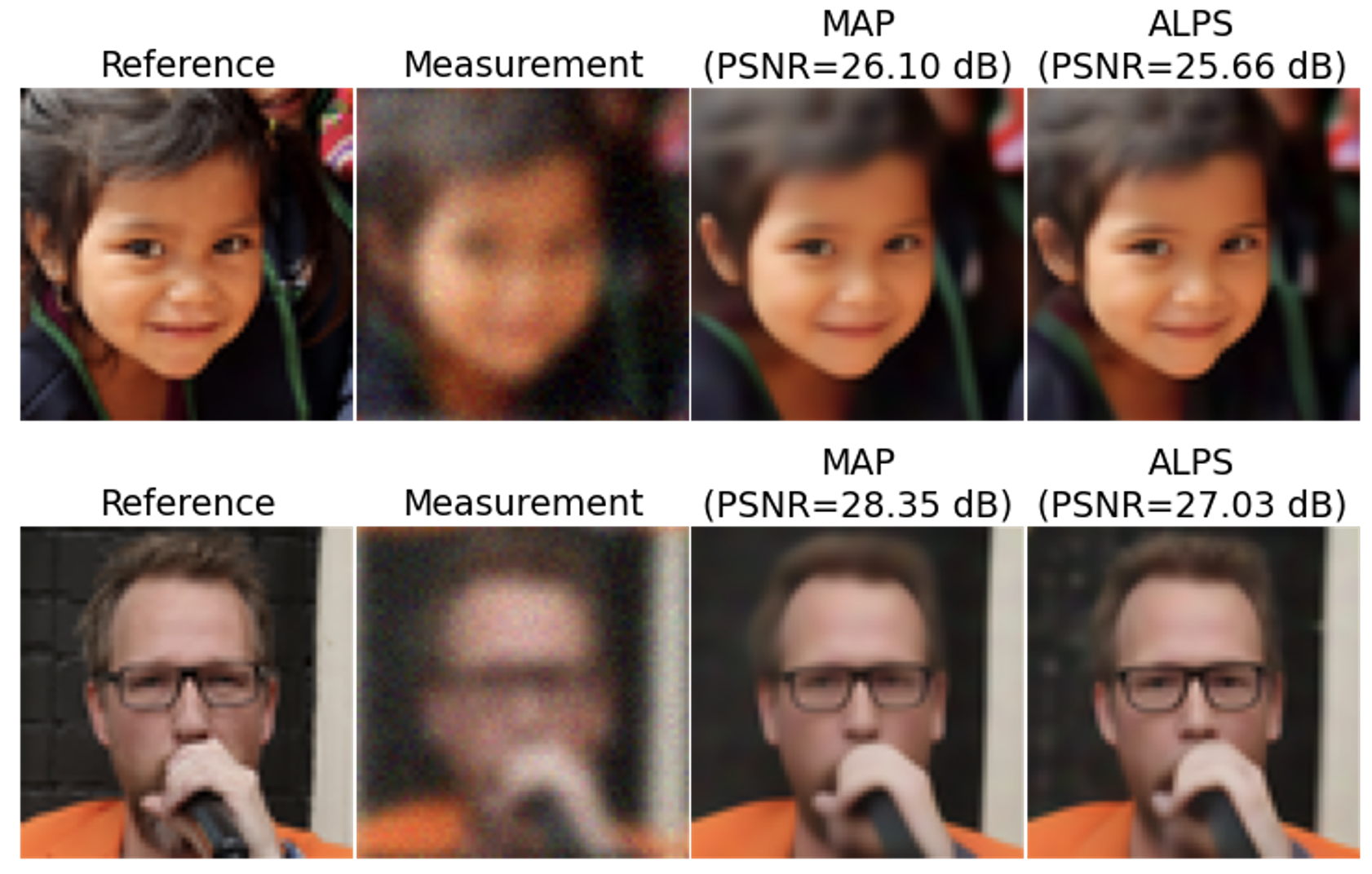}
     \caption{Gaussian Deblurring}
   \end{subfigure}
    \begin{subfigure}[b]{0.48\linewidth}
    \centering
     \includegraphics[width=\linewidth]{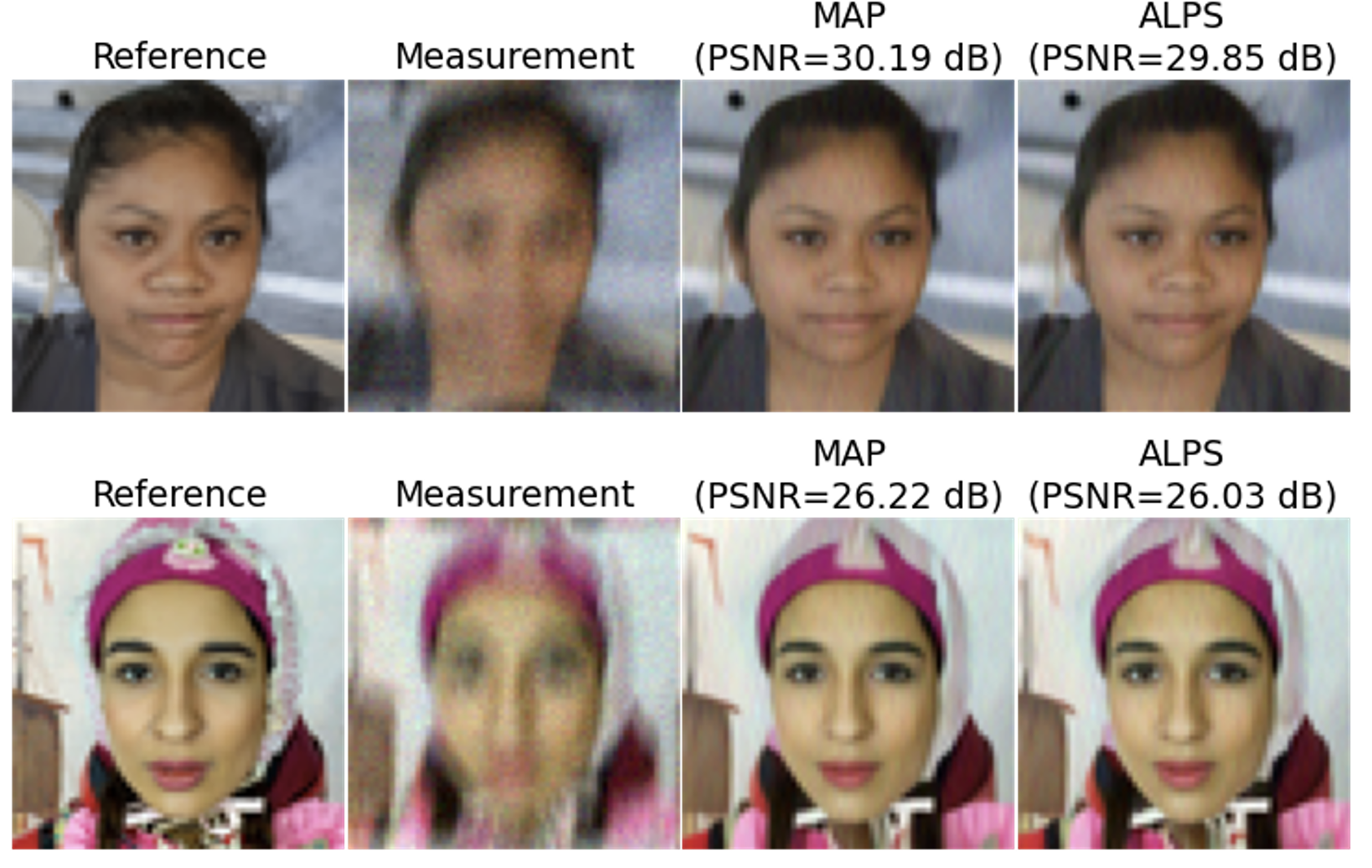}
     \caption{Motion Deblurring}
   \end{subfigure}
\caption{Illustration of performance of multi-scale EBMs on various inverse problems that are evaluated on FFHQ image ($64 \times 64$) dataset. We show the MAP  estimate and a sample from the posterior distribution obtained using the ALPS algorithm. }\label{face_inverse}
\end{figure*}
\begin{figure*}[t!]
   \centering
   \begin{subfigure}[b]{0.65\linewidth}
     \includegraphics[width=1\linewidth]{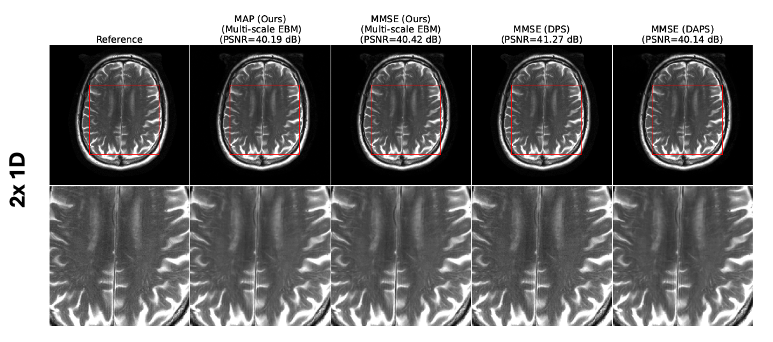}
   \end{subfigure}
    \begin{subfigure}[b]{0.65\linewidth}
     \includegraphics[width=1\linewidth]{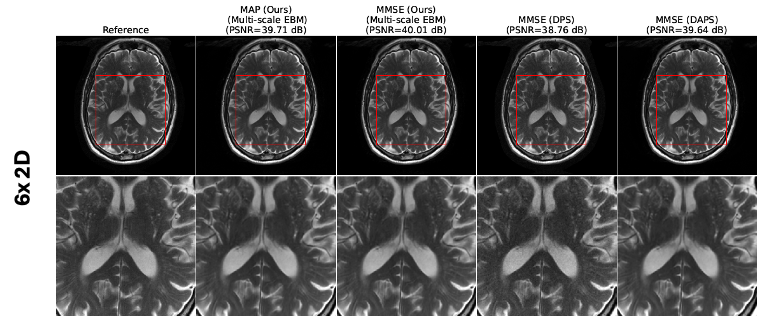}
   \end{subfigure}
 \begin{subfigure}[b]{0.65\linewidth}
     \includegraphics[width=1\linewidth]{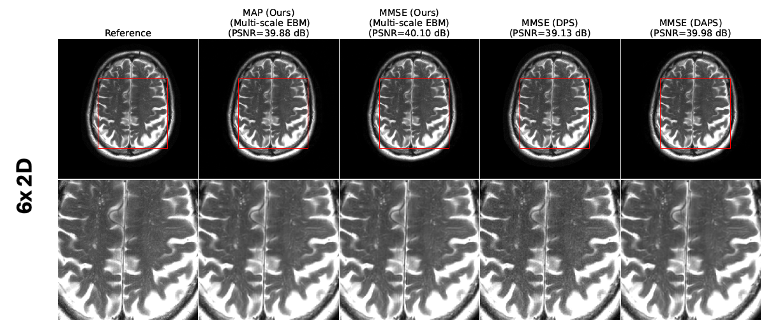}
\end{subfigure}
 \begin{subfigure}[b]{0.65\linewidth}
     \includegraphics[width=1\linewidth]{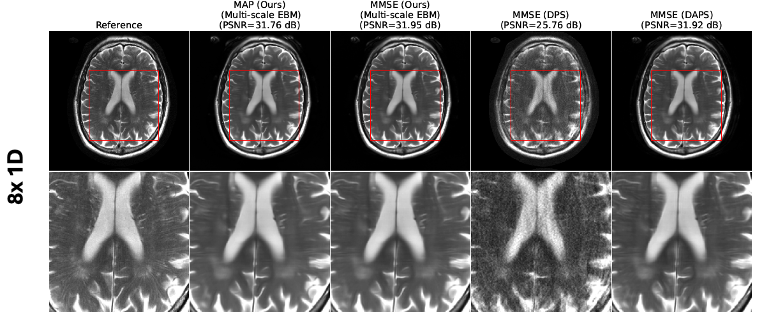}
\end{subfigure}
\caption{{Multi-scale EBM-based algorithms (ALPS and MAP) are compared with diffusion-based algorithms (DAPS and DPS) for MRI image recovery across different accelerations.  }}\label{mri_inverse}
\end{figure*}

\begin{figure*}
 \centering
 \begin{subfigure}[b]{0.65\linewidth}
     \includegraphics[width=\textwidth]{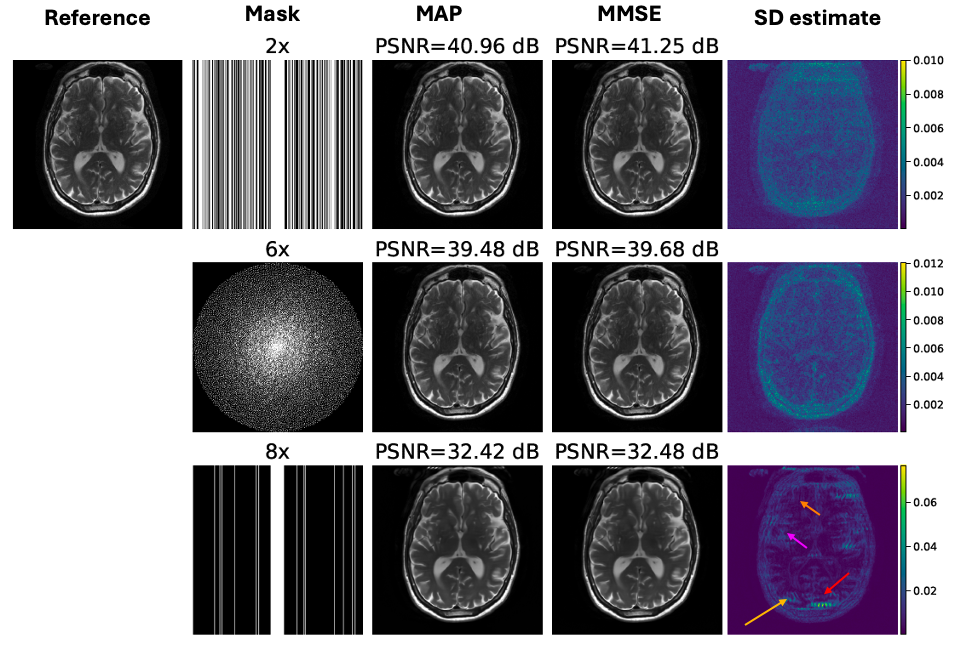}
    \caption{MAP, MMSE, and the standard deviation estimate for different accelerations}
 \end{subfigure}
    \begin{subfigure}[b]{0.65\linewidth}
        \includegraphics[width=\textwidth]{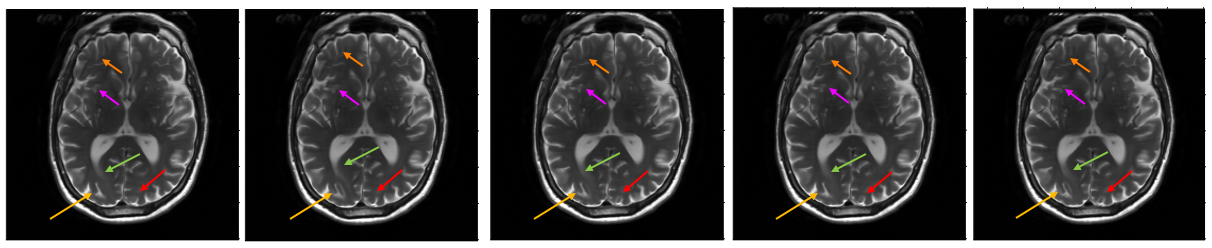}
    \caption{Posterior samples generated using ALPS algorithm for eight-fold acceleration}
    \end{subfigure}
    \caption{(a) Illustration of performance of multi-scale EBM for MRI image recovery for three different accelerations: 2x 1D, 6x 2D, and 8x 1D sampling masks. We show the MAP, MMSE, and the uncertainty estimates. (b) Different posterior consistent samples generated by the ALPS algorithm. The arrow marks highlight the regions which shows differences among the samples. }\label{mri_diff_acc}
\end{figure*}

A larger number of annealing steps provides a more gradual transition between noise levels, rather than making abrupt jumps. Each intermediate scale acts as a bridge between the high-noise initialization and the low-noise posterior, smoothing the optimization landscape and mitigating the risk of getting trapped in sharp local modes. More steps thus make the sampling trajectory more stable and less sensitive to initialization, improving robustness in challenging inverse problems. While more annealing steps can enhance global exploration, the computational cost grows linearly with the number of steps, making it important to balance this factor with the number of Langevin iterations per step for optimal efficiency.

Our experiments show that the impact of the number of ODE steps, which are used in DAPS to transport the samples from the latents to the data distribution,  is less consistent than the choice of $K$ in ALPS. While increasing the number of steps theoretically provides a closer approximation to the samples in the data manifold represented by the current latent sample, the corresponding empirical gains are marginal compared to increasing $K$. As noted in [41], increasing the number of ODE steps help improve the details in the images, often at the expense of PSNR. 

In general, ALPS demonstrates improved convergence than DAPS under comparable settings, requiring fewer function evaluations to achieve similar performance. We attribute this efficiency to ALPS’s problem-aware initialization and null-space noise injection, which together provide a strong starting point and reduce the need for excessive annealing steps. When combined with sufficient Langevin iterations per scale, ALPS achieves a smoother and more direct trajectory toward the posterior, translating to improved reconstruction quality and reduced inference time.

\subsection{Performance comparison with SOTA diffusion solvers}
Table~\ref{performance} summarizes the quantitative results for a range of inverse problems, comparing ALPS with DAPS and DPS using PSNR and SSIM metrics. The MMSE estimates are obtained by computing the average of five different samples. The algorithms are tested on $10$ different slices.  We also report the maximum-a-posteriori (MAP) estimate given by the multi-scale EBM which involves the following update rule (refer to Sec 3.6 in the main paper for more details):
$\bx_{(k+1,i)} 
= \left(\dfrac{\bA^{H}\bA}{\eta^{2}}+\dfrac{L}{{t}^{2}}\bI\right)^{-1}   \left(\dfrac{\bA^{H}\by}{\eta^{2}} + \dfrac{L \bx_{k,t} - \nabla_{\bx_{k,t}} E_{\theta}(\bx_{k,t},{t})}{{t}^{2}}\right) $

The inverse is computed using conjugate gradient method.  For random inpainting tasks, ALPS consistently outperforms DAPS across different corruption levels ($\eta = 0.05, 0.1, 0.5$). The performance gap widens as the noise variance becomes less severe. At high noise levels, the performances are closer. In the less challenging $2\times 1$D setting, DPS is observed to offer a slightly higher PSNR than ALPS and DAPS, but as the complexity increases (e.g., $8\times 1$D and $6\times 2$D), ALPS demonstrates clear superiority. This trend indicates that ALPS scales better with problem complexity, likely due to its ability to exploit forward model structure during sampling. Table \ref{performance} also shows the MAP estimate, which does not require averaging over multiple posterior consistent samples. From the table, it can be observed that MAP performs better than MMSE estimates for the inpainting tasks at all the corruption levels. 

While the table focuses on reconstruction quality, it is important to note that ALPS achieves these results with fewer function evaluations compared to DAPS. This efficiency stems from ALPS's smoother trajectory from initialization to posterior, reducing the need for excessive annealing steps and iterations. Fig. \ref{face_inverse} shows the MAP estimate and posterior sample generated via ALPS algorithm for different inverse problem. We also show the MRI reconstructions at different accelerations in Fig. \ref{mri_inverse}. Fig. \ref{mri_diff_acc} shows the three different estimates that are given by the EBM: MAP, MMSE, and the uncertainty estimates for three different accelerations. One can observe that as the acceleration increases, the uncertainty of the reconstruction also increases.  We also report the average NLPr and negative log-posterior (NLPo) values of the generated samples across different accelerations in Table \ref{mri_values}. One can observe that as the acceleration increases, the mean  NLPr and mean  NLPo values decrease. This occurs because higher acceleration leads to more severe undersampling, making the reconstruction problem increasingly ill-posed. In such settings, the ALPS algorithm relies more heavily on the energy model. As a result, the algorithm outputs reconstructions with lower prior energies, and the posterior energies follow the same trend due to the reduced influence of the negative log-likelihood term.
\begin{table}[h!]
\centering
\begin{tabular}{|p{2cm}|p{3cm}|p{2cm}|}
\hline
\textbf{Acceleration} & \textbf{Mean NLPr}&\textbf{Mean NLPo}\\
&$\bE_{\bth}(\bx;t)/{t^{2}}$ $\times 1e5$ & $C_{t}(\bx)$  $\times 1e5$ \\  \hline
2x & 1.09 & 11.88 \\ \hline
6x & 0.86 & 4.51 \\ \hline
8x & 0.52 & 3.48 \\ \hline
\end{tabular}
\caption{Mean NLPr  and NLPo values for different accelerations for the MRI experiment in Fig. \ref{mri_diff_acc}. We note that as the acceleration increases, the data term becomes less important; the algorithm relies more on the prior and is hence able to converge to solutions with smaller prior values. }\label{mri_values}
\end{table}

\section{EBMs for OOD and model mismatch}
A unique benefit of EBMs is its ability to directly compute the NLPr. By contrast, diffusion models require an ODE integral to compute the NLPr, which is computationally expensive and approximate (involving the Hutchison trace approximation). This NLPr measure can enable a variety of quality checks in inverse problems, which can improve confidence in challenging inverse problems.

\subsection{Alternate quality measure in inverse problems}
The usual quality measures (e.g. PSNR/SSIM/HFEN) that are used in inverse problems require a reference image. Unfortunately, these are not available, when these methods are deployed in practice. The NLPr and NLPo measures can provide a \emph{referenceless} surrogate quality measure. 

\begin{figure*}[!t]
\centering
    \begin{subfigure}[b]{0.48\linewidth}
     \includegraphics[width=\linewidth]{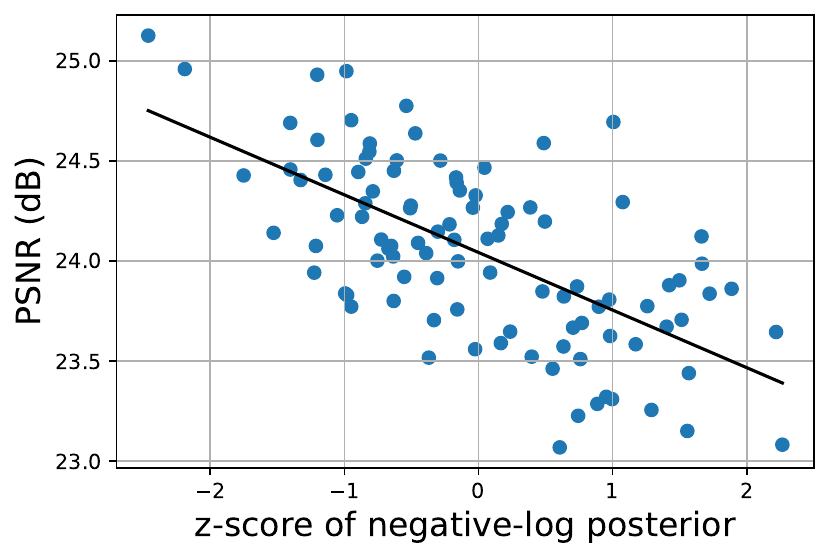}
     \caption{Negative log-posterior}
   \end{subfigure}
       \begin{subfigure}[b]{0.48\linewidth}
     \includegraphics[width=\linewidth]{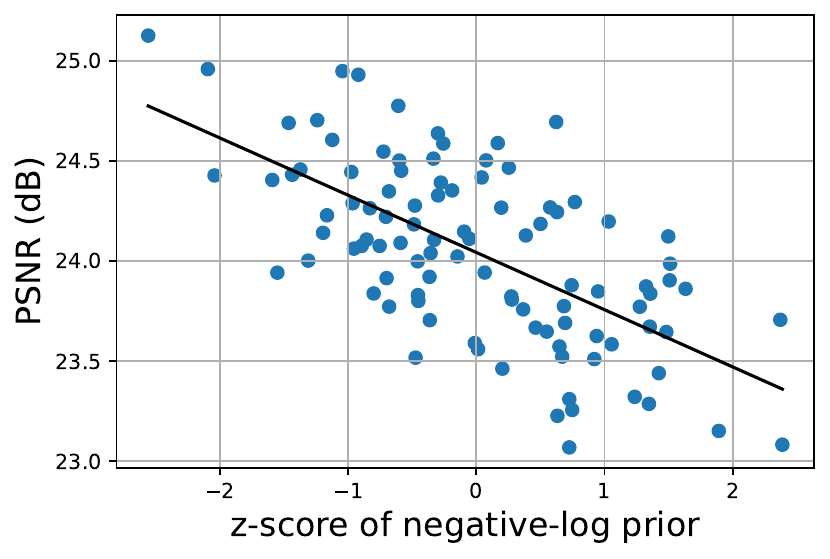}
     \caption{Negative log-prior}
   \end{subfigure}
   \caption{Surrogate for quality measure in the absence of ground truth image using multi-scale EBMs. Scatter plot in (a) and (b) shows that PSNR and the negative log-prior/posterior values have a strong correlation; where images with poor PSNR has high prior/posterior value. }
   \label{psnr-line}
\end{figure*}

We reconstructed 100 different posterior consistent samples using ALPS algorithm for image inpainting task (random) with $\eta=0.05$. The generated samples have different PSNRs (ranging from 23-25 dB) as shown in Fig. \ref{psnr-line}. We computed the z-score of the NLPo and NLPr of the generated samples by first calculating the mean and standard deviation of the NLPr  and NLPo values. We then obtained the z-score by subtracting the corresponding mean from each NLPr and NLPo value and dividing by the standard deviation. Fig. \ref{psnr-line} shows the PSNR vs. z-score of posterior and prior values. From this plot it can be observed that as the posterior/prior value increases the PSNR of the generated samples drop.  This indicates that we can use the log-posterior/prior values to sort the recovered samples based on the quality.

\subsection{Out-of-Domain Detection}
\begin{figure}
    \centering    \includegraphics[width=\linewidth]{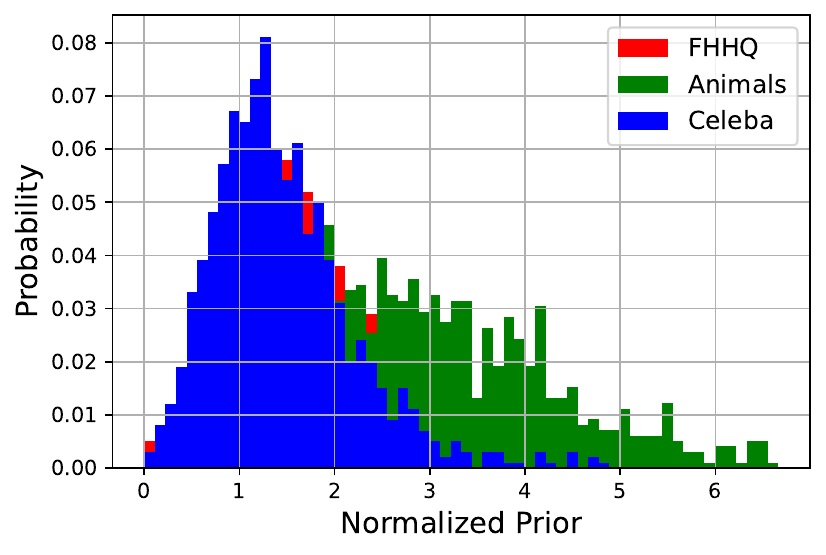}
    \caption{OOD detection using EBMs trained on FFHQ face dataset. The histogram of the normalized prior values computed over $1000$ samples from the dataset of FFHQ, AFHQ (Animals dataset), and another face dataset Celeba is shown in red, green, and blue,respectively. The blue and red curves are centered around lower prior values while the green curves have higher prior values. This enables the usage of EBMs for OOD detection tasks.}
    \label{fig:ood_energy}
\end{figure}

We now apply the trained multi-scale EBM on face images (FFHQ, $64\times64$) and evaluate $E_{\bth}(\bx,t)$ on three different datasets: (i) in-domain face data, (ii) CelebA ($64\times64$), and (iii) animal images (AFHQ). Figure~\ref{fig:ood_energy} shows the histograms of energy values for each dataset, where the $x$-axis represents normalized energy and the $y$-axis represents normalized frequency. The histogram is obtained by computing the energy values for 1000 samples from each dataset. We note that in-domain faces exhibit significantly lower energies, while out-of-domain have higher energies, confirming that EBMs provide a natural OOD score.  The experiments show that EBMs can reliably perform out of domain detection. We also note that the histogram of CelebA and FFHQ significantly overlap which indicates that the trained multi-scale energy model encodes the face manifold.

\begin{figure}
    \centering    \includegraphics[width=\linewidth]{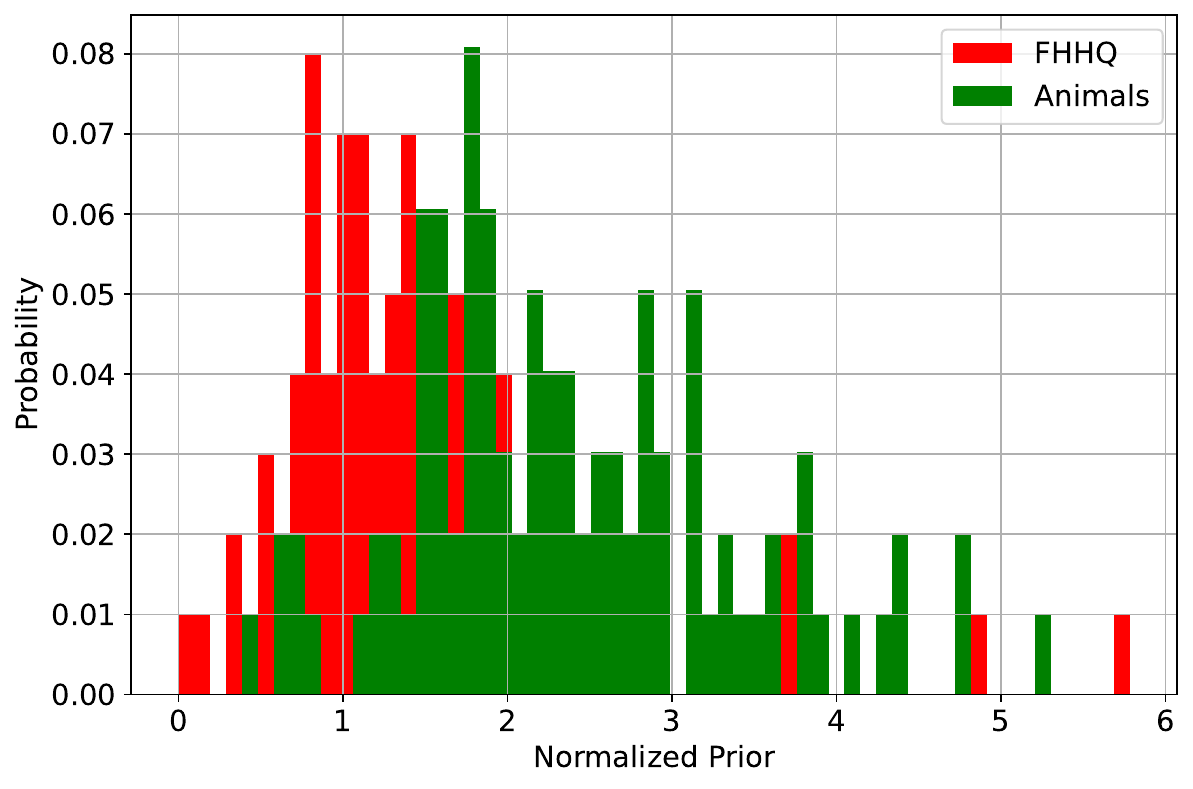}
    \caption{OOD detection using EMBs trained on the FFHQ dataset for the task of recovering Gaussian-blurred ( $\sigma = 0.05$ and $\eta = 0.001$) animal and face images. The red histogram, which shows the normalized prior values for the face recovered images, peaks around 0.9-1, while the green histogram, corresponding to the animal reconstructions, peaks at a larger prior value. }
    \label{fig:ood_inverse}
\end{figure}

\paragraph{OOD Detection in Inverse Problems.}
We further evaluate EBMs in an inverse problem setting (Gaussian deblurring). Using the face-trained EBM, we recover 100 different (a) face images (in-distribution) and (b) animal images (out-of-distribution). Figure~\ref{fig:ood_inverse} compares normalized histograms of NLPr for both cases. From the figure we observe that the recovered face images have low NLPr, whereas the reconstructed animal images exhibit high NLPr. This demonstrates that EBMs can separate plausible reconstructions from implausible ones based on prior consistency.

\subsection{Model Mismatch Sensitivity.}
\begin{figure*}[!t]
\centering
    \begin{subfigure}[b]{0.48\linewidth}
     \includegraphics[width=\linewidth]{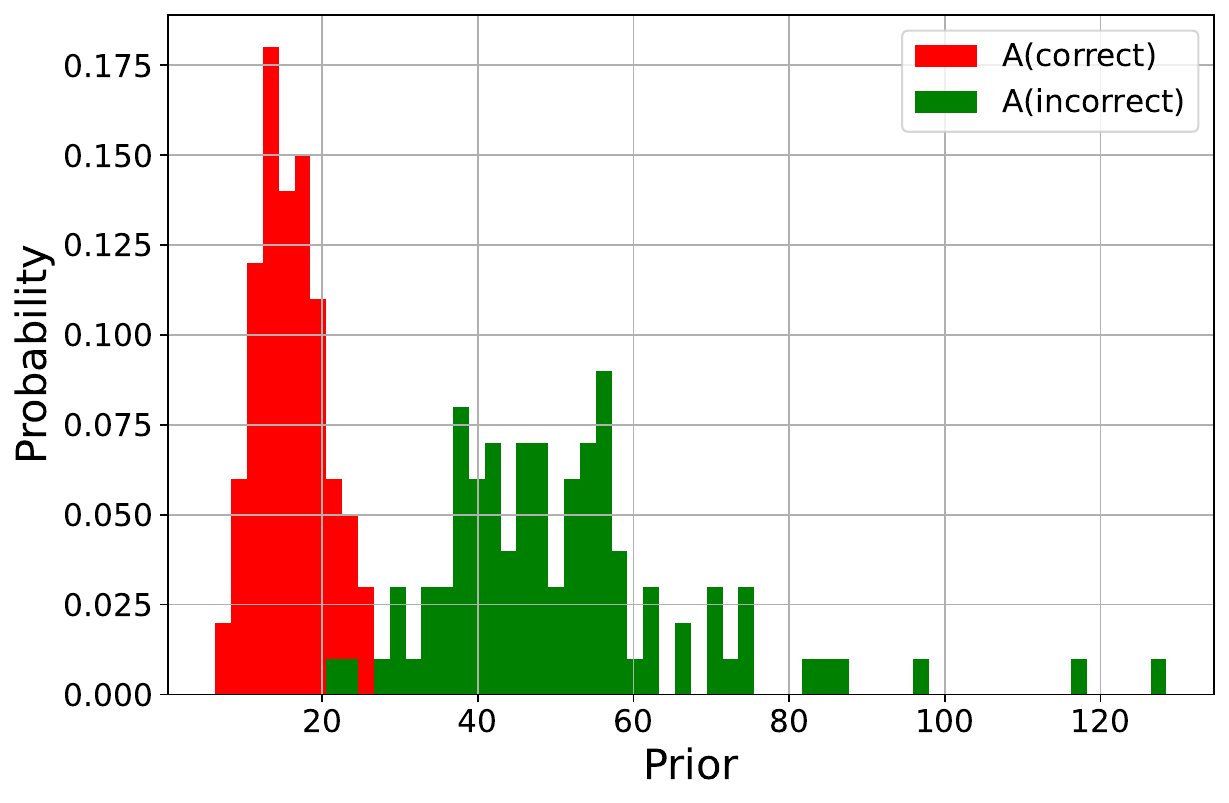}
     \caption{Negative log-prior $\bE_{\bth}(\bx;t)$}
   \end{subfigure}
       \begin{subfigure}[b]{0.48\linewidth}
     \includegraphics[width=\linewidth]{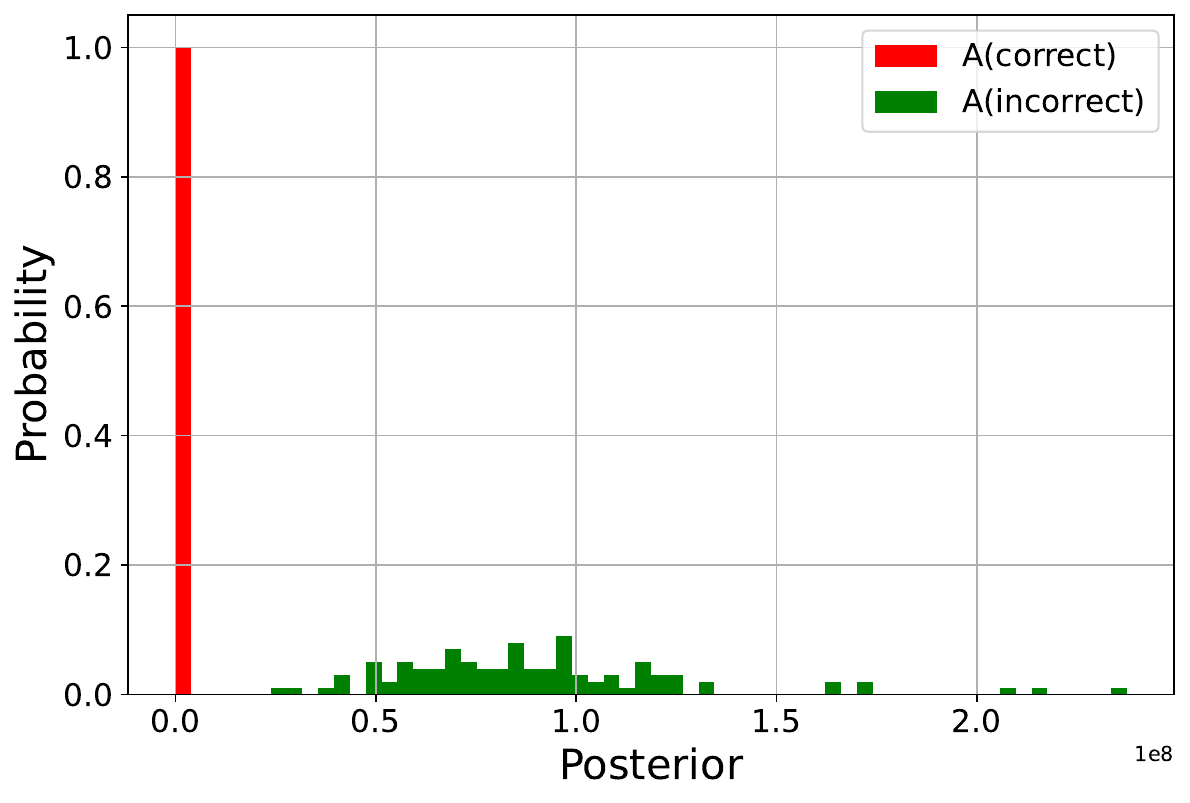}
     \caption{Negative log-posterior $C_{t}(\bx)$ }
   \end{subfigure}
   \caption{Evaluation of multi-scale energy model for model mismatch sensitivity in Gaussian deblurring recovery. The histograms of the recovered images using the correct and incorrect forward operator $\bA$ are shown in red and green, respectively.  Prior and posterior values are not normalized. The images recovered with the incorrect forward operator exhibit higher negative log-prior and negative log-posterior values whereas the reconstructions using the correct forward operator peak at smaller energy values.  }
   \label{fig:model_mismatch}
\end{figure*}

To test whether EBMs reflect forward-model correctness, we simulate measurements using Gaussian blur with $\sigma=0.05$ and attempt recovery assuming $\sigma=0.5$. We compare histograms of NLPr and NLPo values for correct and mismatched models. As shown in Figure~\ref{fig:model_mismatch}, both prior and posterior energies are significantly lower for the images reconstructed using the correct model, indicating that EBMs can serve as a diagnostic tool for model mismatch. This is especially valuable to detect model mismatches (e.g motion, field inhomogeneity distortions etc) in the context of MR image acquisitions; the operator may re-scan the images in this case. 
% \paragraph{Complexity vs. Energy.}
% We hypothesize that prior energy correlates with problem difficulty. For Gaussian deblurring, we vary blur strength ($\sigma\in\{1,3\}$) and plot histograms of log-prior and log-posterior values for recovered samples. Figure~\ref{fig:complexity_energy} shows that prior energies decreases with blur severity, suggesting that EBMs provide a surrogate measure of reconstruction quality without ground truth.  These measures could be used as a surrogate for quality measures. 
% \begin{figure}
%     \centering
%     \includegraphics[width=\linewidth]{cvpr/results/blur_diff_sigma_prior.pdf}
%     \caption{The green and blue clusters show the prior values of the reconstructed images corresponding to Gaussian deblurring tasks with two different sigma values. The green and red curve corresponds to $\sigma=3$ and $\sigma =1 $, respectively.}
%     \label{fig:complexity_energy}
% \end{figure}
% \paragraph{Observations.}
% Across all experiments, EBMs exhibit behaviors expected from a principled probabilistic model:
% \begin{itemize}
%     \item Lower energies for in-domain samples and correct forward models.
%     \item Clear separation of prior energies for OOD cases, even when posteriors are data-consistent.
%     \item Monotonic increase in prior energy with problem difficulty, serving as a proxy for reconstruction fidelity.
% \end{itemize}
% While diffusion models may be used to approximate log prior/posterior values, the complexity in using them is significantly more challenging. \textcolor{red}{It may be good to cite a few}

\section{Preconditioner for MRI inverse problem}
The general preconditioned Langevin dynamics step is specified by:
\begin{equation}
\label{lang}
\bx_{k+1} = \bx_k - \bB ~\nabla_{\bx_{k}} C_{{t}}(\bx_k) + \sqrt{2\bB}\,\xi_k; \quad \forall k=0,\cdots K.
\end{equation}
We chose the preconditioner as $$\bB = \left(\frac{\bA^T\bA}{\eta^2} + \frac{\bI}{t^2} \right)^{-1},$$ which can be computed efficiently for  many linear inverse problems, including inpainting, deblurring, compressed sensing, and single channel MRI. However, in some inverse problems such as multichannel MRI, the second term on the RHS of \eqref{lang} is challenging to compute.  
While the first term in \eqref{lang} can be implemented exactly using conjugate gradients algorithm, the second term can be challenging to implement exactly. We now show that the second term can be approximated using the following approach. 

\subsection{FFT Preconditioner for Multichannel MRI}
 For multichannel MRI with Cartesian undersampling, the system matrix takes the form
 \[
 \bA = \begin{bmatrix} \bS \cdot \mathcal{F}(\bC_1 \cdot x) \\ \vdots \\ \bS \cdot \mathcal{F}(\bC_N \cdot x) \end{bmatrix},
 \]
 where $\mathcal{F}$ denotes the discrete Fourier transform, $\bS$ is a binary sampling mask in k-space, and $\bC_c$ are the coil sensitivity maps. We can approximate $\bA^T\bA \approx \mathcal F^{-1}(S)\mathcal F$. In this case, we choose a diagonal preconditioner in the Fourier domain:
 \[
 \bB = \mathcal{F}^{-1} \cdot \mathrm{diag}\left( \frac{1}{\frac{S(k)}{\eta^2 } + \frac{1}{\sigma_t^2}}  \right) \cdot \mathcal{F},
 \]
 where $S(k)$ is the sampling mask in k-space. 
\begin{figure}[h!]
 \centering
     \includegraphics[width=0.5\textwidth]{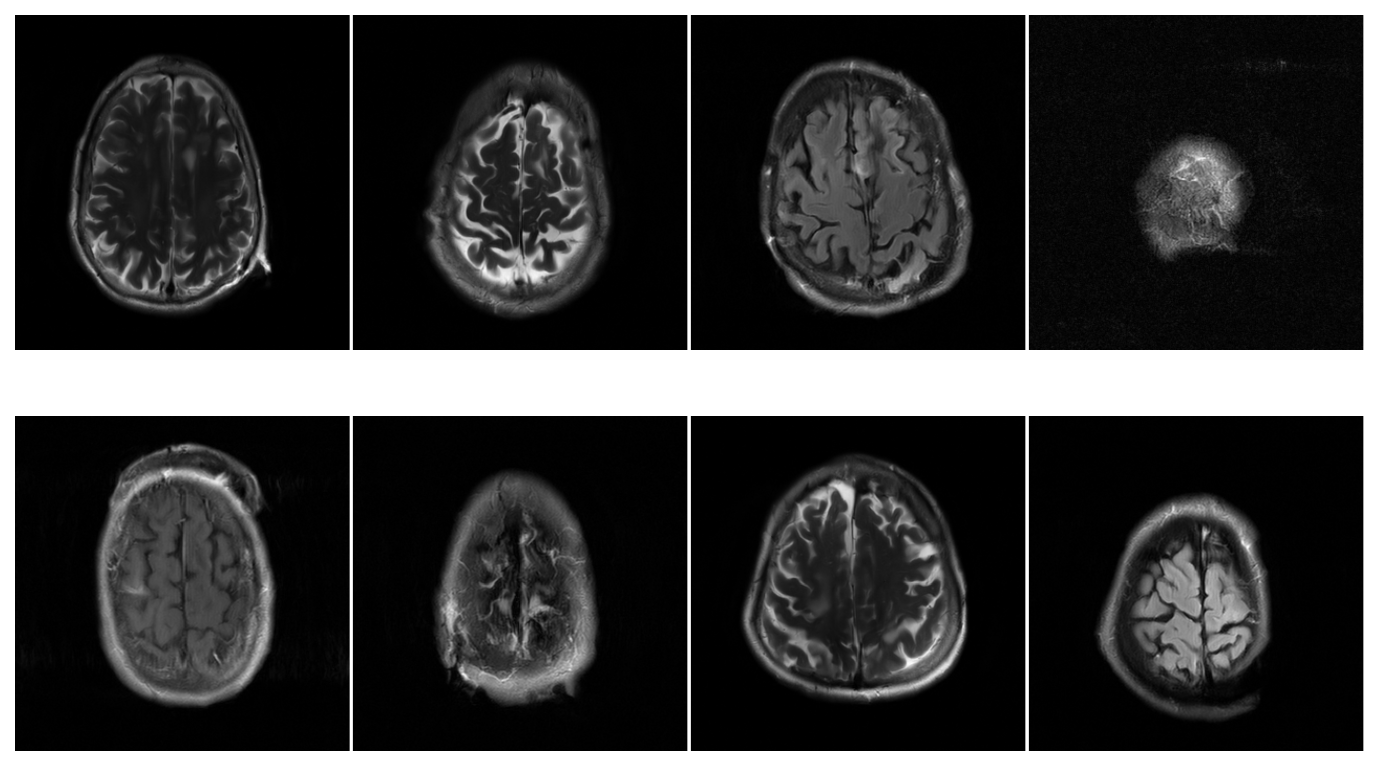}
     \caption{T2-weighted prior samples generated using the conditional multi-scale EBM trained using the score distillation technique. Samples are generated using Heun's method with 18 steps and 35 NFE.}\label{mri_prior}
 \end{figure}

\section{More results on Prior sampling} 
\begin{figure*}[h!]
   \centering
   \begin{subfigure}[b]{0.32\linewidth}
     \includegraphics[width=1\linewidth]{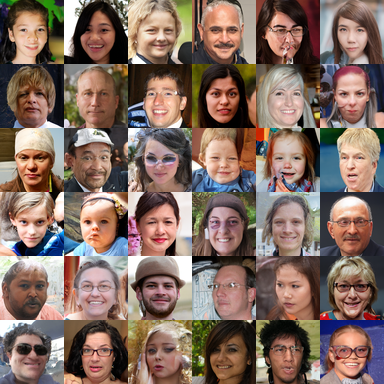}
     \caption{FID 3.76 NFE:79}
   \end{subfigure}
   % \begin{subfigure}[b]{0.32\linewidth}
   %   \includegraphics[width=1\linewidth]{cvpr/results/facedataset_rectifieddistillation.png}
   %   \caption{FID 7.08 NFE:39}
   % \end{subfigure}
   % \begin{subfigure}[b]{0.32\linewidth}
   %   \includegraphics[width=1\linewidth]{cvpr/results/facedataset_singlepotential.png}
   %   \caption{FID 4.07 NFE:79}
   % \end{subfigure}
   \caption {Unconditional generation of multi-scale EBM trained using score distillation }
   \vspace{-5pt}
   \label{face_image}
 \end{figure*}

We show the prior samples generated by training the distilled multi-scale EBMs on the FFHQ dataset ($64 \times 64$) and T2-weighted MRI images ($324 \times 324$) in Fig. \ref{face_image} and Fig. \ref{mri_prior}, respectively. Table \ref{tab:fid_face_uncond} compares the FID score  with other methods.

 \section{Experiment details}
\subsection{Training Protocol}
\textbf{Hardware}: All experiments were conducted on A100 A40 GPUs on a single node of upto 4 GPUs. \\
\textbf{Training parameters}: The optimal weights of the  energy models where obtained by using  Adam optimizer with the learning rate of 0.005. We did not perform augmentation on the training dataset when using score distillation technique to train the EBM. In the case of transport distillation, 50k noise and image pairs were generated using the pre-trained diffusion model.\\

\subsection{Architecture details}
For natural images, we used the \emph{SongNet} NCSNPP  architecture for the denoiser $D_\theta(\bx;t)$; while for MRI we used  \emph{SongNet} DDPM architecture. The details of the network is listed in Table \ref{architecture}. The noise predictor used to realize the single potential is a very small model. It consists of two convolutional layers, global average pooling, and two linear layers outputting a scalar per image.
\begin{table*}[h!]
\caption{Network parameters for CIFAR-10 (NCSN++), FFHQ-64 (NCSN++), and MRI-384 (DDPM++).}
\centering
\scriptsize
\setlength{\tabcolsep}{6pt}
\begin{tabular}{lccc}

\toprule
\textbf{Parameter} &
\textbf{CIFAR-10 (NCSN++)} &
\textbf{FFHQ-64 (NCSN++)} &
\textbf{MRI-324 (DDPM++)} \\
\midrule

normalization          & GroupNorm          & GroupNorm            & GroupNorm \\
nonlinearity           & swish              & swish                & swish \\
nf                     & 128                & 128                  & 128 \\
ch\_mult               & [1,2,2]            & [1,2,2,2]            & [1,2,2,2,2,2] \\
num\_res\_blocks       & 4                  & 4                    & 2 \\
attn\_resolutions      & (16)               & (16)                 & (12) \\

resamp\_with\_conv     & True               & True                 & True \\
fir                    & True               & True                 & True \\
fir\_kernel            & [1,3,3,1]          & [1,3,3,1]            & [1,1] \\

skip\_rescale          & True               & True                 & True \\
resblock\_type         & biggan             & biggan               & ddpmpp \\

progressive            & none               & none                 & none \\
progressive\_input     & residual           & residual             & standard \\
progressive\_combine   & sum                & sum                  & sum \\

attention\_type        & ddpm               & ddpm                 & ddpm \\

embedding\_type        & fourier            & fourier              & positional \\
channel\_mult\_emb     & 4                  & 4                    & 4 \\
channel\_mult\_noise   & 2                  & 2                    & 1 \\

init\_scale            & 0.0                & 0.0                  & 0.0 \\
fourier\_scale         & 16                 & 16                   & --- \\

conv\_size             & 3                  & 3                    & 3 \\
num\_scales            & 15                 & 18                   & 30 \\

dropout                & 0.10               & 0.10                 & 0.10 \\
label\_dropout         & 0.0                & 0.0                  & 0.0 \\

encoder\_type          & residual           & residual             & standard \\
decoder\_type          & standard           & standard             & standard \\
resample\_filter       & [1,3,3,1]          & [1,3,3,1]            & [1,1] \\

\bottomrule
\end{tabular}

\label{architecture}
\end{table*}

\begin{figure*}[t]
\centering
\begin{minipage}[t]{\columnwidth}
\centering
\small
\captionof{table}{FID score on Face dataset}
\label{tab:fid_face_uncond}

\setlength{\tabcolsep}{5.5pt}
\renewcommand{\arraystretch}{1.08}
\begin{adjustbox}{max width=\columnwidth}
\begin{tabular}{@{}l r r@{}}
\toprule
\textbf{Models} & \textbf{Steps} &\textbf{FID} $\downarrow$ \\
\midrule
EDM Diffusion~[19] & 79 &\;\,2.53\\
Distilled-EnergyDiffusion~[35] & 79 &\;\,2.64\\
E-DSM~[35] & 79 &\;\,4.57\\
\midrule
\textbf{EBM-Score Distillation (Ours)} & \textbf{79} &\textbf{\;\,3.76} \\
% \rowcolor{gray!20}\textbf{EBM-Transport Distillation (Ours)} & \textbf{39} &\textbf{\;\,7.08} \\
% \textbf{EBM-Transport Distillation (Ours)} & \textbf{11} &\textbf{\;\,8.61} \\
% \textbf{EBM-Single potential (Ours)} & \textbf{79} &\textbf{\;\,4.07} \\
\bottomrule
\end{tabular}
\end{adjustbox}
\end{minipage}
\end{figure*}

{
    \small
    \bibliographystyle{ieeenat_fullname}
    \bibliography{main}

\begin{thebibliography}{41}
\providecommand{\natexlab}[1]{#1}
\providecommand{\url}[1]{\texttt{#1}}
\expandafter\ifx\csname urlstyle\endcsname\relax
  \providecommand{\doi}[1]{doi: #1}\else
  \providecommand{\doi}{doi: \begingroup \urlstyle{rm}\Url}\fi

\bibitem[Aali et~al.(2025)Aali, Daras, Levac, Kumar, Dimakis, and Tamir]{aali2025ambient}
Asad Aali, Giannis Daras, Brett Levac, Sidharth Kumar, Alexandros~G. Dimakis, and Jonathan~I. Tamir.
\newblock Ambient diffusion posterior sampling: Solving inverse problems with diffusion models trained on corrupted data.
\newblock In \emph{International Conference on Learning Representations (ICLR)}, 2025.

\bibitem[Arbel et~al.(2020)Arbel, Zhou, and Gretton]{arbel2020generalized}
Michael Arbel, Liang Zhou, and Arthur Gretton.
\newblock Generalized energy based models.
\newblock \emph{arXiv preprint arXiv:2003.05033}, 2020.

\bibitem[Baldassari et~al.(2025)Baldassari, Garnier, Solna, and de~Hoop]{baldassari2025preconditioned}
Lorenzo Baldassari, Josselin Garnier, Knut Solna, and Maarten~V. de Hoop.
\newblock Preconditioned langevin dynamics with score-based generative models for infinite-dimensional linear bayesian inverse problems.
\newblock \emph{arXiv preprint arXiv:2505.18276}, 2025.

\bibitem[Chand and Jacob(2024)]{muse}
Jyothi~Rikhab Chand and Mathews Jacob.
\newblock Multi-scale energy (muse) framework for inverse problems in imaging.
\newblock \emph{IEEE Transactions on Computational Imaging}, 10:\penalty0 1250--1265, 2024.

\bibitem[Chung et~al.(2023)Chung, Kim, McCann, Klasky, and Ye]{dps}
Hyungjin Chung, Jeongsol Kim, Michael~T. McCann, Marc~L. Klasky, and Jong~Chul Ye.
\newblock Diffusion posterior sampling for general noisy inverse problems.
\newblock In \emph{International Conference on Learning Representations (ICLR)}, 2023.
\newblock Spotlight.

\bibitem[Daras et~al.(2024)Daras, Chung, Lai, Mitsufuji, Ye, Milanfar, Dimakis, and Delbracio]{daras2024survey}
Giannis Daras, Hyungjin Chung, Chieh{-}Hsin Lai, Yuki Mitsufuji, Jong~Chul Ye, Peyman Milanfar, Alexandros~G. Dimakis, and Mauricio Delbracio.
\newblock A survey on diffusion models for inverse problems.
\newblock \emph{arXiv preprint arXiv:2410.00083}, 2024.

\bibitem[Du and Mordatch(2019)]{du2019implicit}
Yilun Du and Igor Mordatch.
\newblock Implicit generation and modeling with energy based models.
\newblock In \emph{Advances in Neural Information Processing Systems (NeurIPS)}, 2019.

\bibitem[Du et~al.(2020)Du, Li, Tenenbaum, and Mordatch]{du2020improved}
Yilun Du, Shuang Li, Joshua Tenenbaum, and Igor Mordatch.
\newblock Improved contrastive divergence training of energy based models.
\newblock \emph{arXiv preprint arXiv:2012.01316}, 2020.

\bibitem[Feng et~al.(2024)Feng, Yang, An, et~al.]{feng2024rdd}
Weilun Feng, Chuanguang Yang, Zhulin An, et~al.
\newblock Relational diffusion distillation for efficient image generation.
\newblock \emph{arXiv preprint arXiv:2410.07679}, 2024.

\bibitem[Gao et~al.(2020{\natexlab{a}})Gao, Nijkamp, Kingma, Xu, Dai, and Wu]{gao2020flow}
Ruiqi Gao, Erik Nijkamp, Diederik~P Kingma, Zhen Xu, Andrew~M Dai, and Ying~Nian Wu.
\newblock Flow contrastive estimation of energy-based models.
\newblock In \emph{Proceedings of the IEEE/CVF Conference on Computer Vision and Pattern Recognition}, pages 7518--7528, 2020{\natexlab{a}}.

\bibitem[Gao et~al.(2020{\natexlab{b}})Gao, Song, Poole, Wu, and Kingma]{gao2020learning}
Ruiqi Gao, Yang Song, Ben Poole, Ying~Nian Wu, and Diederik~P Kingma.
\newblock Learning energy-based models by diffusion recovery likelihood.
\newblock \emph{arXiv preprint arXiv:2012.08125}, 2020{\natexlab{b}}.

\bibitem[Grathwohl et~al.(2019)Grathwohl, Wang, Jacobsen, Duvenaud, Norouzi, and Swersky]{grathwohl2019your}
Will Grathwohl, Kuan-Chieh Wang, J{\"o}rn-Henrik Jacobsen, David Duvenaud, Mohammad Norouzi, and Kevin Swersky.
\newblock Your classifier is secretly an energy based model and you should treat it like one.
\newblock \emph{arXiv preprint arXiv:1912.03263}, 2019.

\bibitem[Grathwohl et~al.(2020)Grathwohl, Kelly, Hashemi, Norouzi, Swersky, and Duvenaud]{grathwohl2020no}
Will Grathwohl, Jacob Kelly, Milad Hashemi, Mohammad Norouzi, Kevin Swersky, and David Duvenaud.
\newblock No mcmc for me: Amortized sampling for fast and stable training of energy-based models.
\newblock \emph{arXiv preprint arXiv:2010.04230}, 2020.

\bibitem[Guo et~al.(2023)Guo, Ma, Jiang, Yuan, Yu, and Luo]{guo2023egc}
Qiushan Guo, Chuofan Ma, Yi Jiang, Zehuan Yuan, Yizhou Yu, and Ping Luo.
\newblock Egc: Image generation and classification via a diffusion energy-based model.
\newblock In \emph{Proceedings of the IEEE/CVF International Conference on Computer Vision}, pages 22952--22962, 2023.

\bibitem[Habring et~al.(2025)Habring, Holler, Pock, and Zach]{habring2025energy}
Andreas Habring, Martin Holler, Thomas Pock, and Martin Zach.
\newblock Energy-based models for inverse imaging problems.
\newblock \emph{arXiv preprint arXiv:2507.12432}, 2025.

\bibitem[Han et~al.(2020)Han, Nijkamp, Zhou, Pang, Zhu, and Wu]{han2020joint}
Tian Han, Erik Nijkamp, Linqi Zhou, Bo Pang, Song-Chun Zhu, and Ying~Nian Wu.
\newblock Joint training of variational auto-encoder and latent energy-based model.
\newblock In \emph{Proceedings of the IEEE/CVF conference on computer vision and pattern recognition}, pages 7978--7987, 2020.

\bibitem[Hill et~al.(2022)Hill, Nijkamp, Mitchell, Pang, and Zhu]{hill2022learning}
Mitch Hill, Erik Nijkamp, Jonathan Mitchell, Bo Pang, and Song-Chun Zhu.
\newblock Learning probabilistic models from generator latent spaces with hat ebm.
\newblock \emph{Advances in Neural Information Processing Systems}, 35:\penalty0 928--940, 2022.

\bibitem[Hurault et~al.(2022)Hurault, Leclaire, and Papadakis]{hurault2022gradient}
Samuel Hurault, Arthur Leclaire, and Nicolas Papadakis.
\newblock Gradient step denoiser for convergent plug-and-play.
\newblock In \emph{International Conference on Learning Representations (ICLR)}, 2022.

\bibitem[Karras et~al.(2022)Karras, Aittala, Aila, and Laine]{karras2022edm}
Tero Karras, Miika Aittala, Timo Aila, and Samuli Laine.
\newblock Elucidating the design space of diffusion-based generative models.
\newblock In \emph{Advances in Neural Information Processing Systems (NeurIPS)}, 2022.

\bibitem[Kingma et~al.(2021)Kingma, Salimans, Poole, and Ho]{kingma2021variational}
Diederik~P. Kingma, Tim Salimans, Ben Poole, and Jonathan Ho.
\newblock Variational diffusion models.
\newblock In \emph{Advances in Neural Information Processing Systems (NeurIPS)}, pages 21696--21707, 2021.

\bibitem[Kirkpatrick et~al.(1983)Kirkpatrick, Gelatt, and Vecchi]{kirkpatrick1983optimization}
S. Kirkpatrick, C.~D. Gelatt, and M.~P. Vecchi.
\newblock Optimization by simulated annealing.
\newblock \emph{Science}, 220\penalty0 (4598):\penalty0 671--680, 1983.

\bibitem[LeCun et~al.(2006)LeCun, Chopra, Hadsell, Ranzato, and Huang]{lecun2006tutorial}
Yann LeCun, Sumit Chopra, Raia Hadsell, Marc'Aurelio Ranzato, and Fu-Jie Huang.
\newblock A tutorial on energy-based models.
\newblock \emph{Journal of Energy}, 2006.

\bibitem[Lee et~al.(2023)Lee, Jeong, Park, and Shin]{lee2023guiding}
Hankook Lee, Jongheon Jeong, Sejun Park, and Jinwoo Shin.
\newblock Guiding energy-based models via contrastive latent variables.
\newblock \emph{arXiv preprint arXiv:2303.03023}, 2023.

\bibitem[Li et~al.(2023)Li, Chen, and Sommer]{li2023multiscale}
Zengyi Li, Yubei Chen, and Friedrich~T. Sommer.
\newblock Learning energy-based models in high-dimensional spaces with multiscale denoising-score matching.
\newblock \emph{Entropy}, 25\penalty0 (10):\penalty0 1367, 2023.

\bibitem[Meng et~al.(2023)Meng, Rombach, Gao, Kingma, et~al.]{meng2023distillation}
Chenlin Meng, Robin Rombach, Ruiqi Gao, Diederik Kingma, et~al.
\newblock On distillation of guided diffusion models.
\newblock \emph{CVPR}, 2023.

\bibitem[Nijkamp et~al.(2020)Nijkamp, Gao, Sountsov, Vasudevan, Pang, Zhu, and Wu]{nijkamp2020learning}
Erik Nijkamp, Ruiqi Gao, Pavel Sountsov, Srinivas Vasudevan, Bo Pang, Song-Chun Zhu, and Ying~Nian Wu.
\newblock Learning energy-based model with flow-based backbone by neural transport mcmc.
\newblock \emph{arXiv preprint arXiv:2006.06897}, 2, 2020.

\bibitem[Pang et~al.(2020)Pang, Han, Nijkamp, Zhu, and Wu]{pang2020learning}
Bo Pang, Tian Han, Erik Nijkamp, Song-Chun Zhu, and Ying~Nian Wu.
\newblock Learning latent space energy-based prior model.
\newblock \emph{Advances in Neural Information Processing Systems}, 33:\penalty0 21994--22008, 2020.

\bibitem[Salimans and Ho(2022)]{salimans2022progressive}
Tim Salimans and Jonathan Ho.
\newblock Progressive distillation for fast sampling of diffusion models.
\newblock In \emph{International Conference on Learning Representations (ICLR)}, 2022.

\bibitem[Song et~al.(2023{\natexlab{a}})Song, Mardani, Vahdat, and Kautz]{song2023potential}
Jiaming Song, Morteza Mardani, Arash Vahdat, and Jan Kautz.
\newblock It has got potential: Denoising score matching as a regularizer for inverse problems.
\newblock In \emph{International Conference on Learning Representations (ICLR)}, 2023{\natexlab{a}}.

\bibitem[Song and Ermon(2020)]{song2020score}
Yang Song and Stefano Ermon.
\newblock Improved techniques for training score-based generative models.
\newblock In \emph{Advances in Neural Information Processing Systems (NeurIPS)}, pages 12438--12448, 2020.

\bibitem[Song et~al.(2021)Song, Sohl-Dickstein, Kingma, Kumar, Ermon, and Poole]{song2021score}
Yang Song, Jascha Sohl-Dickstein, Diederik Kingma, Abhishek Kumar, Stefano Ermon, and Ben Poole.
\newblock Score-based generative modeling through stochastic differential equations.
\newblock \emph{International Conference on Learning Representations (ICLR)}, 2021.

\bibitem[Song et~al.(2023{\natexlab{b}})Song, Dhariwal, Chen, and Sutskever]{song2023consistency}
Yang Song, Prafulla Dhariwal, Mark Chen, and Ilya Sutskever.
\newblock Consistency models.
\newblock In \emph{International Conference on Machine Learning (ICML)}, 2023{\natexlab{b}}.

\bibitem[Séguin and Kressner(2023)]{seguin2023continuation}
Axel Séguin and Daniel Kressner.
\newblock Continuation methods for riemannian optimization.
\newblock \emph{arXiv preprint arXiv:2106.08839}, 2023.

\bibitem[Thornton et~al.(2025)Thornton, Bethune, Zhang, Bradley, Nakkiran, and Zhai]{thornton2025energy}
James Thornton, Louis Bethune, Ruixiang Zhang, Arwen Bradley, Preetum Nakkiran, and Shuangfei Zhai.
\newblock Composition and control with distilled energy diffusion models and sequential monte carlo.
\newblock \emph{arXiv preprint arXiv:2502.12786}, 2025.

\bibitem[Xiao and Han(2022)]{xiao2022adaptive}
Zhisheng Xiao and Tian Han.
\newblock Adaptive multi-stage density ratio estimation for learning latent space energy-based model.
\newblock \emph{Advances in Neural Information Processing Systems}, 35:\penalty0 21590--21601, 2022.

\bibitem[Xiao et~al.(2020)Xiao, Kreis, Kautz, and Vahdat]{xiao2020vaebm}
Zhisheng Xiao, Karsten Kreis, Jan Kautz, and Arash Vahdat.
\newblock Vaebm: A symbiosis between variational autoencoders and energy-based models.
\newblock \emph{arXiv preprint arXiv:2010.00654}, 2020.

\bibitem[Xie et~al.(2018)Xie, Lu, Gao, and Wu]{coopnets}
Jianwen Xie, Yang Lu, Ruiqi Gao, and Ying~Nian Wu.
\newblock Cooperative learning of energy-based model and latent variable model via mcmc teaching.
\newblock In \emph{AAAI Conference on Artificial Intelligence (AAAI)}, 2018.

\bibitem[Xie et~al.(2021)Xie, Zheng, and Li]{xie2021learning}
Jianwen Xie, Zilong Zheng, and Ping Li.
\newblock Learning energy-based model with variational auto-encoder as amortized sampler.
\newblock In \emph{Proceedings of the AAAI Conference on Artificial Intelligence}, pages 10441--10451, 2021.

\bibitem[Xie et~al.(2022)Xie, Zhu, Li, and Li]{xie2022tale}
Jianwen Xie, Yaxuan Zhu, Jun Li, and Ping Li.
\newblock A tale of two flows: Cooperative learning of langevin flow and normalizing flow toward energy-based model.
\newblock \emph{arXiv preprint arXiv:2205.06924}, 2022.

\bibitem[Zhang et~al.(2025)Zhang, Chu, Berner, Meng, Anandkumar, and Song]{daps}
Bingliang Zhang, Wenda Chu, Julius Berner, Chenlin Meng, Anima Anandkumar, and Yang Song.
\newblock Improving diffusion inverse problem solving with decoupled noise annealing.
\newblock \emph{arXiv preprint arXiv:2511.XXXXX}, 2025.
\newblock Proceedings of CVPR 2025.

\bibitem[Zhao et~al.(2020)Zhao, Xie, and Li]{zhao2020learning}
Yang Zhao, Jianwen Xie, and Ping Li.
\newblock Learning energy-based generative models via coarse-to-fine expanding and sampling.
\newblock In \emph{International Conference on Learning Representations}, 2020.

\end{thebibliography}
}


\end{document}